\documentclass{sig-alternate}
\usepackage{times}
\usepackage{graphicx}
\usepackage{latexsym}
\usepackage{url}
\usepackage{algorithm}
\usepackage{algorithmic}
\usepackage{color}
\usepackage{xspace}
\usepackage{epsfig}
\usepackage{alltt}
\usepackage{balance}  
\usepackage[sort&compress]{natbib}
\usepackage[normalem]{ulem}
\usepackage{listings}
\usepackage{courier}
\usepackage{local}
\usepackage[pdfpagelabels]{hyperref}

\newcommand{\reln}[1]{\dlrel{#1}}
\newenvironment{Dlog}{\vspace{-1mm}\begin{alltt}\scriptsize}{\end{alltt}\vspace{-1mm}}

\newenvironment{compactList}{\begin{list}{$\bullet$}{\leftmargin=0.5em}
    {   \setlength{\itemsep}{0mm}
        \setlength{\topsep}{0mm}}
    }
{\end{list}}

\makeatletter
\let\@copyrightspace\relax
\makeatother

\newenvironment{compactEnumerate}{\begin{enumerate}{\leftmargin=0.5em}
    {   \setlength{\itemsep}{0mm}
        \setlength{\topsep}{0mm}}
    }
{\end{enumerate}}

\newcommand{\Section}[1]{\vspace{-1mm}\section{#1}\vspace{-1mm}}

\newcommand{\Subsection}[1]{\vspace{-1mm}\subsection{#1}\vspace{-1mm}}
\newcommand{\Subsubsection}[1]{\vspace{-1mm}\subsubsection{#1}\vspace{-1mm}}

\begin{document}


\title{Enabling Incremental Query Re-Optimization}

\numberofauthors{3}

\author{
\alignauthor Mengmeng Liu\\
       \affaddr{University of Pennsylvania}\\
       \affaddr{3330 Walnut Street}\\
       \affaddr{Philadelphia, PA 19104}\\
       \email{mengmeng@cis.upenn.edu}
\alignauthor Zachary G. Ives \\
       \affaddr{University of Pennsylvania}\\
       \affaddr{3330 Walnut Street}\\
       \affaddr{Philadelphia, PA 19104}\\
       \email{zives@cis.upenn.edu}
\alignauthor Boon Thau Loo\\
       \affaddr{University of Pennsylvania}\\
       \affaddr{3330 Walnut Street}\\
       \affaddr{Philadelphia, PA 19104}\\
       \email{boonloo@cis.upenn.edu}
}


\maketitle

\begin{abstract}
  As declarative query processing techniques expand in
  scope --- to the Web, data streams, network routers,
  and cloud platforms --- there is an increasing need for
  \emph{adaptive} query processing techniques that can re-plan in the
  presence of failures or unanticipated performance changes.  A status
  update on the data distributions or the compute nodes may have
  significant repercussions on the choice of which query plan should
  be running.  Ideally, new system architectures would be able
  to make \emph{cost-based decisions} about reallocating work,
  migrating data, etc., and react quickly as real-time status information 
  becomes available. Existing cost-based query optimizers are not 
  incremental in nature, and must be run ``from scratch'' upon each 
  status or cost update.  Hence, they generally result in adaptive schemes
  that can only react slowly to updates.

  An open question has been whether it is possible to build a cost-based
  \emph{re-optimization} architecture for adaptive query processing in
  a streaming or repeated query execution environment, e.g., by 
  \emph{incrementally} updating optimizer state given new cost 
  information.  We show that this can be achieved beneficially, 
  especially for stream processing workloads.
  Our techniques build upon the recently proposed approach of
  formulating query plan enumeration as a set of \emph{recursive
    datalog queries}; we develop a variety of novel optimization
  approaches to ensure effective pruning in both static and
  incremental cases. We implement our solution within an existing
  research query processing system, and show that it effectively supports
  cost-based initial optimization as well as frequent adaptivity.
%
\end{abstract}

\Section{Introduction}
\label{sec:intro}

The problem of supporting rapid \emph{adaptation} to runtime conditions during
query processing --- \emph{adaptive query processing}~\cite{aqp-survey} --- is of
increasing importance in today's data processing environments.  Consider declarative cloud data processing systems~\cite{boom-analytics,asterix,DBLP:journals/pvldb/MelnikGLRSTV10} 
and data stream processing~\cite{aurora,stanford-stream,telegraphcq} platforms, where 
data properties and the status of cluster compute nodes may be constantly changing.  Here 
it is very difficult to effectively choose a good plan for query execution: data 
statistics may be unavailable or highly variable; cost parameters may change due to 
resource contention or machine failures; and in fact a \emph{combination} of query plans 
might perform better than any single plan.  Similarly, in conventional DBMSs there may be 
a need to perform \emph{self-tuning} so the performance of a query or set of queries can 
be improved~\cite{db2-mqreopt}.

To this point, query optimization techniques in adaptive query processing systems fall into three general classes: (1) operator-specific techniques that can adapt the order of evaluation for filtering operators~\cite{stanford-aqp}; (2) eddies~\cite{DBLP:conf/sigmod/AvnurH00,stairs} and related flow heuristics, which are highly adaptive but also continuously devote resources to exploring \emph{all} plans and require fully pipelined execution; (3) approaches that use a cost-based query re-optimizer to re-estimate plan costs and determine whether the system should change plans~\cite{DBLP:conf/sigmod/KabraD98,db2-mqreopt,tukwila-04,cape-04}.  Of these, the last is the most flexible, e.g., in that it supports complex query operators like aggregation, as well as expensive adaptations like data repartitioning across a cluster.  Perhaps most importantly, a cost-based engine allows the system to spend the majority of its resources on query execution once the various cost parameters have been properly calibrated.  Put another way, it can be applied to highly complex plans and has the potential to provide significant benefit if a cost estimation error was made, but it should incur little overhead if a good plan was chosen.  Unfortunately, to this point cost-based techniques have not been able to live up to their potential, because the cost-based re-optimization step has been too expensive to perform frequently.

Our goal in this paper is to explore whether \emph{incremental} techniques for re-optimization can be developed, where an optimizer would only re-explore query plans whose costs were affected by an updated cardinality or cost value; and whether such incremental techniques could be used to facilitate more efficient adaptivity.



\Paragraph{Target Domains.} 
Our long-term goal is to develop adaptive techniques for complex OLAP-style queries (which contain operators not amenable to the use of eddies) being executed across a data-partitioned cluster, as in~\cite{boom-analytics,asterix} .  However, in this paper we focus on developing incremental re-optimization techniques that we evaluate within a single-node (local) query engine, in two main contexts. (1) We address the problem of adaptive query processing in \emph{data stream management systems} where data may be bursty, and its distributions may vary over time --- meaning that different query plans may be preferred over different segments. Here it is vital to optimize frequently based on recent data distribution and cost information, ideally as rapidly as possible. (2) We address query re-optimization in traditional OLAP settings when the same query (or highly similar queries) gets executed frequently, as in a prepared statement.  Here we may wish to re-optimize the plan after each iteration, given increasingly accurate information about costs, and we would like this optimization to have minimal overhead.

\eat{ML: Mention our benefits over no re-opt and non-incremental Re-opt. First, we can always adapt to changing conditions in the environment; second,
re-optimizer performance is important because it can save us overall performance by quickly avoiding suffering of bad plans, and enabling more frequent adaptations. Both optimization+execution time and granularity of re-opt is key to the overall performance. We need to say this and show in the experiments. Our solutions work best when most data characteristics remain the same, but some drastic changes to a small amount of data statistics may be important enough to change plans, and the old plan is really bad such as running out-of-memory, or machine faults, or really slow.}

\Paragraph{Approach and Contributions.}
The main contribution of this paper is to show for the first time how an \emph{incremental} re-optimizer can be developed, and how it can be useful in adaptive query processing scenarios matching the application domains cited above.  Our incremental re-optimizer implements the basic capabilities of a modern database query optimizer, and could easily be extended to support other more advanced features; our main goal is to show that an incremental optimizer following our model can be competitive with a standard optimizer implementation for \emph{initial} optimization, and significantly faster for \emph{repeated} optimization.  Moreover, in contrast to randomized or heuristics-based optimization methods, we \textbf{still guarantee the discovery of the best plan} according to the cost model.  Since our work is oriented towards adaptive query processing, we evaluate the system in a variety of settings in conjunction with a basic pipelined query engine for stream and stored data.

We implement the approach using a novel approach, which is based on the observation that query optimization is essentially a recursive process involving the derivation and subsequent pruning of state (namely, alternative plans and their costs).  If one is to build an \emph{incremental} re-optimizer, this requires preservation of state (i.e., the optimizer memoization table) across optimization runs --- but moreover, it must be possible to determine what plans have been \emph{pruned} from this state, and to re-derive such alternatives and test whether they are now viable.

One way to achieve such ``re-pruning'' capabilities is to carefully define a semantics for how state needs to be tracked and recomputed in an optimizer.  However, we observe that this task of ``re-pruning'' in response to updated information looks remarkably similar to the database problem of \emph{view maintenance} through aggregation~\cite{springerlink:10.1007/BFb0014149} and recursion as studied in the database literature~\cite{gms93-dred}.  In fact, recent work~\cite{evita-raced} has shown that query optimization can itself be captured in recursive datalog.  Thus, rather than inventing a custom semantics for incrementally maintaining state within a query optimizer, we instead adopt the approach of developing an incremental re-optimizer expressed \emph{declaratively}.

%


More precisely, we express the optimizer as a recursive datalog
program consisting of a set of rules, and leverage the existing
database query processor to actually execute the declarative
program. In essence, this is optimizing a query optimizer using a
query processor. Our implementation approaches the performance of
conventional procedural optimizers for reasonably-sized queries. Our
implementation recovers the initial overhead during subsequent
re-optimizations by leveraging \emph{incremental view
  maintenance}~\cite{gms93-dred,recursive-views} techniques. It only
recomputes portions of the search space and cost estimates that might
be affected by the cost updates. Frequently, this is only a small
portion of the overall search space, and hence we often see
order-of-magnitude performance benefits.

Our approach achieves pruning levels that rival or best
bottom-up (as in System-R~\cite{systemR}) and top-down (as in
Volcano~\cite{DBLP:journals/debu/Graefe95a,volcano}) plan
enumerations with branch-and-bound pruning.  We develop a variety of
novel \emph{incremental} and \emph{recursive} optimization techniques to capture the kinds of pruning used in a conventional optimizer, and more
importantly, to generalize them to the incremental case. Our techniques are of broader interest to incremental evaluation of recursive queries as well. Empirically, we see updates on only a small portion of the overall search space, and hence we often see order-of-magnitude performance benefits of incremental re-optimization. We also show that our re-optimizer fits nicely into a complete adaptive query processing system, and measure both the performance and quality, the latter demonstrated well in the yielded query plans, of our incremental re-optimization techniques on the Linear Road stream benchmark.
We make the following contributions:


\begin{compactList}
\vspace{-2mm}
\item The first 
  query optimizer that prunes yet supports incremental re-optimization.

\vspace{-2mm}
\item A rule-based, declarative approach to query
  (re)optimization. Our implementation decouples plan enumerations and
  cost estimations, relaxing traditional restrictions on 
  search order and pruning.  


\vspace{-2mm}
\item Novel strategies to prune the state of an executing recursive query, such as a declarative optimizer: \emph{aggregate selection} with \emph{tuple source suppression};
  \emph{reference counting}; and \emph{recursive bounding}.

\vspace{-2mm}
\item A formulation of query re-optimization as an \emph{incremental view
  maintenance} problem, for which we develop novel 
  algorithms.

\vspace{-2mm}
\item An implementation over a query engine developed for recursive
  stream processing~\cite{recursive-views}, with a
  comprehensive evaluation of performance against alternative
  approaches, over a diverse workload.
  
\vspace{-2mm}
\item Demonstration that incremental re-optimization can be incorporated
  to good benefit in existing cost-based adaptive query processing techniques~\cite{tukwila-04,cape-04}.

\vspace{-2mm}
\end{compactList}


\eat{
This paper is structured as follows.  We begin by describing the problem of formulating the query optimization problem in a declarative language in Section~\ref{sec:declarative}.  Section~\ref{sec:sip} 
discusses how to incorporate pruning techniques into the optimizer's
search process, using novel optimization techniques.  Next,
Section~\ref{sec:incremental} describes how the approaches can be
generalized to handle incremental updates to the cost model and
estimates.  We study performance, including against more traditional
implementations, in Section~\ref{sec:eval}, describe related work in
Section~\ref{sec:related}, and conclude in
Section~\ref{sec:conclusion}.
}

\Section{Declarative Query Optimization}
\label{sec:declarative}



Our goal is to develop infrastructure to adapt an optimizer to support
efficient, incrementally maintainable state, and incremental pruning.
Our focus is on developing techniques for \emph{incremental state
  management} for the recursively computed plan costs in the query
optimizer, in response to updates to query plan cost information.
Incremental update propagation is a very well-studied problem for
recursive datalog queries, with a clean semantics and many efficient
implementations.  Prior work has also demonstrated the feasibility of
a datalog-based query optimizer~\cite{evita-raced}.  Hence, rather
than re-inventing incremental recomputation techniques we have built
our optimizer as a series of recursive rules in datalog, executed in
the query engine that already exists in the DBMS. (We could have
further extended to Prolog, but our goal was clean state management
rather than a purely declarative implementation.  Other alternatives
like constraint programming or planning languages do not support
incremental maintenance.)

\begin{figure}[t]
\begin{center}
\includegraphics[width=3.2in]{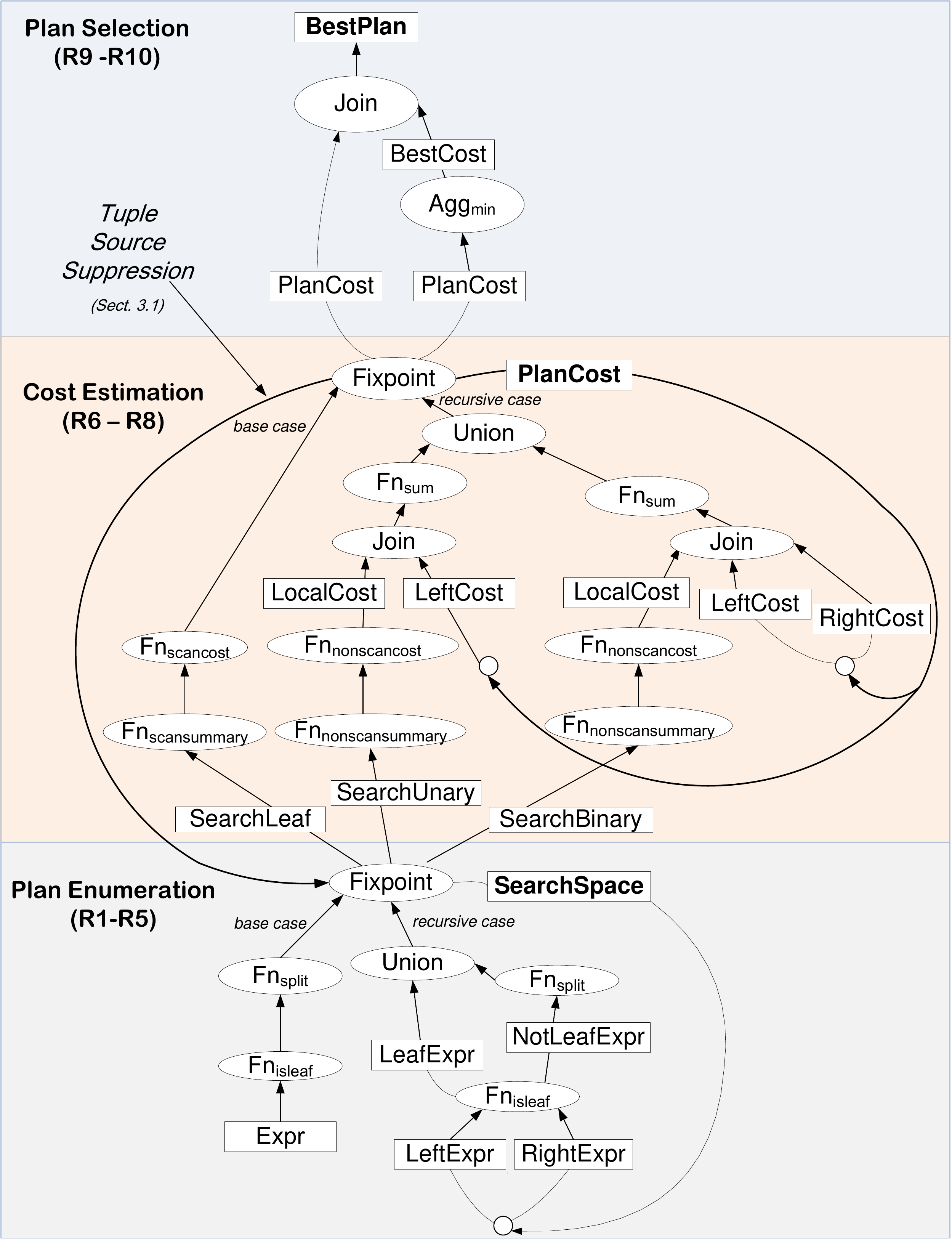}
\vspace{-3mm}
\caption{\label{fig:execution-flow}Query plan of our declarative query optimizer.  Operators are in ellipses; views are in rectangles. Plan enumeration (\dlrel{SearchSpace}) consists of 5 rules, cost estimation (\dlrel{PlanCost}) 3 rules, and plan selection (\dlrel{BestPlan}) 2 rules.  
See Appendix.}

\vspace{-3mm}
\end{center}
\end{figure}

In contrast to work such as~\cite{evita-raced}, our focus is not on formulating every aspect of query optimization in datalog, but rather on formulating those aspects relating to state management and pruning as datalog rules --- so we can use incremental view maintenance (delta rules) and sideways information passing
techniques, respectively.  Other optimizer features that are not reliant on state that changes at runtime, such as cardinality estimation, breaking expressions into subexpressions, etc., are specified as built-in auxiliary functions.  As a result, we specify an entire optimizer in only three stages and 10 rules (dataflow is illustrated in Figure~\ref{fig:execution-flow}).



\eat{In this section, we describe how a query optimizer can be formulated
using datalog rules plus a small set of user-defined functions, and
then we discuss the basic model for executing this datalog program. At
a high level, one can modularize the a query optimizer in terms of the
following three components:}




\vspace{-1mm}
\Paragraph{Plan enumeration (\dlrel{SearchSpace}).} Searching the space of possible plans
has two aspects.  In the {\em logical
  phase}, the original query is recursively broken down into a full
set of alternative relational algebra
subexpressions\footnote{Alternatively, only \emph{left-linear}
  expressions may be considered~\cite{systemR}.}.  The decomposition
is naturally a ``top-down'' type of recursion: it starts from the
original query expression, which it breaks down into subexpressions,
and so on.  The {\em physical phase} takes as input a query expression
from the logical phase, and creates physical plans by enumerating the
set of possible physical operators that satisfy any constraints on the
output \emph{properties}~\cite{volcano} or ``interesting
orders''~\cite{systemR} (e.g., the data must be sorted by a particular
attribute).  Without physical properties, the extension from logical
plans to physical plans can be computed either top-down or bottom-up;
however, the properties are more efficiently computed in goal-directed (top-down) manner.


\vspace{-1mm}
\Paragraph{Cost estimation (\dlrel{PlanCost}).} This phase determines the cost for
each physical plan in the search space, by recursively merging the
statistics and cost estimates of a plan's subplans.  It is
naturally a bottom-up type of recursion, as the plan subexpressions
must already have been cost-estimated before the plan itself.  
Here we can encode in a table the mapping from a plan to its cost.

\vspace{-1mm}
\Paragraph{Plan selection (\dlrel{BestPlan}).} As costs are estimated, the program
produces the plan that incurs the lowest estimated cost.


In our declarative approach to query optimization, we treat optimizer
state as data, use rules to specify what a query optimizer is, and
leverage a database query processor to actually perform the
computation. Figure~\ref{fig:execution-flow} shows
a (simplified) query plan for the datalog rules.  As we can see, the
declarative program is by nature recursive, and is broken into the
three stages mentioned before (with Fixpoint operators between
stages).  Starting from the bottom of the figure, \textbf{plan
  enumeration} recursively generates a \reln{SearchSpace} table
containing plan specifications, by decomposing the query and
enumerating possible output properties; enumerated plans are then fed
into the \textbf{plan estimation} component, \reln{PlanCost}, which
computes a cost for each plan, by building from leaf to complex
expressions; \textbf{plan selection} computes a \reln{BestCost} and
\reln{BestPlan} entry for each query expression and output property,
by aggregating across the \reln{PlanCost} entries.

\eat{
From figure~\ref{fig:execution-flow}, we can see that our
declarative program is by nature recursive, and there are multiple fixpoint operators for different kinds of recursions. The plan enumeration
component (rules R1-R5) 
recursively generates plan tuples, which are then used by the cost estimation component (rules R6-R8) 
to compute the cost for each plan. The computed costs are then used 
by two additional rules (R9 and R10) 
for selecting the best plan.}

\eat{
The total three colored query blocks can be executed in parallel, and sideways information passing can be done asynchronously --- yielding great flexibility and even the potential to distribute and parallelize
the query optimization task (which we consider as future work).}

\vspace{-3mm}
\begin{example}
As our driving example, consider a simplified TPC-H Query 3
with its aggregates and functions removed, called Q3S.
\normalfont
\lstset{language=SQL}
\begin{lstlisting}
SELECT L_orderkey, O_orderdate, O_shippriority
FROM Customer C, Orders O, Lineitem L
WHERE C_mktsegment = 'MACHINERY' and C_custkey = O_custkey and O_orderkey = L_orderkey and O_orderdate < '1995-03-15' and L_shipdate > '1995-03-15'
\end{lstlisting}
\vspace{-3mm}
\end{example}

\eat{We describe the components and dataflow of our optimizer in more
detail next.  Refer to Figure~\ref{fig:execution-flow}
for the dataflow architecture.}


\Subsection{Plan Enumeration}

Plan enumeration takes as input the original query expression as \reln{Expr}, and then generates as output the
set of alternative plans.  As with many optimizers, it is divided into
two levels:

\vspace{-1mm}
\Paragraph{Logical search space.} The logical plan search space
contains all the logical plans that correspond to subexpressions of
the original query expression up to any logically equivalent
transformations (e.g., commutativity and associativity of join
operators). In traditional query optimizers such as
Volcano~\cite{volcano}, a data structure called an {\em and-or-graph}
is maintained to represent the logical plan search space. Bottom-up
dynamic programming optimizers do not need to physically store this
graph but it is still conceptually relevant.


\begin{figure}[t]
\begin{center}
\includegraphics[width=3.5in]{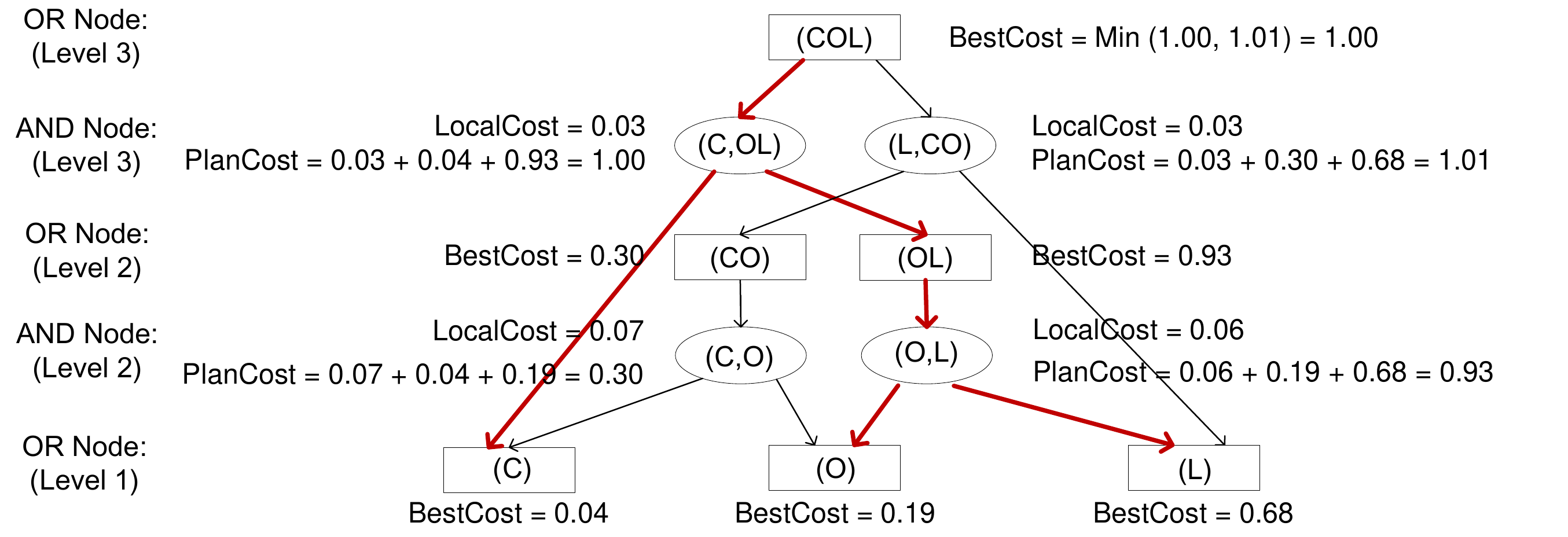}
\vspace{-7mm}
\caption{The and-or-graph for Q3S. Red edges denote the best
  plan. Rectangles and ovals denote ``OR'' and ``AND'' nodes respectively.  
  Each ``OR'' node is labeled with its \reln{BestCost} and each 
  ``AND'' node is labeled with its \reln{LocalCost} and \reln{PlanCost}.
  \label{fig:example-and-or-graph-Q3S}}
\end{center}
\vspace{-2mm}
\end{figure}

\begin{example}
Figure~\ref{fig:example-and-or-graph-Q3S} shows an example and-or-graph for Q3S, which describes a set of alternative subplans and subplan choices using interleaved levels.  ``AND'' nodes represent alternative subplans (typically with join operator roots) and the ``OR'' nodes represent decision points where the cheapest AND-node was chosen.




\eat{There is a single root ``OR'' node, representing the original query expression. The and-or-graph is then recursively defined. For example, each level 3 ``AND'' node represents one way of joining a level 1 ``OR'' node and a level 2 ``OR'' node. Every ``OR'' node at level 2 only has one ``AND'' child, since there is only one possible algebraic rewriting for $C\Join O$ and $O\Join L$ respectively. Finally, at the leaf level, each ``OR'' node represents a base table expression, e.g., $C$, $O$ or $L$.}
\end{example}
\vspace{-2mm}

\eat{ML: This example can be compressed. OR node really represents ALTERNATIVE ways of splitting an expression, and AND node really represents EACH way of splitting. }

In our implementation, we capture each of the nodes in a table called
\reln{SearchSpace}.  In fact, as we discuss next, we supplement this
information with further information about the output properties and
\emph{physical plan}.  (We explain why we combine the results from
both stages in Section~\ref{sec:model}.)

\Paragraph{Physical search space.}  
The physical search space extends the logical one in that it
enumerates all the physical operators for each algebraic logical
operator.  For example, in our figure above, each ``AND'' node denotes
a logical join operator, but it may have multiple physical
implementations such as pipelined-hash, indexed nested-loops, or
sort-merge join. If the physical implementation is not symmetric,
exchanging the left and right child would become a different physical
plan.  A physical plan has not only a root physical operator, but also a
set of physical properties over the data that it maintains or
produces; if we desire a set of output properties, this may constrain
the physical properties of the plan's inputs.

\eat{
Our declarative specification combines both logical and physical plan enumeration (as
explained in Section~\ref{sec:model}), hence the output includes all
physical plans in the search space. (The set of logical plans can be
obtained by projecting certain attributes from the $SearchSpace$
table.)}


\eat{Every logical operator in a logical plan can be implemented by
different physical operators with different physical properties such
as ``interesting orders''.  To encode the physical search space, more
attributes are required. join implementations, and an additional
property attribute representing the interesting orders of each
expression. }


\begin{table*}[ht]
\begin{center}
\footnotesize
\begin{tabular}{|c|c|c|c|c|c|c|c|c|}
  \hline {\bf *Expr} & {\bf *Prop} & {\bf *Index} & {\bf LogOp} & {\bf *PhyOp} & {\bf lExpr} & {\bf lProp} & {\bf rExpr} & {\bf rProp} \\ \hline \hline
  (COL) &  --  & 1 & join & sort-merge & (C) & C\_custkey order & (OL) & O\_custkey order\\ \hline
  (COL) &  --  & 2 & join & indexed nested-loop & (L) & index on L\_orderkey & (CO) &  --  \\ \hline
  (OL) & O\_custkey order & 1 & join & pipelined-hash & (O) & -- & (L) &  --  \\ \hline
  (CO) &  --  & 1 & join & pipelined-hash & (C) & --   & (O) & -- \\ \hline
  (CO) &  --  & 1 & join & sort-merge & (C) & --   & (O) & -- \\ \hline
  (O) & O\_custkey order & -- & scan & local scan & -- &  --  &  --  &  -- \\ \hline
  (L) & index on L\_orderkey & -- & scan & index scan & -- &  --  &  --  &  --  \\ \hline
  (C) & C\_custkey order & -- & scan & local scan & -- &  --  &  --  &  --  \\ \hline
\end{tabular}
\end{center}
\vspace{-3mm}
\caption{A simplified \reln{SearchSpace} relation encoding the and-or-graph for Q3S's search space. Primary keys are denoted by *.}

\label{tab:searchspace}
\end{table*}

\begin{example}
  Table~\ref{tab:searchspace} shows the
  \reln{Search\-Space} content for a subset of
  Figure~\ref{fig:example-and-or-graph-Q3S}. The AND logical operators are either joins
  (with 2 child expressions), or tablescans (with
  selection predicates applied).  Each expression \reln{Expr}
  may have multiple \dlvar{Index}ed alternatives. \dlvar{Prop} and \dlvar{PhyOp} represent the physical properties of a
  plan and its root physical operator, respectively.

  For instance, expression \reln{SearchSpace}$(COL)$ has encodes an ``OR''
  node with two alternatives the first ``AND'' (join) child is \reln{SearchSpace}$(C,OL)$
  and the second is \reln{SearchSpace}$(L,CO)$.  
  For the first \reln{SearchSpace} tuple, the left expression is $C$ and
  the right expression is $OL$. The tuple indicates a Sort-Merge join, whose left and right inputs' physical properties
  require a sort order based on \dlvar{C\_custkey} and \dlvar{O\_custkey},
  respectively. The second alternative uses an Indexed Nested-Loop Join as
  its physical operator. The left expression refers to the inner join
  relation indexed on \dlvar{L\_orderkey}, while there are no ordering
  restrictions on the right expression.
\end{example}
\vspace{-2mm}


Our datalog-based optimizer enumerates both spaces in the same recursive
query (see bottom of Figure~\ref{fig:execution-flow}).  Given an
expression, the \dlrel{Fn_{split}} function enumerates all the algebraically
equivalent rewritings for the given expression, as well as the
possible physical operators and their interesting orders.  Fixpoint is
reached when \reln{Expr} has been decomposed down to leaf-level scan
operations over a base relation (checked using the \dlrel{Fn_{isleaf}}
function).


\Subsection{Cost Estimation and Plan Selection} 

The cost estimation component computes an estimated cost for every 
physical plan.  Given the \reln{SearchSpace} tuples generated in
the plan enumeration phase, three datalog rules (R6 - R8) 
are used to compute
\reln{PlanCost} (corresponding to a more detailed version of the ``AND'' nodes in Figure~\ref{fig:example-and-or-graph-Q3S}, with physical operators considered and all costs enumerated), and two additional rules (R9 - R10) 
select the
\reln{BestPlan} (corresponding to an ``OR'' node). Cost estimates are recursively computed by summing up the children costs and operation costs. The computed sum for each physical plan is stored in \reln{PlanCost}.

In addition to the search space, cost estimation requires a set of
\emph{summaries} (statistics) on the input relations and indexes, e.g.,
cardinality of a (indexed) relation, 
selectivity of operators, data distribution, etc. These summaries are
computed using functions \dlrel{Fn_{scansummary}} and \dlrel{Fn_{nonscansummary}}.
The former computes the leaf level summaries over base tables, and the
latter computes the output summaries of an operator based on input
summaries.  
Given the statistics, the cost of a plan can be computed by combining factors
such as CPU, I/O, bandwidth and energy into a single cost metric. We compute the cost of each physical operator using functions \dlrel{Fn_{scancost}} and \dlrel{Fn_{nonscancost}} respectively. 

Given the above functions, cost estimation becomes a recursive
computation that sums up the cost of children expressions and the root
cost, to finally compute a cost for the entire query plan.  At
each step, \dlrel{Fn_{sum}} is used to sum up the \reln{PlanCost} of its
child plans with \reln{LocalCost}. The particular form of the operation depends on whether the plan root is a leaf node, a unary or a binary operator.

\begin{example}
  To illustrate the process of cost estimation, we revisit
  Figure~\ref{fig:example-and-or-graph-Q3S}, which shows a simplified
  logical search space (omitting physical operators and properties) for our
  simplified TPC-H Q3S. For every ``AND'' node, we compute the
  plan cost by summing up the cost of the join operator, with the best costs of
  computing its two inputs (e.g., the level 2 ``AND'' node $(C,O)$ sums
  up its local cost 0.07, its left best cost 0.04, and its right best
  cost 0.19, and gets its plan cost 0.30).  For every ``OR'' node, we
  determine the alternative with minimum cost among its ``AND'' node
  children (e.g., the level 3 ``OR'' node $(COL)$ computes a minimum
  over its two input plan costs 1.00 and 1.01, and gets its best cost
  1.00). After the best cost is computed for the root ``OR'' node in the
  graph, the optimization process is done, and an optimal plan tree
  is chosen.
\end{example}
\vspace{-1mm}

Once the \reln{PlanCost} for every ``AND'' node are generated, the final
two rules compute the \reln{BestCost} for every ``OR'' node by computing a
min aggregate over \reln{PlanCost} of its alternative ``AND'' node
derivations, and output the \reln{BestPlan} for each ``AND'' node by
computing a join between \reln{BestCost} and \reln{PlanCost}.


\Subsection{Execution Strategy}
\label{sec:model}

Given a query optimizer specified in datalog, the natural question is
how to actually execute it.  We seek to be general enough to
incorporate typical features of existing query optimizers, to rival
their performance and flexibility, and to only develop implementation
techniques that generalize.  We adopt two strategies:

\vspace{-1mm}
\Paragraph{Merging of logical and physical plan enumeration.} The
logical and physical plan enumeration phases are closely related, and
in general one can think of the physical plan as an elaboration of the
logical one.  As both logical and physical enumeration are top-down
types of recursion, and as there is no pruning information from the
physical stage that should be propagated back to the logical one, we
can merge the logical and physical enumeration stages into a single
recursive query.

As we enumerate each logical subexpression, we simultaneously join
with the table representing the possible physical operators that can
implement it.  This generates the entire set of possible physical
query plans. To make it more efficient to generate multiple physical
plans from a single logical expression, we use \emph{caching} to
memoize the results of \dlrel{Fn_{nonscansummary}} and \dlrel{Fn_{split}}. 

\vspace{-1mm}
\Paragraph{Decoupling of cost estimation and plan enumeration.} 
Cost estimation requires bottom-up evaluation: a cost estimate can
\emph{only} be obtained once cost estimates and statistics are
obtained from child expressions. The enumeration stage naturally
produces expressions in the order from parent to child, yet estimation
must be done from child to parent.  We \emph{decouple} the execution
order among plan enumeration and cost estimation, making the
connections between these two components flexible.  For example, some
cost estimates may happen before all related plans have been
enumerated.  Cost estimates may even be used to \emph{prune} portions
of the plan enumeration space (and hence also further prune cost
estimation itself) in an opportunistic way.

\eat{On the other hand, the search order of plan enumeration and cost
estimation itself is highly flexible as our declarative model does not
require how to implement the search order.}

In subsequent sections, we develop techniques to prune and maintain
the optimizer state that has no constraints on enumeration order,
search order or pruning frequency.  Our approach relaxes the
traditional restrictions on the search order and pruning techniques in
either Volcano's~\cite{volcano} top-down traversal or System
R's~\cite{systemR}'s bottom-up dynamic programming approaches.  For
example, a top-down search may have a depth-first, breadth-first or
another order.

We leverage a pipelined
push-based query processor to execute the rules in an
incremental fashion, which simultaneously explores many expressions.
We pipeline the results of plan enumeration to the cost estimation
stage without synchronization or blocking.  All of our techniques
extend naturally to a distributed or parallel setting, which we hope
to study in future work. 



\eat{Interestingly, Volcano's~\cite{volcano} depth-first pre-order
traversal search order with branch-and-bounding can be emulated by the
execution strategy of tuple-based fixpoint propagation; and System
R's~\cite{systemR} dynamic programming approach can be emulated by the
execution strategy of stratum-based fixpoint propagation. These can be
implemented using different data structures and different recursive
query processing algorithms.}


\eat{
Indeed, System R~\cite{systemR} and Volcano~\cite{volcano} differ most
in the search order of these two. System R enumerates plans and
estimates costs in the same order of \emph{bottom-up iterative
  deepening}. This order is effective for dynamic programming as best
sub-plans are always computed before the parent plan; however, it has
no pruning at all, because every child node is already cost-estimated
when a parent node is decided to have been pruned. On the other hand,
Volcano enumerates plans in depth-first pre-order traversal of the
search space, and estimates costs in depth-first post-order traversal
of the search space. Here cost estimation is a bottom-up traversal,
however, it is called via recursive functions hence it must have the
same depth-first order as plan enumeration. Volcano's search order
needs to achieve dynamic programming via memorization; on the other
hand, it is effective in exploiting branch-and-bounding because
depth-first traversal ensures that a complete plan from the root is
computed early enough that it can be used as a bound to potentially
prune others.  
}


\eat{The output of this Fixpoint operator, which is the output of the
  plan enumeration part, is sent to the cost estimator. As we have
  discussed, Rule 6-8 correspond to the zero-input logical operator
  (such as Scan), one-input logical operator (such as Function and
  Aggregate) and two-input logical operator (such as Join)
  respectively. The zero-input $SearchSpace$ tuple is sent to
  $F\_Scan\_Summary$ and $F\_Scan\_Cost$ to compute its scan cost; the
  one-input $SearchSpace$ tuple is sent to $F\_Non\_Scan\_Summary$ and
  $F\_Non\_Scan\_Cost$ to compute its local cost, which is sent to
  $F\_Sum$ to sum up with the $PlanCost$ of its input expression (the
  left expression); the two-input $SearchSpace$ tuple is also sent to
  $F\_Non\_Scan\_Summary$ and $F\_Non\_Scan\_Cost$ to compute its
  local cost, but then it is is sent to $F\_Sum$ to sum up with both
  the $PlanCost$ of its left expression and right expression.}







\eat{
Next, Rule 9 computes a $BestCost$ for every ``OR'' node. This is done
by computing a minimum cost of every ``AND'' node associated with this
``OR'' node. Below are all the $BestCost$ tuples we maintain for the
example query.

\begin{Dlog}
BestCost('COL', 'None', 1.00)
BestCost('OL', 'Sort on O_custkey', 0.93)
BestCost('CO', 'None', 0.30)
BestCost('C', 'Sort on C_custkey', 0.04)
BestCost('L', 'None', 0.68)
BestCost('O', 'Sort on O_custkey', 0.19)
\end{Dlog}

Finally, Rule 10 is to return the $BestPlan$ plan chosen for the
original query. It uses a join between $BestCost$ and $PlanCost$. For
every ``OR'' node, it records which ``AND'' node child it finally
picks and its logical and physical operators.

\begin{Dlog}
BestPlan('COL', 'None', 1, 'Join', 'Sort-Merge Join', 
'C', 'Sort on C_custkey', 'OL', 'Sort on O_custkey', 1.00)
BestPlan('OL', 'Sort on O_custkey', 1, 'Join',
'Pipelined-Hash Join', 'O', 'Sort on O_custkey', 
'L', 'None', 0.93)
BestPlan('C', 'Sort on C_custkey', 1, 'Scan', 'Index-Scan', 
null, null, null, null, 0.04)
BestPlan('L', 'None', 1, 'Scan', 'Local-Scan', 
null, null, null, null, 0.68)
BestPlan('O', 'Sort on O_custkey', 1, 'Scan', 'Index-Scan', 
null, null, null, null, 0.19)
\end{Dlog}

}

\eat{

\Subsection{Discussion}


\reminder{Placeholder text. To be heavily condensed.}

Logical plan enumeration explores all possible logically equivalent
relational algebra subexpressions of a query, which construct a search
space of logical plans. Based on these subexpressions, 

physical plan enumeration extends that search
space to physical plans, adding different implementations of a logical
operator, and possible ``interesting orders''.

Figure~\ref{fig:execution-flow} depicts the execution flow of a query
processor executing the rules in Appendix
Section~\ref{sec:append-declarative}. This graph shows the logical
dependencies among different components, and is not subject to
implementation details. It has four main components: plan enumeration
(corresponding to Rule 1-5), cost estimation (Rule 6-8), best cost
(Rule 9) and best plan (Rule 10).

A query optimizer maps a declarative query to the most efficient plan
tree, and a cost-based query optimizer normally involves four
components: logical plan enumeration, physical plan enumeration, plan
cost estimation and optimal plan selection. Logical plan enumeration
explores all possible logically equivalent relational algebra
subexpressions of a query, which construct a search space of logical
plans; physical plan enumeration extends that search space to physical
plans, adding different implementations of a logical operator, and
possible ``interesting orders''; a plan cost estimator computes a cost
for each physical plan in the search space, usually based on
histograms estimating the cardinality and selectivity of each
operator; and optimal plan selection simply picks the physical plan
that is associated with the smallest estimated cost.

Now let us examine the query re-optimization process. Suppose the cost
of a subexpression is somehow proved to be wrongly estimated after
query execution starts, and therefore gets updated. The
state-of-the-art adaptive query processing systems all re-optimize the
query from scratch~\cite{tukwila-04,cape-04}. However, they can be
done more smartly. Obviously we can re-use the logical plan and
physical plan search space, unless some useful plans are pruned by
procedures such as ``branch-and-bounding''~\cite{volcano}. On the
other hand, although the re-costing of any single subexpression may
potentially affect the cost of any ancestor expressions, it would not
have any effect on the costing of expressions that have no dependency
on it. Unless it affects the costing of root-level expressions, it may
not change the ultimate optimal plan even if some intermediate costs
are updated.

The goal of this work is to provide an efficient incremental solution
to query re-optimization. Interestingly, if we formulate a query
optimizer as a series of \emph{declarative rules}, this problem is in
essence an ``incremental view maintenance'' problem. In particular, if
we consider the ultimate optimal plan as a ``view'' over 1) the query
expression, 2) the summaries (e.g., histograms) of a physical plan,
and 3) the cost estimators of a physical plan, then a query
re-optimization originates from the insertion, deletion or update of
the inputs triggers an incremental maintenance of the view. In this
paper, we consider the scenarios where the summaries or the cost
estimations of a set of plans in the search space are inserted,
deleted or updated. We assume that possible query rewrites and
``interesting orders'' remain unchanged during query
re-optimization. This covers most of the important applications of
adaptive query processing, as a re-optimization process is normally
triggered by correcting the inaccurate statistics summaries or wrong
cost estimations~\cite{aqp-survey}.

Indeed, there are many intermediate ``views'' inside a query
optimizer: logical search space, physical search space, plan cost
estimations, and even pruned search space via ``branch-and-bounding''
can all be regarded as views. In a typical query optimizer such as
Volcano/Cascades~\cite{volcano}, it maintains a MEMO table to record
all viable plans in the search space. This can be regarded as
materializing the view of unpruned physical search space. However, in
a query re-optimization process, this MEMO table may be frequently
updated. Some ``views'' are essential to other views, such as logical
search space to physical search space. Other information, such as the
cardinality and selectivity of a plan, is only used once inside a cost
estimator.

We take a data-centric point view of query optimization: the
intermediate state that could potentially be shared by multiple
computations (e.g., plan search space, plan cost) in the query
optimizer can be modeled as data, and the query optimization process
can be expressed as a set of recursive rules over these data. This
approach decouples the specification of a query optimizer from the
execution strategy of implementing the query optimizer. It is a
paradigm shift from any procedural query optimizers~\cite{systemR,
  volcano}. There are many ways of designing a declarative query
optimizer based on different abstraction levels. In particular, recent
work~\cite{evita-raced} specify Volcano-style query optimizer and
System R-style query optimizer in different set of datalog rules based
on different enumeration algorithms. In this work, we choose a higher
abstract level that is extensible and customizable to most existing
query optimizers. In particular, it is independent of the search order
(e.g., top-down or bottom-up or A* planning).

There are many benefits of this declarative approach: 1) query
re-optimization and optimization is based upon the same rules, just
with the extension of supporting a richer data model of Insertion,
Deletion and Updates; 2) the execution orders of different components
of a query optimizer is highly customizable, for example, the cost of
a parent expression can be computed before its children has completed
finding its optimal cost, and may be later on updated until the
children optimal plan is determined; 3) plan enumeration and cost
computation can be decoupled: there is no need to obey the strict
order of Volcano-style or System R-style search order, essentially any
planning algorithm on the search space is allowed; 4) optimization
strategies in datalog programming such as Magic
Sets~\cite{DBLP:conf/pods/BancilhonMSU86} and Sideways Information
Passing~\cite{zives-sip} can be leveraged to optimize the execution
flow of this meta-optimizer. This approach also opens a new angle
towards parallel and distributed query optimization as partitioning
techniques for executing datalog rules can be leveraged. We leave this
problem for future work.
}



\eat{

Next, we demonstrate our design of the declarative query optimizer. We
formulate the query optimizer into 10 datalog rules, with 7 auxiliary
functions. For reference, the full datalog rules are shown in Appendix
Section~\ref{sec:append-declarative}. Semantically, Rule 1-5 refer to
the logical and physical plan enumeration, Rule 6-8 refer to the cost
estimation, Rule 9 maintains a minimum cost for each expression, and
Rule 10 returns the best plan associated with this minimum cost.

Table~\ref{table:functions} shows a list of functions in our
declarative optimizer. The choice of specifying them as functions
versus queries over data is based on interoperability: if the internal
state of a computation is not used by other component, then these
self-contained computations form a function. For example, testing
whether an expression is a scan expression over base relation is a
function ($F\_IsScan$); computing the summary metadata based on input
summaries is a function ($F\_Non\_Scan\_Summary$); and estimating the
local cost of a non-scan expression is a function
($F\_Non\_Scan\_Cost$). Note that we regard the process of splitting
an ``OR'' node expression into ``AND'' node children as a function
$F\_Split$, this is because we think the process of splitting a query
is not used anywhere else other than the splitting process.

\begin{table}
\begin{tabular}{|p{8cm}|}
\hline
Functions (an attribute without a \& is an input attribute; an attribute with a \& is an output attribute) \\
\hline
\hline
F\_IsScan(expr, \&isScan) \\
\hline
F\_Split(expr, prop, \&index, \&logOp, \&phyOp, \&lExpr, \&lProp, \&rExpr, \&rProp) \\
\hline
F\_Scan\_Summary(expr, prop, \&metadata) \\
\hline
F\_Non\_Scan\_Summary(expr, prop, index, logOp, lMetadata, rMetadata, \&metadata) \\
\hline
F\_Scan\_Cost(expr, prop, phyOp, metadata, \&cost) \\
\hline
F\_Non\_Scan\_Cost(expr, prop, index, logOp, phyOp, lExpr, lProp, rExpr, rProp, metadata, \&localCost) \\
\hline
F\_Sum(leftCost, rightCost, localCost, \&planCost) \\
\hline
\end{tabular}
\caption{\label{table:functions} Functions used in our declarative program}
\end{table}

}

\Section{Achieving Pruning}
\label{sec:sip}
In this section, we describe how we can incorporate \emph{pruning} of the
search space into pipelined execution of our query optimizer.  To achieve
this, we use techniques based on the idea of \emph{sideways information passing}, in which the computation of one portion of the query plan may be made more efficient by filtering against information computed elsewhere, but not connected directly by pipelined dataflow.  Specifically, we incorporate the technique of aggregate selection~\cite{sudarshan91aggregation} from the deductive database literature, which we briefly review; we extend it to perform further pruning; and we develop two new techniques for recursive queries that enable tracking of dependencies and computation of bounds.  Beyond query optimization, our techniques are broadly useful in the evaluation of recursive datalog queries.  In the next section we develop novel techniques to make these strategies \emph{incrementally maintainable}.

Section~\ref{subsec:agg-sel} reviews aggregate selection, which
removes non-viable plans from the optimizer state if they are not
cost-effective, and shows how we can use it to achieve the similar effects to dynamic programming in System-R. There we also introduce a novel technique called \emph{tuple source suppression}.  Then in the remainder of the section we show how to introduce two familiar notions into datalog execution:  namely, \emph{reference counting} that enables us to removes plan subexpressions once all of their parent expressions have been pruned (Section~\ref{subsec:refcount}), and \emph{recursive bounding}, which lets the datalog engine incorporate branch-and-bound pruning as in a typical Volcano-style top-down query optimizer (Section~\ref{subsec:branch-and-bounding}). Our solutions are valid for any execution order, take full advantage of the parallel
exploration provided by pipelining, and are extensible to parallel or distributed architectures.

\eat{
\begin{example}
  To illustrate our pruning and incremental computation techniques, we use a
  more complex query, Q10S, as an example. It is a four-way join query    
  simplifying TPCH Q10 as shown below. The and-or-graph graph for this Q10S 
  query is shown in Figure~\ref{fig:example-and-or-graph-Q10S}.
\small
\begin{Dlog}
SELECT C_custkey, C_name, L_extendedprice, L_discount, 
C_acctbal, N_name, C_address, C_phone, C_comment 
FROM Customer C, Orders O, Lineitem L, Nation N
WHERE C_custkey = O_custkey and L_orderkey = O_orderkey 
and O_orderdate >= '1993-06-01' and O_orderdate < '1993-09-01' 
and L_returnflag = 'R' and C_nationkey = N_nationkey;
\end{Dlog}
\end{example}
\begin{figure}
\begin{center}
\includegraphics[width=3.25in]{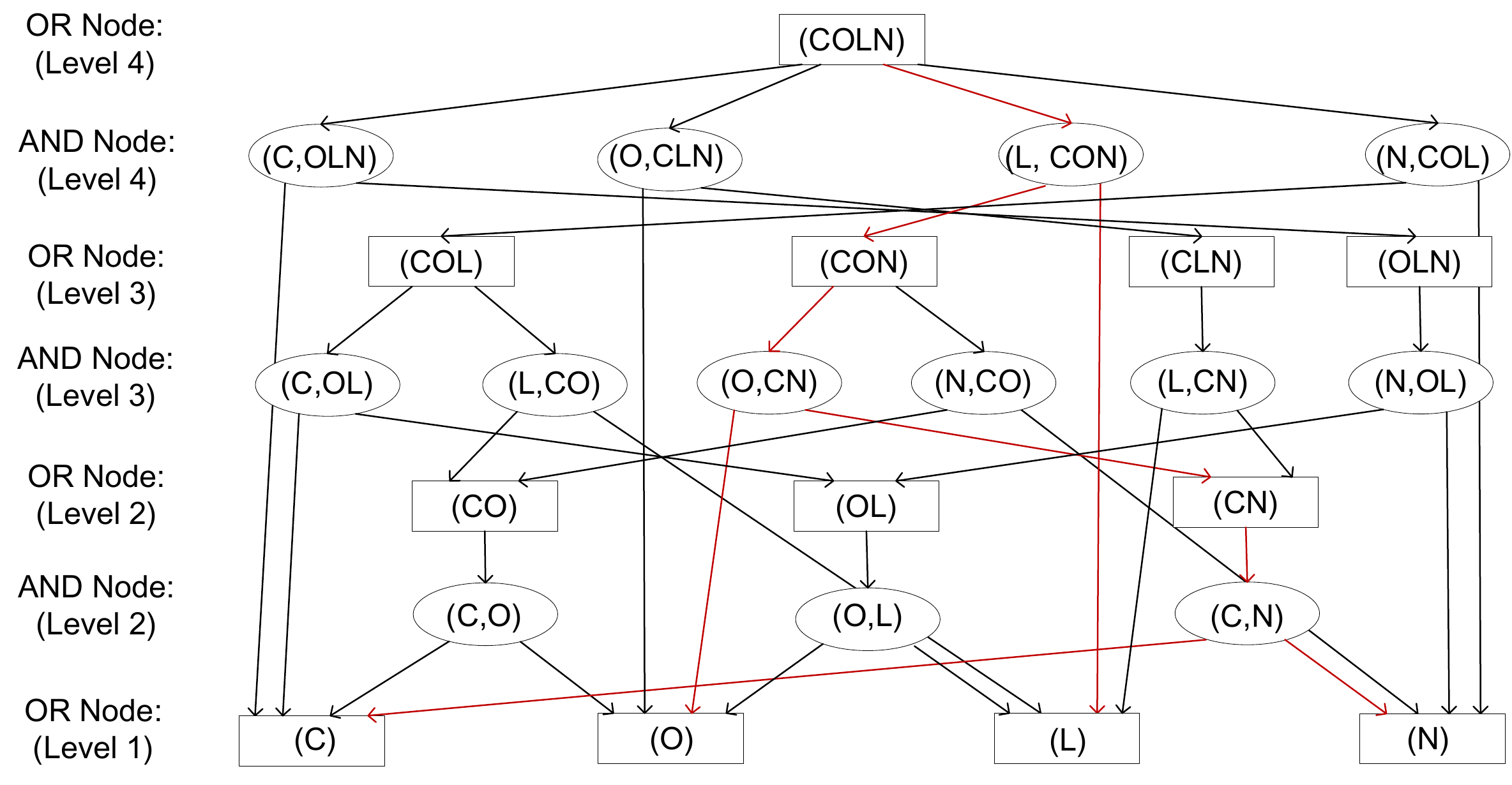}
\end{center}
\vspace{-3mm}
\caption{And-or-graph for four-way join query TPCH-Q10S\label{fig:example-and-or-graph-Q10S},
The red subtree is the optimal plan tree for the root expression.
The blue subtree is a suboptimal plan tree for a subexpression.}
\end{figure}
}

\Subsection{Pruning Suboptimal Plan Expressions}
\label{subsec:agg-sel}

Dating back to System-R~\cite{systemR}, every modern query optimizer
uses dynamic programming techniques (although some via
memoization~\cite{volcano}).  Dynamic programming is based on the
\emph{principle of optimality}, i.e. an optimal plan can be decomposed
into sub-plans that must themselves be optimal solutions.  This
property is vital to optimizer performance, because the same
subexpression may appear repeatedly in many parent expressions. Formally:

\vspace{-2mm}
\begin{proposition} \label{prop:viability} Given a query expression
  $E$ and property $p$, consider a plan tree $T\langle E, p
  \rangle$ that evaluates $E$ with output property $p$. 
  For this and other propositions, we assume that plans have distinct costs.
  Here one such $T$ will have the minimum cost: call that $T_{OPT}$.  		
  Suppose $E^s$ is a
  subexpression of $E$, and consider a plan tree
  $T^s\langle E^s, p^s \rangle$ that evaluates $E^s$ with output property 
  $p^s$. Again one such $T^s$ will have the minimum cost: call that 
  $T^s_{OPT}$. If $T^s$ is \emph{not} $T^s_{OPT}$, then $T^s$ must \emph{not} 
  be the subtree of $T_{OPT}$. 
\end{proposition}
\vspace{-2mm}

This proposition ensures that we can safely discard suboptimal
subplans without affecting the final optimal plan. Consider the
and-or-graph of the example query Q3S
(Figure~\ref{fig:example-and-or-graph-Q3S}). The red (bolded) subtree
is the optimal plan for the root expression $(COL)$. The subplan of
the level 3 ``AND'' node $(L, CO)$ has suboptimal cost 1.00. If there
exists a super-expression containing $(COL)$, then the only viable
subplan is the one marked in the figure. State for any alternative
subplan for $(COL$) may be pruned from \reln{SearchSpace} and
\reln{PlanCost}.
We achieve pruning over both relations as follows.

\eat{We take advantage of the proposition to optimize our declarative program in two ways. First, the suboptimal plans in \reln{PlanCost} are pruned. Second, the suboptimal plans in \reln{SearchSpace} are pruned. We describe each below.}

\paragraph{Pruning \reln{PlanCost} via aggregate selection.} Refer
back to Figure~\ref{fig:execution-flow}: each \reln{BestCost} tuple
encodes the minimum cost for a given query expression-property pair,
over all the plans associated with this pair in \reln{PlanCost}. To
avoid enumerating unnecessary \reln{PlanCost} tuples, one can wait
until the \reln{BestCost} of subplans are obtained before computing a
\reln{PlanCost} for a root plan. This is how System R-style dynamic
programming works.  However, this approach constrains the order of evaluation.


We instead extend a logic programming optimization technique called
\emph{aggregate selection}~\cite{sudarshan91aggregation}, to achieve
dynamic programming-like benefits for any arbitrary order of
implementation. In aggregate selection, we ``push down'' a selection
predicate into the input of an aggregate, such that we can prune
results that exceed the current minimum value or are below the current
maximum value.  In our case (as shown in the middle box of
Figure~\ref{fig:execution-flow}), the current best-known cost for any
equivalent query expression-property pair is maintained within our
Fixpoint operator (which also performs the non-blocked \textbf{min} aggregation).
We only propagate a newly generated \reln{PlanCost} tuple if its cost
is smaller than the current minimum. This does not affect the
computation of \reln{BestCost}, which still outputs the minimum cost
for each expression-property pair.  Since pruning bounds are updated
upon every newly generated tuple, there is no restriction on
evaluation order.  As with pruning strategies used in Volcano-style
optimizers, the amount of state pruned varies depending on the order
of exploration: the sooner a min-cost plan is encountered, the more
effective the pruning is.

\paragraph{Pruning \reln{SearchSpace} via tuple source suppression.}
Enumeration of the search space will generally happen in parallel with
enumeration of plans.  Thus, as we prune tuples from \reln{PlanCost},
we may be able to remove related tuples (e.g., representing 
subexpressions) from \reln{SearchSpace}, possibly preventing enumeration 
of their subexpressions and/or costs.  We achieve such pruning through \emph{tuple source suppression}, along the arcs
indicated in Figure~\ref{fig:execution-flow}. Any \reln{PlanCost}
tuples pruned by aggregate selection should also trigger cascading deletions to the \emph{source tuples} by which they were derived from the \reln{SearchSpace} relation. 
\eat{\footnote{We could model the same
  semantics in the datalog program by adding~\reln{PlanCost} and
  \reln{BestCost} as extra atoms to the datalog rules computing
  \reln{SearchSpace}.  However, this is a highly unusual
  \emph{noninflationary semantics} for datalog (see~\cite[Chapter
  14]{ahv95}), with mutual dependencies among \reln{SearchSpace} and
  \reln{PlanCost}, which is undesirable.  Our mechanism essentially
  optimizes this approach by exploiting over-derivation followed by
  correction of \reln{SearchSpace} with correctness guarantees.}}
To achieve this, since \reln{PlanCost} contains a superset of the attributes in
\reln{SearchSpace}, we simply project out the \reln{cost} field and
propagate a deletion to the corresponding \reln{SearchSpace} tuple.

\eat{ML: The resolution of mutual dependency is important and novel. Should emphasize more here.}

\Subsection{Pruning Unused Plan Subexpressions}
\label{subsec:refcount}

The techniques described in the previous section remove
\emph{suboptimal plans} for specific expression-property pairs.
However, ultimately some \emph{optimal plans} for certain expressions may be
unused in the final query execution plan.
Consider in Figure~\ref{fig:example-and-or-graph-Q3S} the level 2
``AND'' node $(C,O)$: this node is not in the final plan because
its ``OR'' node parent expression $(CO)$ does not appear in the final
result. In turn, this is because $(CO)$'s parent ``AND'' nodes
(in this example, just a single plan $(L, CO)$) do not contribute to
the final plan. Intuitively, we may prune an ``AND'' node if all of
its parent ``AND'' nodes (through only one connecting ``OR'' node)
have been pruned.

We would like to remove such plans once they are discovered to not appear
in the final optimal plan, which requires a form of \emph{reference counting}
within the datalog engine executing the optimizer.  To achieve this,
every tuple in \reln{SearchSpace} is annotated with a \emph{count}.
This count represents the number of \textbf{parent plans} still present in the
\reln{SearchSpace}. For example, in Table~\ref{tab:searchspace}, the
plan entry of $O\Join L$ has reference count of 1, because it only has
one parent plan, which is $C\Join OL$; on the other hand, the plan
entry of $(O)$ has reference count of 2, because it has two parent
plans, which are $O\Join L$ and $C\Join O$. Below is a proposition
about reference counting:

\eat{
During execution, in order to \emph{prune} an entry from the search
space according to the techniques in Section~\ref{subsec:agg-sel}, we
treat the removal as a \emph{deletion} from the \reln{SearchSpace}
table, and use \emph{delta rules} to propagate the effects of the
deletion recursively to the tuples representing its subexpressions and
their properties.  Each propagated deletion decrements the count of a
subexpression, in effect dropping its reference count.}

\vspace{-2mm}
\begin{proposition} \label{prop:refcount} Given a query expression $E$
  with output property $p$: let $T^s$ be a plan tree for $E$'s 
  subexpression $E^s$ with property $p^s$. If $T^s$ has reference count of 
  zero, then $T^s$ must not be a subtree of the optimal plan tree for the 
  query $E$ with property $p$. 
\end{proposition}
\vspace{-2mm}

The proposition ensures that a plan with a reference count of zero can
be safely deleted. Note that a deleted plan may make more reference
counts to drop to zero, hence the deletion process may be recursive.
Our reference counting scheme is more efficient than the \emph{counting}
algorithm of~\cite{gms93-dred}, which uses a count representing the
\emph{total number of derivations} of each tuple in bag semantics.  Our
count represents the number of \emph{unique parent plans from which a subplan may be derived}, and can typically be incrementally updated in a single recursive step (whereas \textbf{counting} often requires multiple recursive steps to compute the whole derivation count).

Our reference counting mechanism complements the pruning techniques discussed in Section~\ref{subsec:agg-sel}.  Following an insertion (exploration) or deletion (pruning) of a \reln{SearchSpace} tuple, we update the reference counts of relevant tuples accordingly; cascading insertions or deletions of \reln{SearchSpace} (and further \reln{PlanCost}) tuples may be triggered because their reference counts may be raised above zero (or dropped to zero). Finally, the optimal plan computed by the query optimizer is unchanged, but more
tuples in \reln{SearchSpace} and \reln{PlanCost} are pruned. Indeed,
by the end of the process, the combination of aggregate selection and
reference counts ensure \reln{SearchSpace} and \reln{PlanCost}
\emph{only} contain those plans that are on the final optimal plan tree.
Such ``garbage collection'' greatly reduces the optimizer's state and the number of data items that must be updated incrementally, as described in Section~\ref{sec:incremental}.

\eat{ML: We should emphasize the importance of this garbage collecting of unused plans: their bounds do not need to updated when they are pruned which is a huge saver!}

\Subsection{Full Branch-and-Bound Pruning}
\label{subsec:branch-and-bounding}

\begin{figure}[t]
\small
\linespread{0.9} {
%
%
%
\begin{datalog}
\heading{r1:} \head{ParentBound}{lExpr, lProp, bound-rCost-localCost} \dlog \\
\>\>\atom{Bound}{expr, prop, bound}, \atom{BestCost}{rExpr, rProp, rCost}, \\
\>\>\dlrel{LocalCost}\dlvar{(expr, prop, index, lExpr, lProp, rExpr,}\\
\>\>\hspace{+0.15in}\dlvar{rProp, -, localCost)}; \\
\heading{r2:} \head{ParentBound}{rExpr, rProp, bound-lCost-localCost} \dlog \\
\>\>\atom{Bound}{expr, prop, bound}, \atom{BestCost}{Expr, lProp, lCost}, \\
\>\>\dlrel{LocalCost}\dlvar{(expr, prop, index, lExpr, lProp, rExpr, }\\
\>\>\hspace{+0.15in}\dlvar{rProp, -, localCost)};\\
\heading{r3:} \head{MaxBound}{expr, prop, max<bound>} \dlog \\
\>\>\atom{ParentBound}{expr, prop, bound}; \\
\heading{r4:} \head{Bound}{expr, prop, min<minCost, maxBound>} \dlog \\
\>\>\atom{BestCost}{expr, prop, minCost}, \\
\>\>\atom{MaxBound}{expr, prop, maxBound};
\end{datalog}
}
\vspace{-3mm}
\caption{\label{fig:bounds} Datalog rules to express bounds computation}
\end{figure}

Our third innovation is to implement the full effect of
\emph{branch-and-bound pruning}, as in top-down optimizers like
Volcano, during cost estimation of physical plans.  Branch-and-bound
pruning uses \emph{prior exploration} of related plans to prune the
exploration of new plans: a physical plan for a subexpression is
pruned if its cost already exceeds the cost of a plan for the
equivalent subexpression, \emph{or} the cost of a plan to any parent,
grandparent, or other ancestor expression of this subexpression.
Unfortunately, branch-and-bound pruning assumes a single-recursive
descent execution thread its enumeration.  Our ultimate goal is to
find a branch-and-bounding solution independent of the search order,
and able to support parallel enumeration.

Previous work~\cite{evita-raced} has shown that it is possible to do a
limited form of branch-and-bound pruning in a declarative optimizer,
by initializing a bound based on the cost of the parent expression,
and then pruning subplan exploration whenever the cost has exceeded an
\emph{equivalent} expression. This can actually be achieved by our
aggregate selection approach described in
Section~\ref{subsec:agg-sel}.

\eat{This is achieved through a form of
\emph{aggregate selection}~\cite{sudarshan91aggregation}: in the
declarative query, to find the best plan for an expression-property
pair $\langle E, p \rangle$, we take a \textbf{min} operation across
several equivalent alternatives (this is an ``OR'' node in an and-or-graph
like Figure~\ref{fig:example-and-or-graph-Q3S}).  If we use the
current minimum as a threshold during enumeration of alternatives then
we can prune many alternative plans.}

We seek to generalize this to prune against the \emph{best} known
bound for an expression-property pair --- which may be from a plan for
an equivalent expression, or from any ancestor plan that
\emph{contains} the subplan corresponding to this expression-property
pair.  (Recall that there may be several parent plans for a subplan:
this introduces some complexity as each parent plan may have different
cost bounds, and at certain point in time we may not know the costs
for some of the parent plans.)  The bound should be continuously
\emph{updated} as different parts of the search space are explored via
pipelined execution.  In this section, we assume that bounds are
initialized to infinity and \emph{monotonically decreasing}.  In
Section~\ref{sec:incr-bound} we will relax this requirement.

Our solution, \emph{recursive bounding}, creates and updates a single
recursive relation \reln{Bound}, whose values form the \emph{best-known} bound
on each expression-property pair (each ``OR'' node).  This bound is
the minimum of (1) known costs of any equivalent plans; (2) the
highest bound of any parent plan's expression-property pair, which in
turn is defined recursively in terms of parents of this parent plan.
Figure~\ref{fig:bounds} shows how we can express the bounds table
using recursive datalog rules.  \reln{ParentBound} propagates cost
bounds from a parent expression-property pair to child
expression-property pairs, through \reln{LocalCost}, while the
child bound also takes into account the cost of the local
operator, and the best cost from the sibling side. \reln{MaxBound}
finds the highest of bounds from parent plans, and \reln{Bound}
maintains the minimum bounding information derived from
\reln{BestCost} or \reln{MaxBound}, allowing for more strict pruning.

Given the definition of \reln{Bound}, we can reason about the viability of
certain physical plans below:

\vspace{-2mm}
\begin{proposition} \label{prop:bound} Given a query expression $E$
  with desired output property $p$: let $T^s$ be a plan tree that
  produces $E$'s subexpression $E^s$ and yields property $p^s$. If
  $T^s$ has a cumulative cost that is larger than $Bound$ $\langle E^s,
  p^s \rangle$, then $T^s$ cannot be a sub-tree of the optimal
  plan tree for the query $E$, for property $p$.
\end{proposition}
\vspace{-2mm}


Based on Proposition~\ref{prop:bound}, recursive bounding may safely
remove any plan that exceeds the bound for its expression-property
pair. Indeed, with our definition of the bounds, this strategy is a
generalization of the aggregate selection strategy discussed in
Section~\ref{subsec:agg-sel}. However, bounds are recursively defined
here and a single plan cost update may result in a number of changes
to bounds for others.

Overall the execution flow of pruning \reln{PlanCost} and
\reln{SearchSpace} via recursive bounding is similar to that described in
Section~\ref{subsec:agg-sel}. Specifically, \reln{PlanCost} is pruned
inside the Fixpoint operator, where an additional comparison check
\reln{PlanCost} $<$ \reln{Bound} is performed before propagating a
newly generated \reln{PlanCost}. Updates over other \reln{Bound}
tuples derived from a given \reln{PlanCost} tuple are computed
separately.  \reln{SearchSpace} is again pruned via sideways
information passing where the pruned \reln{PlanCost} tuples are
directly mapped to deletions over \reln{SearchSpace}.

\eat{
\reminder{should discuss these in Sec 4. In addition, Bound is not monotonically goes down --- suppose the first BestCost is not obtained, then Bound is equal to MaxBound which may monotonically goes up. We should emphasis that query pre-optimizatoin and re-optimization uses the same incremental algorithms, or we can say the latter has more cases than the previous.}

We initially set \reln{BestCost} for each expression-property
combination to infinity.  
From this we can compute all of the
relations (views) in the datalog program.  Now, we update a tuple in
\reln{BestCost}. we can use standard incremental maintenance
techniques~\cite{gms93-dred} to propagate the effects recursively on
the \reln{Bound} and \reln{MaxBound} tables --- with the added wrinkle
that we only make (and propagate) an update if the aggregate (maximum
or minimum) value changes during re-evaluation.  During the
maintenance process, \reln{BestCost} monotonically goes down; \reln{MaxBound}
monotonically goes up; and \reln{Bound} monotonically goes down after an initial \reln{BestCost} has been obtained. 
}


\eat{
Recall that we have defined relations \reln{BestPlan}, capturing the
best plan for the pair of query expression $E$ and property $p$, by
computing a \textbf{min} over all plans for $\langle E, p \rangle$;
and \reln{SearchSpace}, capturing the set of enumerated physical plans
which remain viable.  Intuitively, we can consider that there to be a
join between \reln{SearchSpace} and its cost estimate in
\reln{PlanCost}, followed by a \emph{semijoin} with the cost of the
minimum-cost plan, in \reln{BestCost} --- such that \reln{SearchSpace}
tuples whose costs exceed the \reln{BestCost} for the equivalent
$\langle E, p \rangle$ are removed.  We could model this by adding
\reln{PlanCost} and \reln{BestCost} as extra atoms to the datalog rule
computing \reln{SearchSpace}.  However, this is a highly unusual
\emph{noninflationary semantics} for datalog (see~\cite[Chapter
14]{ahv95}), with mutual dependencies among relations, which is undesirable.

We instead adopt an approach that relies on \emph{over-derivation}
followed by \emph{correction}.  In essence, we first over-derive all
the tuples of \reln{SearchSpace}, and \emph{defer} the join with
\reln{PlanCost} and semijoin with \reln{BestPlan}.  Each
expression-property-plan triple $\langle E, p, T\rangle$ in
\reln{SearchSpace} gets propagated to the cost enumeration stage
\reln{PlanCost}, where $T$'s cost is estimated; then \reln{BestCost}
is computed by finding the cost of the cheapest $T$ for the pair
$\langle E, p \rangle$.  Finally the query processor applies the
semijoin between (\reln{SearchSpace} $\Join$ \reln{PlanCost}) and
\reln{BestCost}, determining which $\langle E, p, T \rangle$ triples
are at the minimum cost.  Any \reln{SearchSpace} entry not satisfying
the semijoin is then \emph{deleted}\footnote{The discussion assumes
  there is a single minimal-cost plan.  Otherwise, we arbitrarily
  break ties and remove all alternative plans except the first one.}.
We refer to this process as \emph{deferred aggregate selection}, as we
defer the evaluation of the predicate until after we have propagated
the plan to the cost enumeration phase.  When deferred aggregate
selection removes a tuple in \reln{SearchSpace}, the deletion also gets propagated downstream to remove the \reln{PlanCost}, and also recursively to
remove any further derived plan state for this particular physical
plan --- achieving pruning.
}

\eat{
We accomplish this (and other) optimizations by modifying the query processor to support \textbf{direct processing of deltas}: we could propagate a deletion of the previous $PlanCost$ after an insertion of the new $PlanCost$; instead, we may just construct a delta tuple with the same expression and property, but with a delta cost, that is, the difference between the new cost and the old cost, and propagate this delta cost to the next aggregate operator. This delta cost should always be negative, because the new cost is always smaller than the old one.}

\eat{
We accomplish this (and other) optimizations by modifying the query
processor to support \textbf{direct processing of deltas}: instead of
processing standard tuples, the operators instead process updates.  An
update to relation \reln{R} may be an insertion (\ins{R}{x}), deletion
(\del{R}{x}), or replacement (\upd{R}{x}{x'}).  The \emph{state}
maintained in each operator is annotated with positive or negative
\emph{counts}\footnote{Formally, these are
  $\mathbb{Z}$-relations~\cite{DBLP:conf/icdt/GreenIT09}.},
representing the cumulative total of how many times the tuples have
been inserted or deleted, respectively, in the update stream.  A tuple
exists in the state if an only if the count is positive in the view.
Results are processed according to a slight modification of the
standard rules of incremental maintenance~\cite{gms93-dred}, as we
discuss later.  Counts are guaranteed to remain finite despite
recursion, because a query has a finite number of subexpressions.  We
also ensure that the tuple counts will always be nonnegative, by
limiting the number of deletions and replacements in the stream.
}

\eat{
Stateful query operators maintain not only tuples, but
also the counts of how many times they have been added and/or deleted;
these counts are adjusted by summing in the incoming deltas'
counts\footnote{Tuple counts can briefly become negative if deltas are
  received out of order, but by the end will always be zero or
  positive.}.}
  
\eat{
Returning to the mutual recursion problem, after we have over-derived
the tuples of \reln{SearchSpace} corresponding to $\langle E, p
\rangle$, and computed the current \reln{BestPlan}, we apply the
semijoin between \reln{SearchSpace} and \reln{BestPlan} and propagate
deletions by injecting a (\del{R}{x}) delta with a negative value
equal to the tuple's count in \reln{SearchSpace}.  This deletion gets
propagated downstream to remove any further derived plan state for
this particular physical plan --- achieving pruning.}

\eat{
Deferred aggregate selection is, not surprisingly, an extended version
of \emph{aggregate selection}~\cite{sudarshan91aggregation}, which
pushes down \textbf{min/max} aggregates to earlier operations as
filter predicates (comparing newly arriving tuples against the current
min or max).  We in fact use ``standard'' aggregate selection during
execution as well, to ensure that we do not enumerate \reln{PlanCost}s
for plans unless they are better than the current \reln{BestPlan}
entry for the equivalent expression.}

\eat{We may use this proposition to derive our first optimization strategy
in a strong flavor of Magic Sets~\cite{magic-sets} and Sideways
Information Passing~\cite{zives-sip}:

Optimization Strategy 1: Delete the state of the plan with suboptimal
plan cost.  As can be inferred from proposition 1, the suboptimal plan
of a subexpression $E^S$ cannot be a subtree of the optimal plan of
the original expression $E$. Hence, the existence of this suboptimal
plan of a subexpression has no effect on the ultimate optimal plan of
the entire expression. We may freely delete this suboptimal sub-plan.}

\eat{
In datalog rules, this strategy is expressed as follows:
\begin{verbatim}
SearchSpace(expr, prop, index, lExpr, lProp, 
rExpr, rProp) :- existing body, BestPlan
(expr, prop, index, lExpr, lProp, rExpr, rProp,
-, -)
\end{verbatim}

If a physical plan is not selected as the best plan of its expression,
then this physical plan can be removed from the optimizer state. In
other words, $SEARCHSPACE$$($$expr$, $prop$, $index$, $lExpr$,
$lProp$, $rExpr$, $rProp$$)$ exists if and only if
$BESTCOST$$($$expr$, $prop$, $index$, $lExpr$, $lProp$, $rExpr$,
$rProp$, $metadata$, $cost$$)$ exists. Hence, the relation
$SEARCHSPACE$ has a dependency on relation $BESTPLAN$.

On the other hand, $BESTPLAN$ obviously has a dependency over
$SEARCHSPACE$. This incurs a mutually recursive program. As a feasible
execution strategy to actually implement this program, we propose the
following solution.

Suppose we have an update semantics over the relations in the datalog
program. Every tuple in a relation may have either an insertion or a
deletion operation associated with it. In the paper thereafter, we use
symbol $+$ to denote insertions; $-$ to denote deletions. Then we can
exploit delta rules. 
}

\eat{
 as insertions, then
propagate to $PlanCost$ and $BestPlan$ via positive delta rules, then
use $BestPlan$ as a filter predicate to delete unnecessary
$SearchSpace$ via negative delta rules. The process is shown as
follows,

\begin{align*}
& SearchSpace+ \Rightarrow PlanCost+ \Rightarrow BestPlan+\\
& \Rightarrow SearchSpace- \Rightarrow PlanCost- \\
\end{align*}

Because of Proposition 1, this over-derive and then-delete strategy would not affect the results because those tuples that we delete do not contribute to the final result. Meanwhile this strategy can minimize the number of $SEARCHSPACE$ tuples in the optimizer state based on the observation of Proposition 1. }

\eat{
. Our program can also
benefit from aggregate selections, because we have an aggregate of
computing min costs.  Essentially, aggregate selection corresponds to
the following program:

\begin{verbatim}
PlanCost(expr, prop, index, lExpr, lProp, rExpr, 
rProp, min, cost) :- existing body, BestPlan
(expr, prop, index, lExpr, lProp, rExpr, rProp,
min, cost)
\end{verbatim}

In other words, $PlanCost$$($$expr$, $prop$, $index$, $lExpr$, $lProp$, $rExpr$, $rProp$, $metadata$, $cost$$)$ exists if and only if $BestPlan$$($$expr$, $prop$, $index$, $lExpr$, $lProp$, $rExpr$, $rProp$, $metadata$, $cost$$)$ exists.

Similarly, the execution strategy to achieve aggregate selection through delta rules is as follows:
\begin{align*}
& SearchSpace+ \Rightarrow PlanCost+ \Rightarrow BestPlan+ \\
& \Rightarrow PlanCost-\\
\end{align*}

Through rule unfolding, we can see that our sideways information passing strategy includes the aggregate selection strategy implicitly. }

\eat{
The deletion of a $SearchSpace$ tuple may trigger more deletions because some nodes in the search space may be obsolete after some initial deletions. Hence here we develop a reference counting algorithm as a further optimization strategy to garbage collect all the obsolete $SearchSpace$ tuples.

Suppose in a typical search space AND-OR graph, we maintain a reference count $RefCount(expr, prop)$ for every OR node as the number of ancestor AND nodes it has. Here, each OR node is distinguishable by pair $(expr, prop)$, and each AND node is distinguishable by triple $(expr, prop, index)$. Every AND node may have up to two OR node children: $(lExpr, lProp)$ and $(rExpr, rProp)$. 

By definition, $RefCount(expr, prop)$ maintains a count for every OR node that is equal to the number of distinct OR node ancestors in the AND-OR graph. 

Here is the specification of $RefCount$ in datalog rules:

\begin{Dlog}
LeftRefCount(expr, prop, distinct count<pExpr, pProp>) :- 
SearchSpace(pExpr, pProp, -, -, -, expr, prop, -, -);

RightRefCount(expr, prop, distinct count<pExpr, pProp>) :- 
SEARCHSPACE(pExpr, pProp, -, -, -, -, -, expr, prop);

RefCount(expr, prop, lCount+rCount) :- LeftRefCount(expr, 
prop, lCount), RightRefCount(expr, prop, rCount);
\end{Dlog}

One can easily maintain $RefCount$ using the following procedure:

Initially, for every OR node, $RefCount(expr, prop) = 0$.

Then, every time we insert a new $SearchSpace$ tuple to the Fixpoint operator in the plan enumeration phase, we increment the count of its left and right OR node: $RefCount(lExpr, lProp)++$, $RefCOunt(rExpr, rProp)++$.

Next, every time we delete a $SearchSpace$ tuple via sideways information passing, we decrement the count of its left and right OR node child: $RefCount(lExpr, lProp)--$, $RefCount(rExpr, rProp)--$

After every update, if there exists an OR node $(expr, prop)$ that $RefCount(expr, prop) == 0$, then for every AND node children this this OR node, we decrement the $RefCount$ of their left and right OR nodes: $RefCount(lExpr, lProp)--$, $RefCount(rExpr, rProp)--$ If any of these makes its own $RefCount$ drop to zero, then it will recursively trigger this garbage collection procedure. 

Here is a proposition about $RefCount$:

\vspace{-1mm}
\begin{proposition} \label{prop:refcount} A plan $P(E^S)$ of a
  subexpression $E^S$ has reference count of zero, $RefCount(E)= 0$,
  then this plan should not be the sub-tree of the optimal plan tree
  of the original query expression $P_{OPT}{(E)}$.
\end{proposition}
\vspace{-1mm}

According to Proposition 2, by deleting the OR node whose $RefCount$ is zero, it minimizes the number of $SearchSpace$ tuples while also ensuring correct results.}

\eat{
An important property of our reference count maintenance scheme is
that it always maintains the correct information regardless of the
order in which updates are applied or propagated, due to the use of
$\mathbb{Z}$-relations.\eat{ in whatever order the updates might be.
  For example, we could decrement one's reference count before we
  increment its reference count. $RefCount$ ultimately maintains the
  correct answers no matter what implementation strategy we choose to
  execute the logic program.}  This makes it easy to ensure
correctness even with multiple execution threads.}


\eat{
--

First, we define a loose bound for every OR node, 
$LeftBound$ $(lExpr, lProp)$ $=$ $Max_{(expr, prop)}$ $(Bound(expr, prop)$ 
$-$ $BestCost$ $(rExpr, rProp)$ $-$ $LocalCost$ $(expr$, $prop$, $index))$

$RightBound$ $(rExpr, rProp)$ $=$ $Max_{(expr, prop)}$ $(Bound(expr, prop)$ 
$-$ $BestCost$ $(lExpr, lProp)$ $-$ $LocalCost$ $(expr$, $prop$, $index))$

$MaxBound(expr, prop)$ $=$ $Max$ $(LeftBound$ $(expr, prop)$, $RightBound$ $(expr, prop))$

$Bound(expr, prop)$ $=$ $Min$ $(BestCost(expr, prop))$, $MaxBound(expr, prop))$

This can also be written in a datalog program:}

\eat{
We maintain the bounds through the following process:

Step 1: Initially, $\forall (expr, prop)$, $Bound(expr, prop)$ $=$ $\inf$,
$MaxBound$ $(expr, prop)$ $=$ $\inf$.

Step 2: $\forall (expr, prop)$, when we update $BestCost$ $(expr, prop)$,
if $BestCost$ $(expr, prop)$ $<$ $Bound(expr, prop)$ then we update 
$Bound(expr, prop)$ $=$ $BestCost$ $(expr, prop)$.

Step 3: $\forall (expr, prop)$, when we update $Bound(expr, prop)$,
we check the following two conditions:
If $Bound$ $(expr, prop)$ $-$ $BestCost$ $(rExpr, rProp)$ $-$ $LocalCost$ $(expr$, $prop$, $index)$ $>$ $MaxBound$ $(lExpr, lProp)$, then update $MaxBound$ $(lExpr, lProp)$ $=$ $Bound$ $(expr, prop)$ $-$ $BestCost$ $(rExpr, rProp)$ $-$ $LocalCost$ $(expr$, $prop$, $index)$.
If $Bound$ $(expr, prop)$ $-$ $BestCost$ $(lExpr, lProp)$ $-$ $LocalCost$ $(expr$, $prop$, $index)$ $>$ $MaxBound$ $(rExpr, rProp)$, then update $MaxBound$ $(rExpr, rProp)$ $=$ $Bound$ $(expr, prop)$ $-$ $BestCost$ $(lExpr, lProp)$ $-$ $LocalCost$ $(expr$, $prop$, $index)$.

Step 4: If $\forall (expr, prop)$, when we update $MaxBound$ $(expr, prop)$,
if $MaxBound$ $(expr, prop)$ $<$ $Bound$ $(expr, prop)$ and $MaxBound$ has visited all the parents of $expr$ then we update 
$Bound$ $(expr, prop)$ $=$ $MaxBound$ $(expr, prop)$. Then we go to Step 3 again. This is a recursive process.

Here is a proposition on the relationships between $BestCost$ and $Bound$.}

\eat{

Optimization Strategy 3: Delete the state of the plan if the minimum cost of this plan's expression is larger than its bound.

If we write this optimization strategy into datalog rules, it would be:

\begin{verbatim}
SearchSpace(expr, prop, index, logOp, phyOp,
lExpr, lProp, rExpr, rProp) :- existing body, 
PlanCost(expr, prop, index, logOp, phyOp,
lExpr, lProp, rExpr, rProp, -, cost), Bound
(expr, bound), cost <= bound
\end{verbatim}

In English, this rule demonstrates that $SearchSpace$ $($ $expr$, $prop$, $index$, $lExpr$, $lProp$, 
$rExpr$, $rProp)$ exists if and only if $PlanCost$ $($ $expr$, $prop$, $index$, $lExpr$, $lProp$,
$rExpr$, $rProp$, $-$, $cost)$ exists and $cost$ $<=$ $bound$. 

Next section we will discuss incremental maintenance of branch-and-bounding via sideways information passing, where $PlanCost$ and $Bound$ may be updated, and the value would either go up or go down. 

\begin{align*}
& SearchSpace+ \Rightarrow PlanCost+ \Rightarrow BestPlan+ \\
& \Rightarrow BOUND\# \\
& PLANCOST+ > BOUND \Rightarrow SEARCHSPACE- \\
& \Rightarrow PLANCOST- \Rightarrow BESTPLAN-\\
\end{align*}
}


\Section{Incremental Re-Optimization}
\label{sec:incremental}
\eat{ML: Do incremental view maintenance techniques discussed in this section apply to general logic programming? Anything special our techniques are for this particular incremental re-opt problem? }

The previous section described how we achieve pruning at a level
comparable to a conventional query optimizer, without being
constrained to the standard data and control flow of a top-down or
bottom-up procedural implementation. In this section, we discuss
\emph{incremental} maintenance during both query optimization and
re-optimization. In particular, we seek to incrementally update not
only the state of the optimizer, but also the state that affects
pruning decisions, e.g., reference counts and bounds.

Initial query optimization takes a query expression and metadata like
summaries, and produces a set of tables encoding the plan search space
and cost estimates.  During execution, pruning bounds will always be
monotonically decreasing.  Now consider \emph{incremental}
re-optimization, where the optimizer is given updated cost (or
cardinality) estimates based on information collected at runtime after
partial execution.  This scenario commonly occurs in adaptive query
processing, where we monitor execution and periodically re-optimize
based on the updated status.  (We reiterate that our focus in this
paper is purely on incremental re-optimization techniques, not the
full adaptive query processing problem.)  For simplicity, our
discussion of the approaches assumes that a single \emph{cost
  parameter} (operator estimated cost, output cardinality) changes.
Our implementation is able to handle multiple such changes
simultaneously.

Given a change to a cost parameter, our goal is in principle to
re-evaluate the costs for all affected query plans.  Some of these
plans might have previously been pruned from the search space, meaning
they will need to be re-enumerated.  Some of the pruning bounds might
need to be adjusted (possibly even raised), as some plans become more
expensive and others become cheaper.  As the bounds are changed, we
may in turn need to re-introduce further plans that had been
previously pruned, or to remove plans that had previously been viable.
This is where our declarative query optimizer formulation is extremely
helpful: we use \emph{incremental view maintenance} techniques to only
recompute the necessary results, while guaranteeing correctness.


\Paragraph{Incremental maintenance enabled via datalog.}
From the declarative point of view, initial query optimization and
query re-optimization can be considered roughly the same task, if the
data model of the datalog program is \textbf{extended} to include
updates (insertions, deletions and replacements). Indeed, incremental
query re-optimization can be specified using a delta rules formulation
like~\cite{gms93-dred}.  This requires several extensions to the
database query processor to support \textbf{direct processing of
  deltas}: instead of processing standard tuples, each operator in the
query processor must be extended to process delta tuples encoding
changes.  A delta tuple of a relation \reln{R} may be an insertion
(\ins{R}{x}), deletion (\del{R}{x}), or update (\upd{R}{x}{x'}). For
example, a new plan generated in \reln{SearchSpace} is an insertion; a
pruned plan in \reln{PlanCost} is a deletion; an updated cost of
\reln{BestCost} is an update.

We extend the query processor following standard conventions from
continuous query systems~\cite{liu_diff_eval} and stream management
systems~\cite{stanford-stream}.  The extended query operators consume
and emit deltas largely as if they were standard tuples.  For stateful
operators, we maintain for each encountered tuple value a (possibly
temporarily negative) \emph{count},
representing the cumulative total of how many times the tuple has been
inserted and deleted. Insertions increment the count and deletions
decrement it; counts may temporarily become negative if a deletion is
processed out of order with its corresponding insertion, though
ultimately the counts converge to nonnegative values, since every
deletion is linked to an insertion.  A tuple only affects the output
of a stateful operator if its count is positive.

\eat{In representing the internal state of counts, we use the formulation
of $\mathbb{Z}$-relations~\cite{DBLP:conf/icdt/GreenIT09}, in which
negative counts are (temporarily) allowed, such that $(A-B)+B \equiv
(A+B)-B$: under our problem formulation the final number of deletions
of a tuple can never exceed the final number of insertions, and it is
advantageous to ensure that any arbitrary ordering of deletions and
insertions will result in the same final answer.}

Upon receiving a series of delta tuples, every query operator (1) updates
its corresponding state, if necessary; (2) performs any internal
computations such as predicate evaluation over the tuple or against
state; (3) constructs a set of output delta tuples.  Joins follow the
rules established in~\cite{gms93-dred}.  For aggregation operators
that compute minimum (or maximum) values, we must further extend the
internal state management to keep track of \emph{all} values
encountered --- such that, e.g., we can recover the
``second-from-minimum'' value.  If the minimum is deleted, the
operator should propagate an update delta, replacing its previous
output with the next-best-minimum for the associated group (and
conversely for maximum).

\Paragraph{Challenge: recomputation of pruned state.}
While datalog allows us to propagate of updates through rules,
a major challenge is that the pruning strategies of Section~\ref{sec:sip} 
are achieved \emph{indirectly}.  In this section we detail how we incrementally re-create pruned state as necessary.
Section~\ref{sec:incr-search} shows how we incrementally maintain the
output of aggregate selection and ``undo'' tuple source suppression.
Section~\ref{sec:incr-count} describes how to incrementally adjust the
reference counts and maintain the pruned plans. Finally,
Section~\ref{sec:incr-bound} shows how we can incrementally modify the
pruning bounds and the affected plans.


\eat{
In general, to support this direct processing of deltas, one needs to implement the following logic into an operator in a query processor:
\begin{enumerate}
\vspace{-3mm}
\item Initially every count is set to zero in the state.
\vspace{-3mm}
\item Process a delta tuple and update the count in the state. \ins{R}{x}: increment x's count by 1; \del{R}{x}: decrement x's count by 1; \upd{R}{x}{x'}: decrement x's count by 1 and increment the count of x' by 1.
\vspace{-3mm}
\item If the count of x is updated from 0 to 1, then \ins{R}{x} is
  processed; if the count of x is updated from 1 to 0, then \del{R}{x}
  is processed; if the count of x is updated from 1 to 0, and the
  count of x' is updated from 0 to 1, then \upd{R}{x}{x'} is
  processed.  \vspace{-3mm}
\item For certain operators, e.g., joins, one need to carefully design
  the operations of joining a delta tuple with the corresponding
  tuples in the state, and generating the output delta tuple to the next     
  operator.
\end{enumerate}
}

\Subsection{Incremental Aggregate Selection}
\label{sec:incr-search}

Aggregate selection~\cite{sudarshan91aggregation} prunes state against
bounds and does not consider how incremental maintenance might change
the bound itself. Our incremental aggregate selection algorithm is a generalization of
the non-incremental case we describe in Section~\ref{subsec:agg-sel}.
Recall that we push down a selection predicate, \reln{PlanCost} $<$
\reln{BestCost}, within the Fixpoint operator that generates
\reln{PlanCost}.  To illustrate how this works, consider how we
may revise \reln{BestCost} and \reln{BestPlan} after encountering
an insertion, deletion or update to \reln{PlanCost}.  There are four
possible cases:

\begin{compactEnumerate}
\vspace{-3mm}
\item Upon an \emph{insertion} \ins{\reln{PlanCost}}{c}, set \reln{BestCost} to \textbf{min} $(c,$ current \reln{BestCost}$)$.
\vspace{-3mm}
\item Upon a \emph{deletion} \del{\reln{PlanCost}}{c}, set \reln{BestCost} to the next-best \reln{PlanCost} iff the current \reln{BestCost} is equal to $c$.
\vspace{-3mm}
\item Upon a \emph{cost update} \upd{\reln{PlanCost}}{c}{c'}, if $c < c'$, set  \reln{BestCost} to \textbf{min} $(c'$, next-best \reln{PlanCost} $)$ iff the current \reln{BestCost} is equal to $c$. 
\vspace{-3mm}
\item Upon a \emph{cost update} \upd{\reln{PlanCost}}{c}{c'}, if $c > c'$, if the current \reln{BestCost} is equal to $c$, then set \reln{BestCost} to $c'$; else set \reln{BestCost} to \textbf{min} $(c',$ current \reln{BestCost}$)$.
\vspace{-3mm}
\end{compactEnumerate}

Recall that each \reln{PlanCost} tuple denotes a newly computed cost
associated with a physical plan, and a \reln{BestCost} tuple denotes
the best cost that has been computed so far for this physical plan's
expression-property pair. We update \reln{BestCost} based on the
current state of \reln{PlanCost}. In Cases 1 and 4, we can directly
compute updates to \reln{BestCost}.  In Cases 2 and 3, we rely on the
fact that the aggregate operator preserves all the computed, even
pruned \reln{PlanCost} tuples (as described previously), so it can
find the ``next best'' value even if the minimum is removed.  In our
implementation we use a priority queue to store the sorted tuples.



We may also need to re-introduce tuples in \reln{SearchSpace} that
were suppressed when they led to \reln{PlanCost} tuples that were pruned,
we achieve this by propagating an insertion (rather than deletion as in
Section~\ref{subsec:agg-sel}) to the previous stage.

\Subsection{Incremental Reference Checking}
\label{sec:incr-count}

Once we have updated the set of viable plans for given expressions in
the search space, we must consider how this impacts the viability of
their subplans: we must incrementally update the reference counts on
the child expressions to determine if they should be left the same,
re-introduced, or pruned.  As before, we simplify this process and
make it order-independent through the use of incremental maintenance
techniques.

\eat{
In Section~\ref{subsec:refcount} we explained that the
\reln{SearchSpace} relation is defined recursively, where each
subexpression is defined as the result of running a split operation on
the parent expression, then a join with the set of possible plan
properties.Reference counts were originally maintained as
$\mathbb{Z}$-relation annotations over streams of insertions and
deletions to this relation.  As we re-insert a tuple into
\reln{SearchSpace}, or remove a tuple, applying incremental
maintenance techniques to the recursive view will \emph{automatically}
trigger adjustments to the reference counts of subexpressions,
re-introducing tuples if their reference count moves from zero to a
positive value, and removing tuples if their reference count moves to
zero.  (If a tuple is further deleted, we save the tuple with a
negative count, but do not propagate the changes ``downstream'' until
we reach a positive value.)}

We incrementally and recursively maintain the reference counts for
each expression-property pair whenever an
associated plan in the \reln{PlanCost} relation is inserted, deleted
or updated.  When a new entry is inserted into \reln{PlanCost}, we
increment the count of each of its child expression-property pairs; similarly, whenever an existing entry is deleted from \reln{PlanCost}, we decrement each child reference count. Replacement values for \reln{PlanCost} entries do not change the reference counts, but may recursively affect the \reln{PlanCost}
entries for super-expressions. Whenever a count goes from 0 to 1 (or drops from 1 to 0) we recompute (prune, respectively) all of the physical plans associated with this expression-property pair.

\eat{
When maintaining \reln{SearchSpace} state, we assume that tuples not
currently in the table have a count of 0, and an initial insert introduces
the tuple into the table with count 1. Subsequent inserts and deletes of the tuple increment and decrement this value, respectively. }

If we combine this strategy with aggregate selection, only the
best-cost plan needs to be pruned or re-introduced (all others are
pruned via aggregate selection). Similar to
Section~\ref{sec:incr-search}, the aggregate operators internally
maintain a record of all \reln{PlanCost} tuples they have received as
input, so ``next-best'' plans can be retrieved if the best-cost entry
gets deleted or updated to a higher cost value.  During incremental
updates, we only propagate changes affecting the old and new best-cost
plan and all recursively dependent plans.

\Subsection{Incremental Branch-and-bounding}
\label{sec:incr-bound}

\eat{ML: Market this section more. Should emphasize on the delta update part of branch-and-bounding: we are not just bookkeeping all stuff to preserve MIN/COUNT maintenance, we are enabling delta-based state propagation that bounds can be updated and the condition of cost <= bound is incrementally maintained as well! This is certainly non-trivial, important and novel. But we should emphasize more to avoid people like reviewer1 from thinking this is naive! Discuss memory-saving data structures like heaps and show in the experiments that memory overhead is neglegible.}

Perhaps the most complex pruning technique to adapt to incremental maintenance is 
the branch-and-bound pruning structure of Section~\ref{subsec:branch-and-bounding}:  as new costs for
any operation are discovered, we must recursively recompute the bounds for
all super-expressions.  As necessary we then
update \reln{PlanCost} and \reln{SearchSpace} tuples
based on the updated bounds.  Recall from Figure~\ref{fig:bounds} that
the \reln{Bound} relation's contents are computed recursively based on
the \textbf{max} bounds derived from parent plans; and also based on
the \textbf{min} values for equivalent plan costs. Hence, an update to
\reln{LocalCost} or \reln{BestCost} may affect the entries in
\reln{Bound}.  Here we again rely in part on the fact that \reln{Bound} 
is a recursive query and we can incrementally maintain it, then use
its new content to adjust the pruning.
We illustrate the handling of cost updates by looking at what happens
when a cost \emph{increases}.

\eat{
\begin{compactEnumerate}
\vspace{-3mm}
\item Upon an updated cost on a plan \upd{\reln{LocalCost}}{c}{c'}, if $c < c'$, then \reln{ParentBound} of this plan's children expression-property pairs may increase, in turn \reln{MaxBound} of those children may increase, and \reln{Bound} of those children may potentially increase. Ultimately, \reln{Bound} of the updated plan's descendants may increase.
\vspace{-3mm}
\item Upon an update cost on an expression-property pair's \upd{\reln{BestCost}}{c}{c'}, if $c < c'$, then \reln{Bound} of its children expression-property pairs may increase, in turn \reln{Bound} of this expression-property pair's descendants may increase. Meanwhile, \reln{ParentBound} of the sibling expression-property pairs may decrease, in turn their \reln{MaxBound} and potentially \reln{Bound} may decrease. Hence, \reln{Bound} of the sibling expression-property pair's descendants may ultimately decrease.
\vspace{-3mm}
\end{compactEnumerate}

\noindent
For Case 1, s}

Suppose a plan's \reln{LocalCost} increases.  As a consequence of the
rules in Figure~\ref{fig:bounds}, the \reln{ParentBound} of this
plan's children may increase due to rules r1 and r2. \reln{MaxBound}
is then updated by r3 to be the maximum of the \reln{ParentBound}
entries: hence it may also increase.  As in the previous cases, the
internal aggregate operator for \reln{ParentBound} maintains all input
values; thus, it can recompute the new minimum bound and output a
corresponding update from old to new value.  Finally, as a result of
the updated \reln{ParentBound}, \reln{Bound} in r4 may also increase.
The process may continue recursively to this plan's descendant
expression-property pairs, until \reln{Bound} has converged to the
correct bounds for all expression-property pairs.

Alternatively, suppose an expression-property pair's \reln{BestCost}
estimate increases (e.g., due to discovering the machine is heavily
loaded).  This may trigger an update to the corresponding entry in
\reln{Bound} (via rule r4).  Moreover, via rules r1 and r2, an update
to this bound may affect the bounds on the parent expression, i.e.,
\reln{ParentBound}, and thus affecting any expression whose costs were pruned
via \reln{ParentBound}.

The cases for handling cost \emph{decreases} are similar (and
generally simpler).  Sometimes, in fact, we get simultaneous changes
in different directions.  Consider, for instance, that an expression's
cost bound may increase, as in the previous paragraph.  At the same
time, perhaps the expression-property pair's \reln{ParentBound} may
decrease.  Any equivalent plan (sibling expression) for our original
expression-property pair is bounded \emph{both} by the bounds of
sibling expressions and parents.  As \reln{ParentBound} decreases,
\reln{MaxBound} and \reln{Bound} may also potentially decrease through
r3 and r4.  The results are guaranteed to converge to the best of the
sibling and parent bounds.

So far we focused only on how to update bounds given updated cost
information; of course, there is the added issue of updating the
pruning results. Recall in Section~\ref{subsec:branch-and-bounding}
that we evaluate the following predicate $\phi$ before propagating a
newly generated \reln{PlanCost} value: if \reln{PlanCost} $<$
\reln{Bound} then set \reln{Bound} to \reln{PlanCost}. When
\reln{PlanCost} or \reln{Bound} is updated, we can end up in any of 3
cases:

\begin{compactEnumerate}
\vspace{-3mm}
\item Upon an update to a plan cost entry, i.e.,
  \ins{\reln{PlanCost}}{c},\protect\linebreak\del{\reln{PlanCost}}{c}
  or \upd{\reln{PlanCost}}{c}{c'}: if predicate $\phi$'s result
  changes from false to true, then emit an insertion of the
  \reln{PlanCost} tuple; otherwise if $\phi$'s result changes from
  true to false, then emit a deletion.  Incrementally update the
  corresponding \reln{Bound} entry, including its aggregated cost
  value, as a result.  \vspace{-3mm}
\item Upon an update on \upd{\reln{Bound}}{b}{b'} where $b < b'$: for those
  tuples $t$ in \reln{PlanCost} where $b < t.cost < b'$, re-insert $t$
  into \reln{PlanCost} and re-insert $t$'s counterpart in \reln{SearchSpace}  
  to \emph{undo} tuple source suppression.  
  \vspace{-3mm}
\item Upon an update on \upd{\reln{Bound}}{b}{b'} where $b > b'$: for those
  tuples $t$ in \reln{PlanCost} where $b > t.cost > b'$, prune tuple
  $t$ from \reln{PlanCost} and delete $t$'s counterpart from \reln{SearchSpace}
  via tuple source suppression.  
\vspace{-3mm}
\end{compactEnumerate}

Indeed, the first step is similar to incremental aggregate selection
(Section~\ref{sec:incr-search}). The main difference is that here the
condition check is not on \reln{BestCost} but rather on
\reln{Bound}. Essentially we want to incrementally update \reln{Bound}
based on the current bounding status, hence a sorted list of
\reln{PlanCost} tuples needs to be maintained.

An interesting observation of Cases 2 and 3 is that an update on
\reln{Bound} may affect the pruned or propagated plans as well. If a
bound is raised, it may re-introduce previously pruned plans; if a
bound is lowered, it may incrementally prune previously viable
plans. If incremental aggregate selection is used, then only the
optimal plan among the pruned plans needs to be revisited.
\reln{SearchSpace} is again updated via sideways information passing.


\eat{
\begin{align*}
& SEARCHSPACE+, SEARCHSPACE- \\
& PLANCOST+, PLANCOST- \\
& PlanCost\#(cost<cost'), PlanCost\#(cost>cost') \\
\end{align*}

\paragraph{PLANCOST : SEARCHSPACE }

\begin{align*}
& SEARCHSPACE+ \Rightarrow PLANCOST\#(cost<=cost') \\
& SEARCHSPACE- \Rightarrow PLANCOST\#(cost>=cost') \\
\end{align*}

\paragraph{ BESTCOST(min<cost>) : PLANCOST }

\begin{align*}
& PLANCOST\#(cost<cost') \Rightarrow BESTCOST\#(cost<=cost') \\
& PLANCOST\#(cost>cost') \Rightarrow BESTCOST\#(cost>=cost') \\
\end{align*}

\paragraph{ BESTPLAN : PLANCOST, BESTCOST }

\begin{align*}
& PLANCOST+ \Rightarrow BESTCOST\#(cost>=cost') \\
& \Rightarrow BESTPLAN\#(other,this) \\
& PLANCOST- \Rightarrow BESTCOST\#(cost<=cost') \\
& \Rightarrow BESTPLAN\#(this,other) \\
& PLANCOST\#(cost<cost') \Rightarrow BESTPLAN\#(this,other) \\
& PLANCOST\#(cost>cost') \Rightarrow BESTPLAN\#(other,this) \\
\end{align*}

Three optimization rules:

\paragraph{ SEARCHSPACE : BESTPLAN }

\begin{align*}
& BESTPLAN+ \Rightarrow SEARCHSPACE+ \\
& BESTPLAN- \Rightarrow SEARCHSPACE- \\
& BESTPLAN\#(other,this) \Rightarrow SEARCHSPACE-(other), \\
& SEARCHSPACE+(this) \\
\end{align*}

\paragraph{Incremental updates.}
In principle, the ``base values'' in our declarative query optimizer
are the input query expression and the cost and cardinality estimate
values for each subexpression, so one might imagine these will be the
values that are incrementally updated.  However, in our target problem
space, the query expression itself will never change: only the cost
estimates will be revised (manifesting themselves as updates to the
contents of the \reln{PlanCost} table).  In turn, updates to tuples
\reln{PlanCost} should trigger updates to other \reln{PlanCost} tuples
(e.g., for super-plans of the newly re-estimated plans), to the
pruning \reln{Bound} relation described in the previous section, and
to the contents of the \reln{BestCost} and \reln{BestPlan} for
each query expression.  These updates will also ultimately result in
insertions, deletions or updates into the \reln{SearchSpace} and
\reln{PlanCost} tables as a result of changes to pruning bounds.

There we assumed that we were
doing the \emph{initial} optimization of a query plan, and that all
pruning bounds advanced in a \emph{monotonic} fashion: pruning bounds
moved from infinity towards the best possible cost, and we sought to
discard portions of the search space as we encountered plans that
exceeded the bounds.
}

\eat{
Applications: adaptive query processing where query optimization and
query execution is interleaved. Relation statistics or plan estimators
or system run-time conditions might change.
}

\eat{
Following the techniques of Section~\ref{subsec:agg-sel}, during its
initial execution the query optimizer will have pruned portions of the
\reln{SearchSpace} table: for a particular query expression $E$ and
property $p$, any query plan for $\langle E,p \rangle$ other than the
one with minimum cost, i.e., the entry stored in relation
If a plan $T$ is associated with $\langle E,p \rangle$ in
\reln{BestPlan}, then it is maintained in \reln{SearchSpace};
otherwise it is removed.

Of course, if we change a given cost parameter, then the
\reln{BestPlan} entry for any $\langle E, p \rangle$ may change.
Given an update to one of the cost parameters to $\langle E, p
\rangle$, the query processor does the following:

\begin{enumerate}

\vspace{-3mm}
\item Re-estimate the \reln{PlanCost} for the current plan $T$
  associated with $\langle E, p \rangle$ in \reln{BestPlan}, and set
  \reln{BestCost} to that value.

\vspace{-3mm}
\item Incrementally re-enumerate every possible plan $T'$ other than
  $T$ for $\langle E, p \rangle$, by taking expression $E$ and
  recomputing the plan enumeration stage for it, to produce all plans
  $T' \in (\cal P - \{ P \})$.

\vspace{-3mm}
\item For each \reln{SearchSpace} tuple representing $\langle E, p, P'
  \rangle$, compute a \reln{PlanCost} associating a cost to the plan.
  Incrementally update the \reln{BestCost} and \reln{BestPlan} as each
  new plan is evaluated for $\langle E, p$.  If a \reln{PlanCost} for
  $\langle E, p, P \rangle$ exceeds the current \reln{BestCost} for
  $\langle E, p \rangle$, then remove $\langle E, p, P' \rangle$ from
  the \reln{SearchSpace}.

\end{enumerate}

\vspace{-1mm}
\begin{proposition} \label{prop:viability-2} Given a query expression
  $E$ and property $p$.  Then the incremental update procedure
  described above computes a new plan $T$, such that (1) the $T$ is
  the plan associated with $\langle E, p \rangle$ in \reln{BestPlan};
  (2) $T$'s cost appears with $\langle E, p \rangle$ in
  \reln{BestCost}; (3) $T$ is the only plan associated with $\langle
  E, p \rangle$ in \reln{SearchSpace} and \reln{PlanCost}; (4) $T$ is
  the plan of minimal cost for $\langle E, p \rangle$.
\end{proposition}
}
\vspace{-1mm}


\eat{
re-evaluate the semijoin between \reln{SearchSpace} and
\reln{BestPlan}, using a delta rules formulation
like~\cite{gms93-dred}.  Suppose that the \reln{BestPlan} for
expression/property pair $(E,p)$ is replaced, such that plan $P'$ is
preferred instead of the old plan $P$.  Then after incremental
re-evaluation, the tuple in \reln{SearchSpace} corresponding to
$(E,p,P)$ will be replaced with one matching $(E,p,P')$.  Similarly, a
delete of $(E,p,P)$ from \reln{BestPlan} will trigger a deletion in
\reln{SearchSpace}, and likewise for an insertion.

\reminder{Where can we have an insertion into BestPlan, given that
  there should already be one entry for each plan?  Presumably only
  during initial evaluation, which isn't really the target here...?
  And how do we re-introduce SearchSpace entries to re-estimate their
  costs?  Perhaps we should formulate this in terms of PlanCost
  instead of SearchSpace??}
}

\eat{
\paragraph{ COUNT : SEARCHSPACE }

\begin{verbatim}
LEFTREFCOUNT(count<count>) :- SEARCHSPACE(parent)
RIGHTREFCOUNT(count<count>) :- SEARCHSPACE(parent)
REFCOUNT :- LEFTREFCOUNT, RIGHTREFCOUNT
SEARCHSPACE: REFCOUNT > 0
\end{verbatim}
  
\begin{align*} 
& SEARCHSPACE+(parent) \Rightarrow LEFTREFCOUNT\# \\
& (count<=count'), RIGHTREFCOUNT\#(count<=count') \\
& \Rightarrow REFCOUNT\#(count<=count') \\
& \Rightarrow SEARCHSPACE+ \\
& SEARCHSPACE-(parent) \Rightarrow LEFTREFCOUNT\# \\
& (count>=count'), RIGHTREFCOUNT\#(count>=count') \\
& \Rightarrow REFCOUNT\#(count>=count') \\
& \Rightarrow SEARCHSPACE- \\
\end{align*}
}

\eat{
\paragraph{ BOUND: BESTCOST, LOCALCOST }

\begin{verbatim}
LEFTBOUND(max<bound-rCost-localCost>) :- 
BOUND(parent), BESTCOST(sibling), LOCALCOST(parent)

RIGHTBOUND(max<bound-lCost-localCost>) :- 
BOUND(parent), BESTCOST(sibling), LOCALCOST(parent)

BOUND :- BESTCOST, LEFTBOUND, RIGHTBOUND, min
(bestCost, max(lBound,rBound))

SEARCHSPACE :- PLANCOST > BOUND
\end{verbatim}
	 
parent node:	 
\begin{align*}
& BOUND\#(parent, bound>=bound') \Rightarrow \\
& LEFTBOUND\#(bound>=bound'), RIGHTBOUND\# \\
& (bound>=bound') \Rightarrow BOUND\# \\
& (bound>=bound') \\
& BOUND\#(parent, bound<=bound') \Rightarrow \\
& LEFTBOUND\#(bound<=bound'), RIGHTBOUND\# \\
& (bound<=bound') \Rightarrow BOUND\#(bound<=bound') \\
\end{align*}

sibling node:
\begin{align*}
& BESTCOST\#(sibling, cost>=cost') \Rightarrow \\
& LEFTBOUND\#(bound<=bound'), RIGHTBOUND\# \\
&(bound<=bound') \Rightarrow BOUND\#(bound<=bound') \\
& BESTCOST\#(sibling, cost<=cost') \Rightarrow \\
& LEFTBOUND\#(bound>=bound'), RIGHTBOUND\# \\
& (bound>=bound') \Rightarrow BOUND\#(bound>=bound') \\
\end{align*}

this node:	 
\begin{align*}
& BESTCOST\#(cost>=cost') \Rightarrow BOUND\# \\
& (bound>=bound') \\
& BESTCOST\#(cost<=cost') \Rightarrow BOUND\# \\
& (bound<=bound') \\
& BOUND\#(bound>=bound') \Rightarrow SEARCHSPACE+ \\
& (planCost, bound'<=planCost<=bound) \\
& BOUND\#(bound<=bound') \Rightarrow SEARCHSPACE- \\
& (planCost, bound<=planCost<=bound') \\
& PLANCOST\#(cost>=cost') \Rightarrow SEARCHSPACE- \\
& PLANCOST\#(cost<=cost') \Rightarrow SEARCHSPACE+ \\
\end{align*}
}

\eat{
After a cost
parameter update, we will either receive a modification to
\reln{BestCost} or to a \reln{LocalCost}.
We must propagate these
using incremental maintenance techniques, through joins using standard
delta-rules formulations~\cite{gms93-dred} and also through
\textbf{min} or \textbf{max} aggregation computations.

We propagate an individual update through the aggregate operator as
follows, extending the incremental maintenance-through-aggregation
techniques developed in~\cite{recursive-views} to support replacement
updates:

\begin{enumerate}
\vspace{-3mm}
\item Suppose $x$ is a tuple with grouping key $k_x$ and cost
  $cost(x)$.  The update \upd{R}{x}{x'} may have \emph{no effect}, if
  $x$ is a tuple other than the current best tuple, and $x$'s cost
  value $cost(x)$ is dominated by the current best tuple $m$ (e.g., if
  the operator is \textbf{min}, and $cost(x) > cost(m)$).

\vspace{-3mm}
\item The update may \emph{result in a new aggregate value for the
    same plan}, e.g., if the operator is \textbf{min}, $x$ is the
  minimum-cost plan, and $cost(x') < cost(m)$.  Here we propagate the
  update \upd{}{($k_x$,cost(x))}{($k_x$,cost(x'))} to the next
  operator.  Note also that $cost(x')$ can be greater than $cost(x)$
  if $x$ corresponds to a plan that remains the best.  (As
  in~\cite{recursive-views} the aggregate operator must maintain all
  of the alternative tuples in its internal state, e.g., using a
  priority queue.)

\vspace{-3mm}
\item The update may \emph{result in a new plan}, e.g., if the
  operator is \textbf{min}, $x$ is a plan that was not previously of
  minimum cost, but $x'$ costs less than the current best plan $m$.  Here
  we propagate the update \upd{}{($k_m$,cost(m))}{($k_{x'}$,cost(x'))} through the aggregate operator,
    where $m$ was the previous minimum-scoring plan.

\end{enumerate} }

\eat{
We repeat this propagation recursively to recompute \reln{Bound}.  As
we do this, we must re-evaluate entries in \reln{SearchSpace},
comparing them against the matching tuples in \reln{Bound}.  Those
that now exceed the bounds should be removed, and those that are at or
below the bounds should be propagated.
}

\Section{Evaluation}
\label{sec:eval}
\begin{figure*}[t]
\begin{minipage}[t]{2,25in}
\begin{center}
\includegraphics[width=2.25in,height=1.25in]{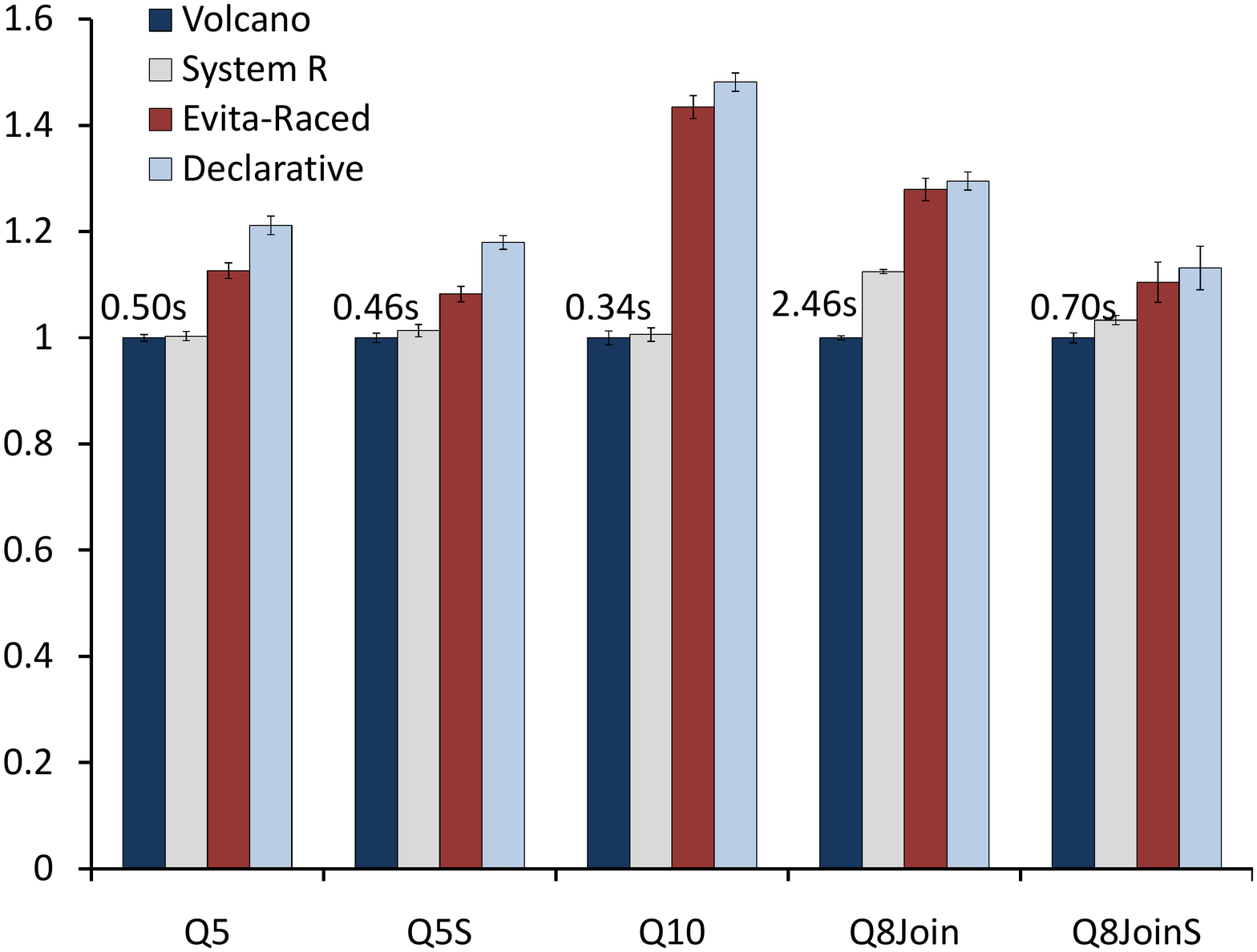}
\vspace{-2mm}
\small (a) Execution time (normalized to Volcano)
\end{center}
\end{minipage}
\hfill
\begin{minipage}[t]{2.25in}
\begin{center}
\includegraphics[width=2.25in,height=1.25in]{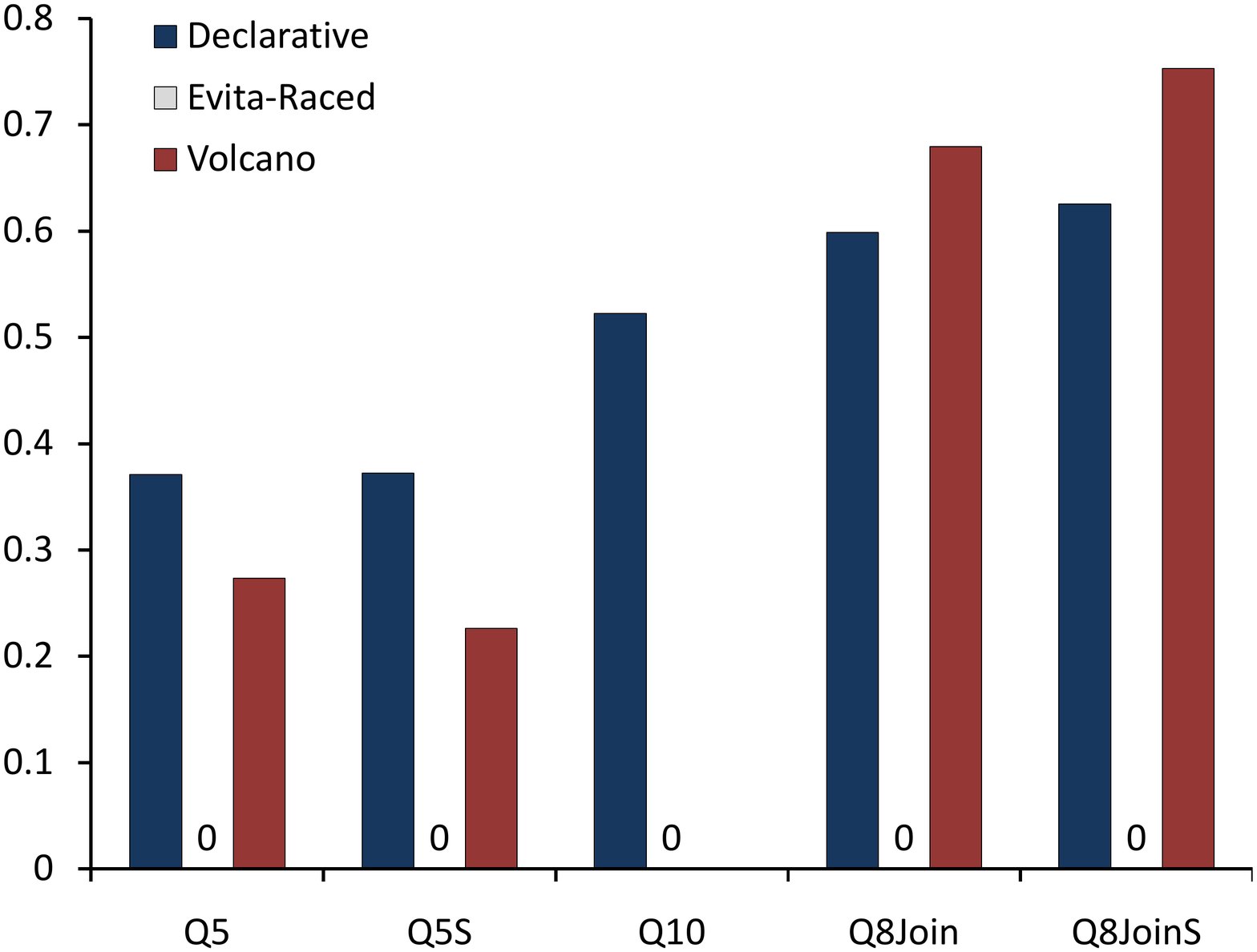}
\vspace{-2mm}
\small (b) Pruning ratio: plan table entries
\end{center}
\end{minipage}
\hfill
\begin{minipage}[t]{2.25in}
\begin{center}
\includegraphics[width=2.25in,height=1.25in]{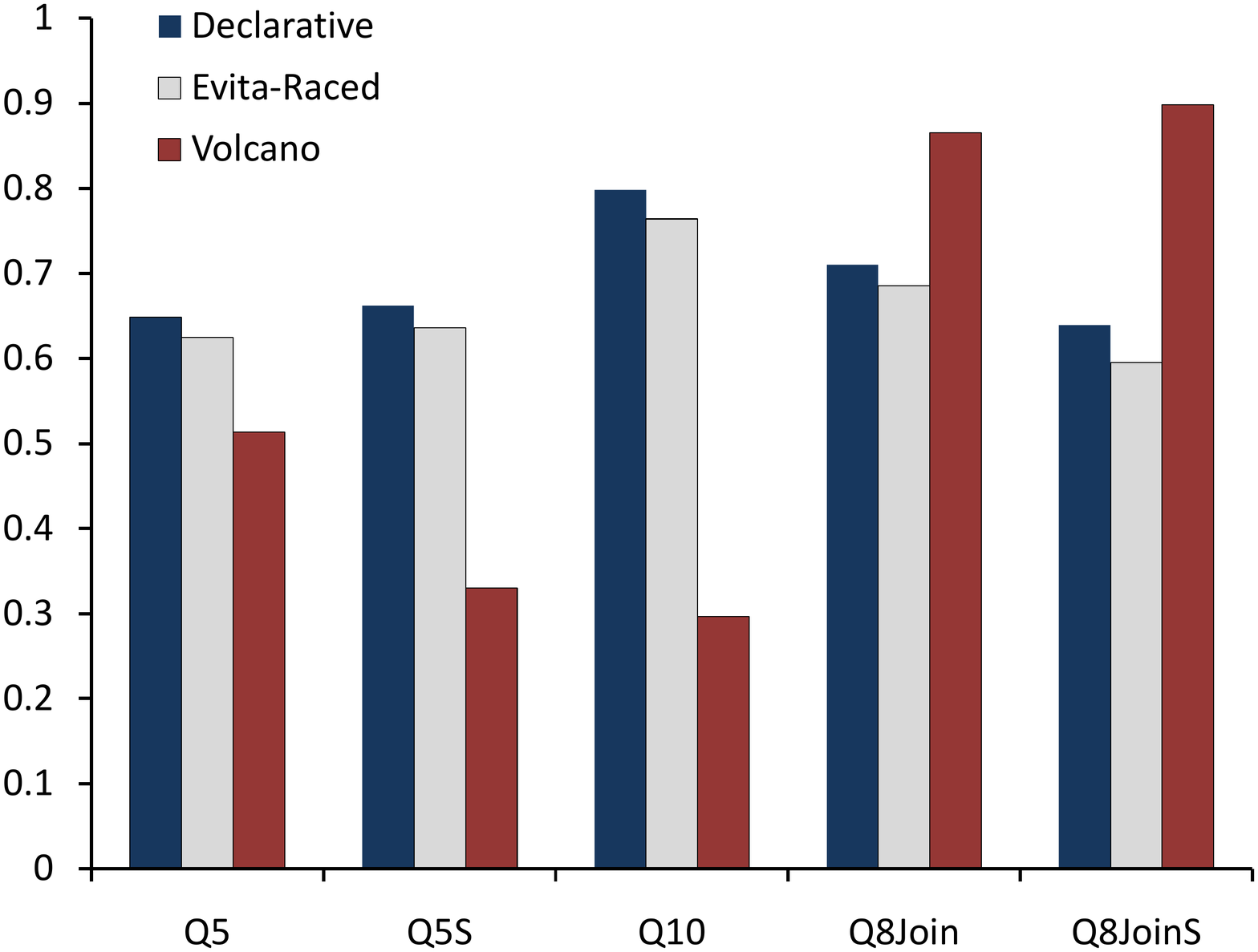}
\vspace{-2mm}
\small (c) Pruning ratio: plan alternatives
\end{center}
\end{minipage}
\caption{\small Performance comparison for initial query optimization, across different optimizer architectures \label{fig:declarative-procedural}}
\end{figure*}

\eat{
\begin{figure*}[t]
\begin{minipage}[t]{2.25in}
\begin{center}
\includegraphics[width=2.25in]{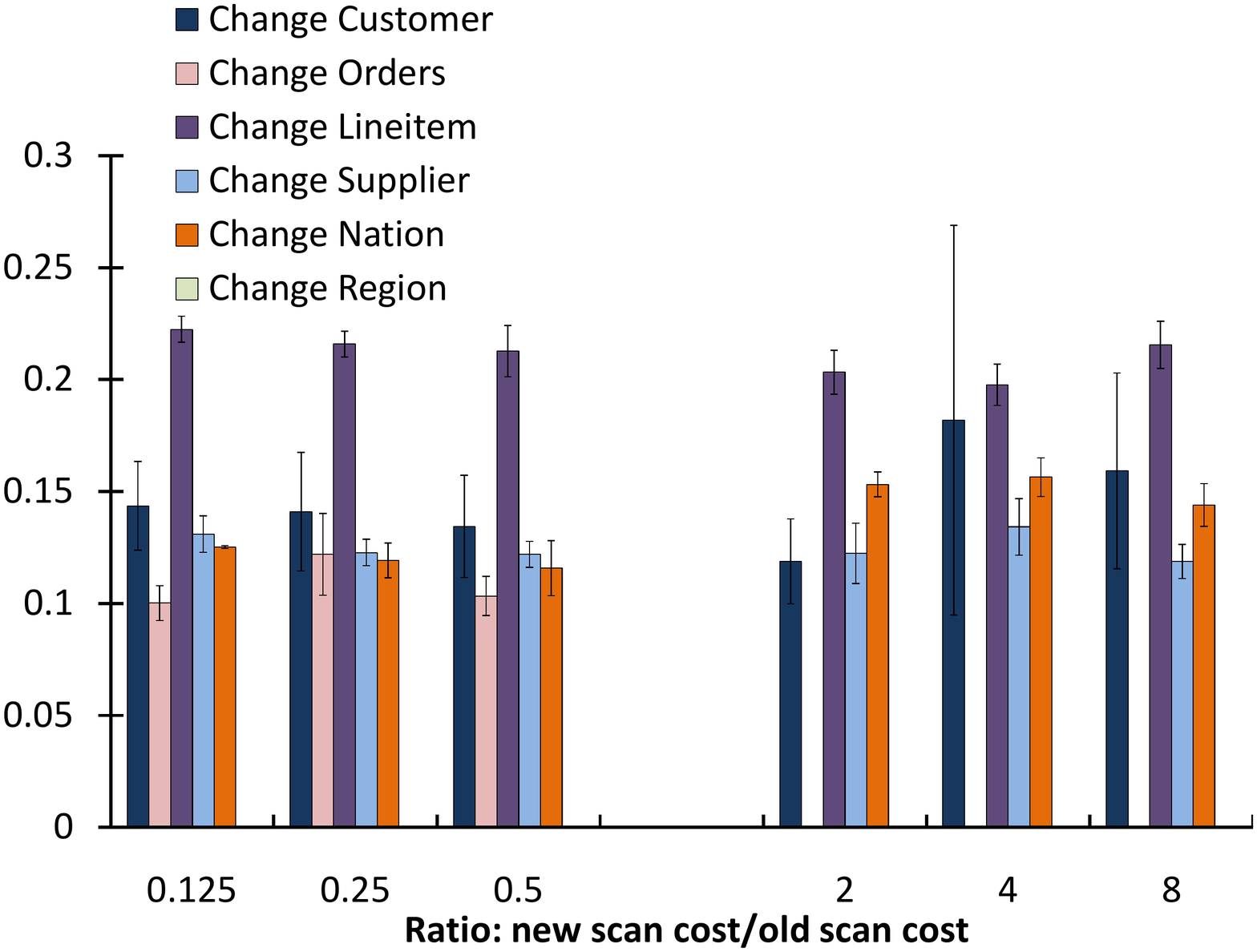}
\vspace{-2mm}
\small (a) Execution time (normalized to Volcano)
\end{center}
\end{minipage}
\hfill
\begin{minipage}[t]{2.25in}
\begin{center}
\includegraphics[width=2.25in]{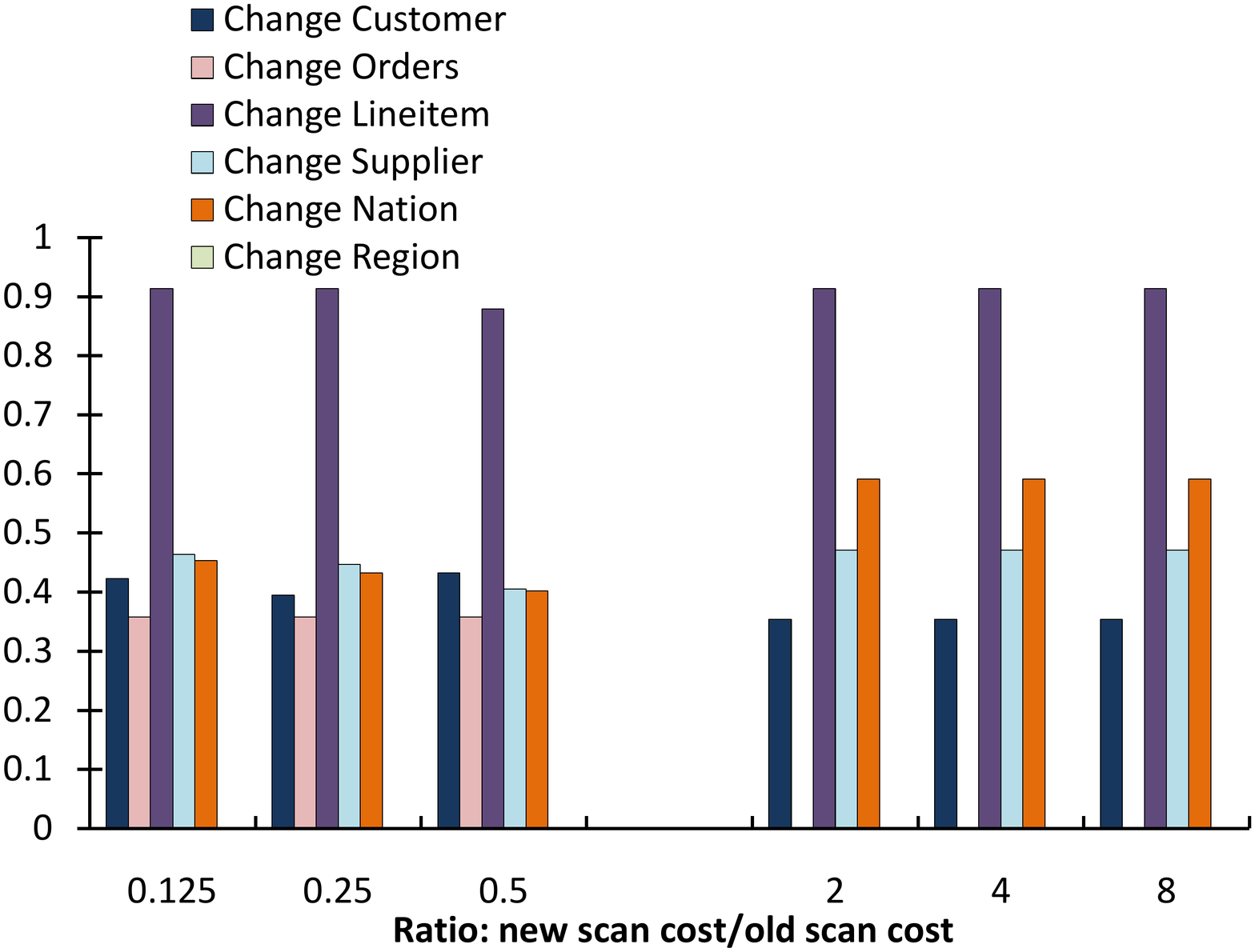}
\vspace{-2mm}
\small (b) Update ratio: plan table entries
\end{center}
\end{minipage}
\hfill
\begin{minipage}[t]{2.25in}
\begin{center}
\includegraphics[width=2.25in]{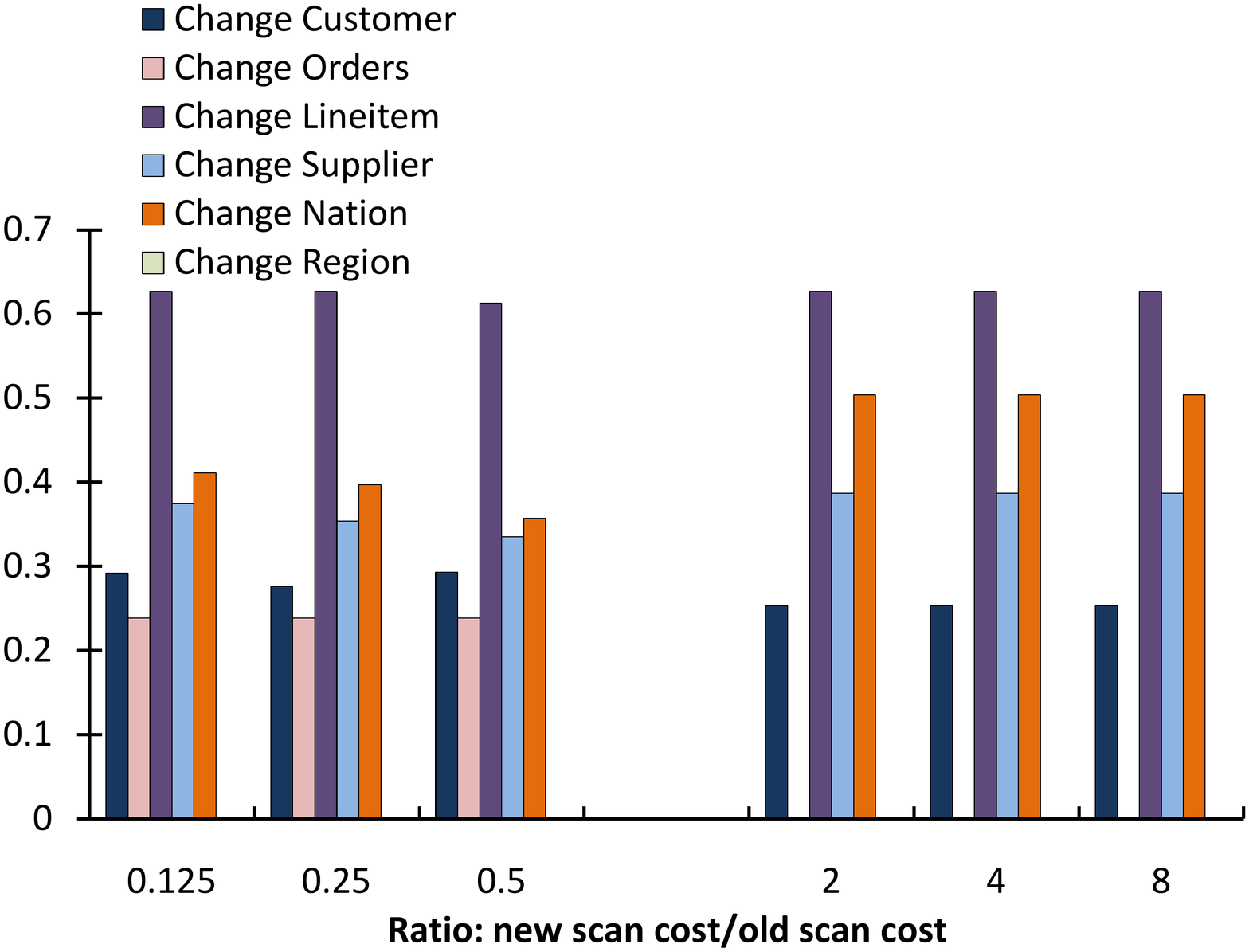}
\vspace{-2mm}
\small (c) Update ratio: plan alternatives
\end{center}
\end{minipage}
\caption{\small Performance during incremental re-optimization of TPC-H Q5 --- change to leaf operator cost \label{fig:incremental-Q5-scan-cost}}
\end{figure*}}



\begin{figure*}[t]
\begin{minipage}[t]{2.25in}
\begin{center}
\includegraphics[width=2.25in,height=1.25in]{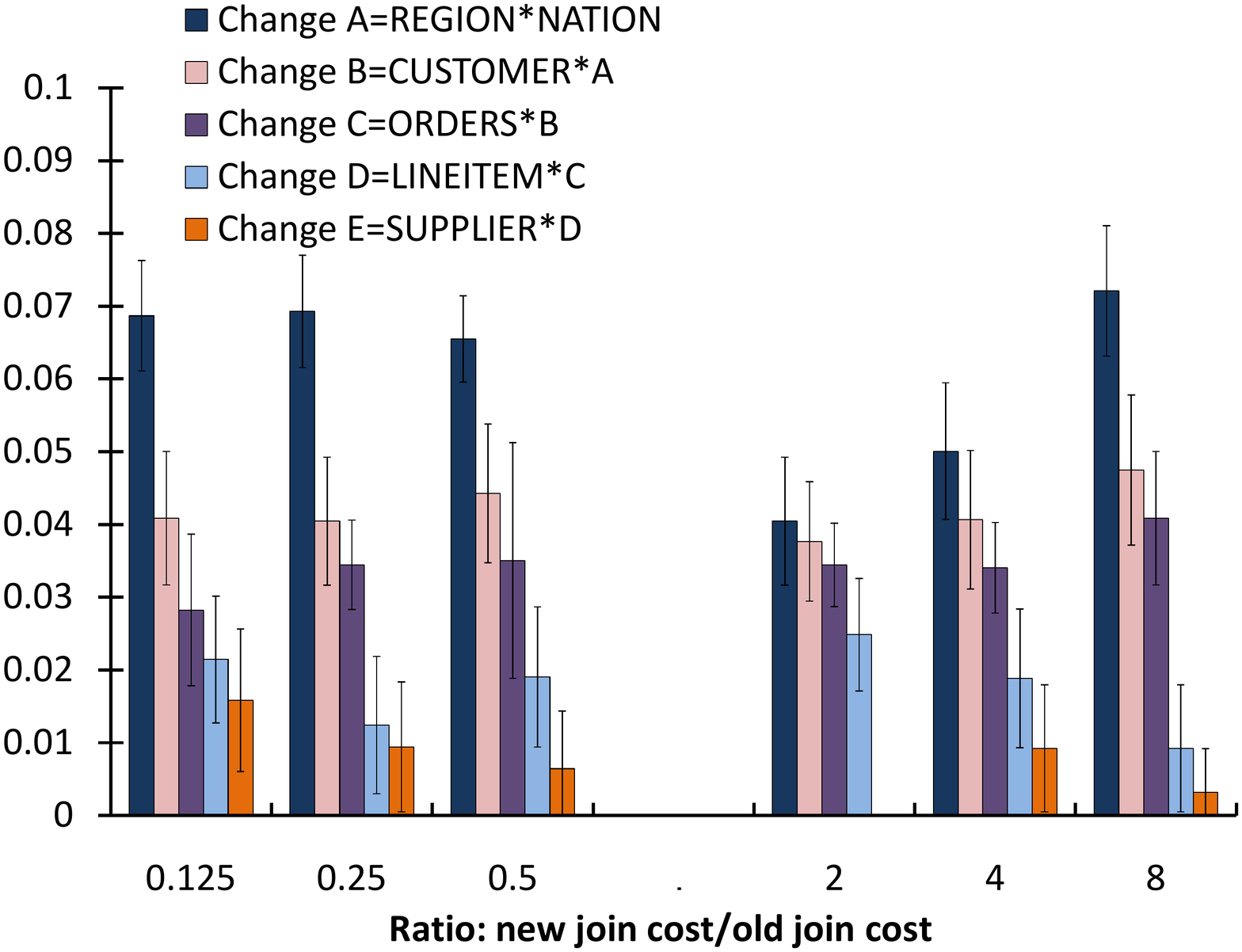}
\vspace{-2mm}
\small (a) Execution time (normalized to Volcano)
\end{center}
\end{minipage}
\hfill
\begin{minipage}[t]{2.25in}
\begin{center}
\includegraphics[width=2.25in,height=1.25in]{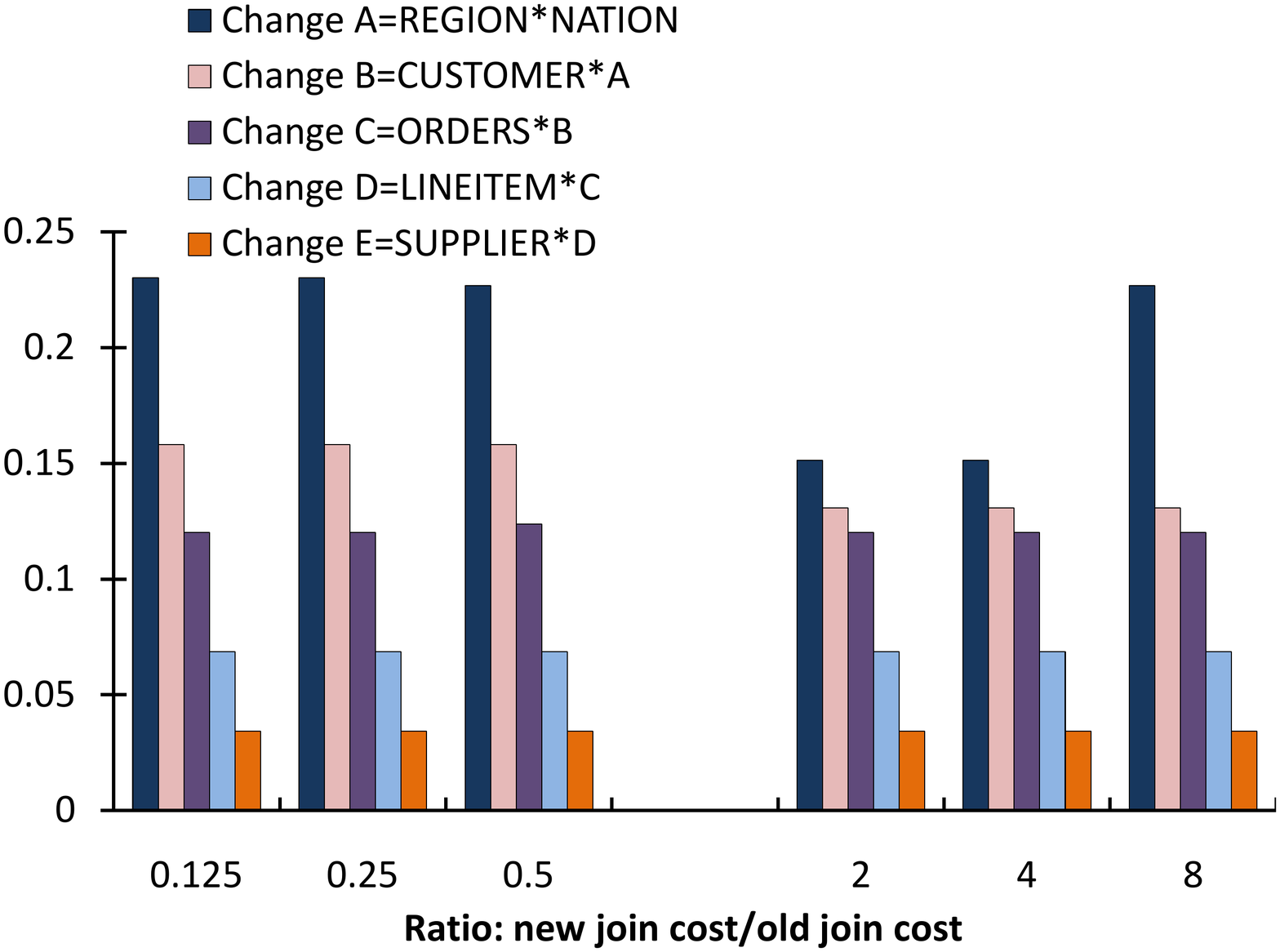}
\vspace{-2mm}
\small (b) Update ratio: plan table entries
\end{center}
\end{minipage}
\hfill
\begin{minipage}[t]{2.25in}
\begin{center}
\includegraphics[width=2.25in,height=1.25in]{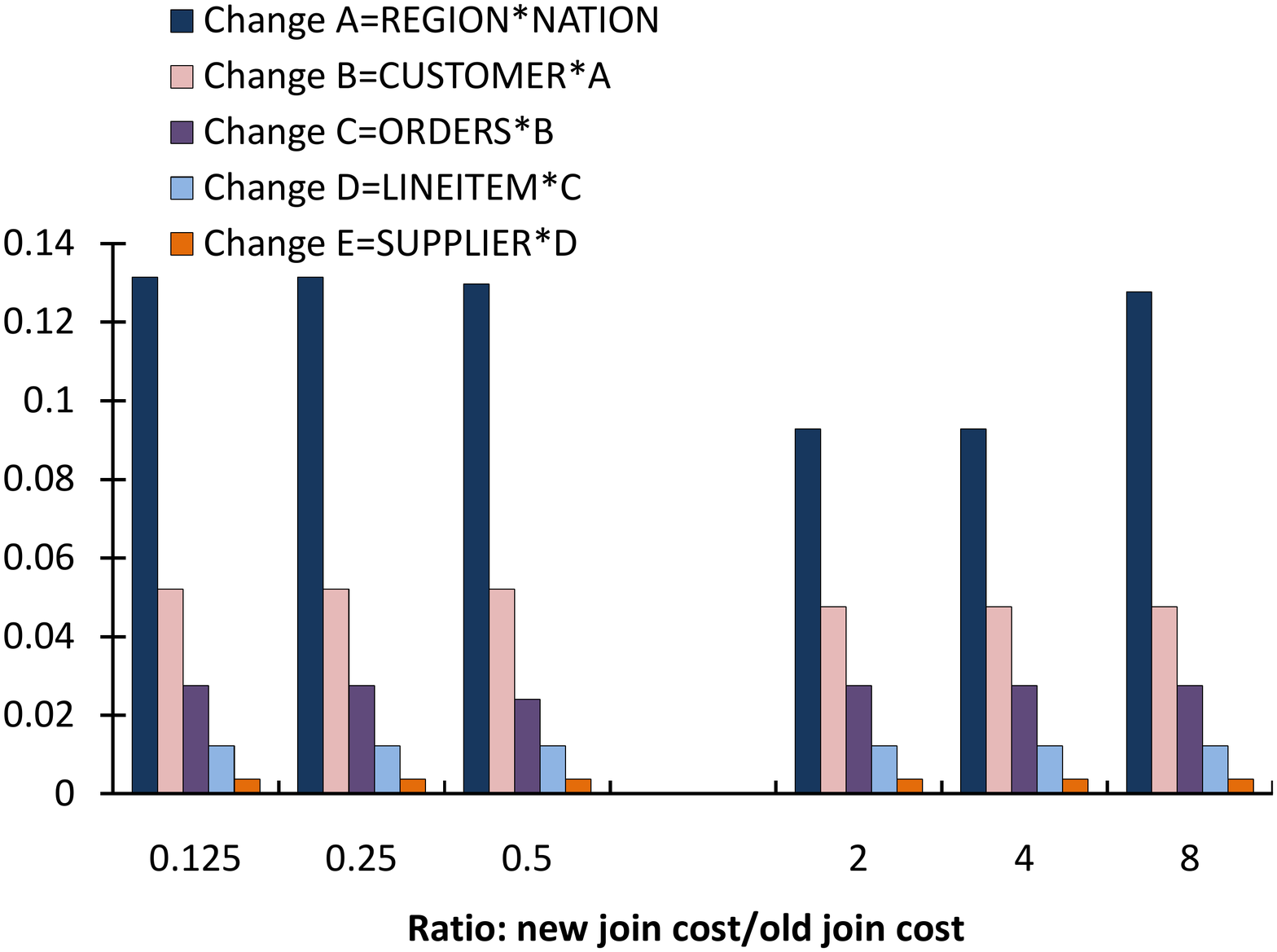}
\vspace{-2mm}
\small (c) Update ratio: plan alternatives
\end{center}
\end{minipage}
\caption{\small Performance during incremental re-optimization of TPC-H Q5 --- change to join selectivity estimate \label{fig:incremental-Q5-join-selectivity}}
\end{figure*}

\begin{figure*}[t]
\begin{minipage}[t]{2.25in}
\begin{center}
\includegraphics[width=2.25in,height=1.25in]{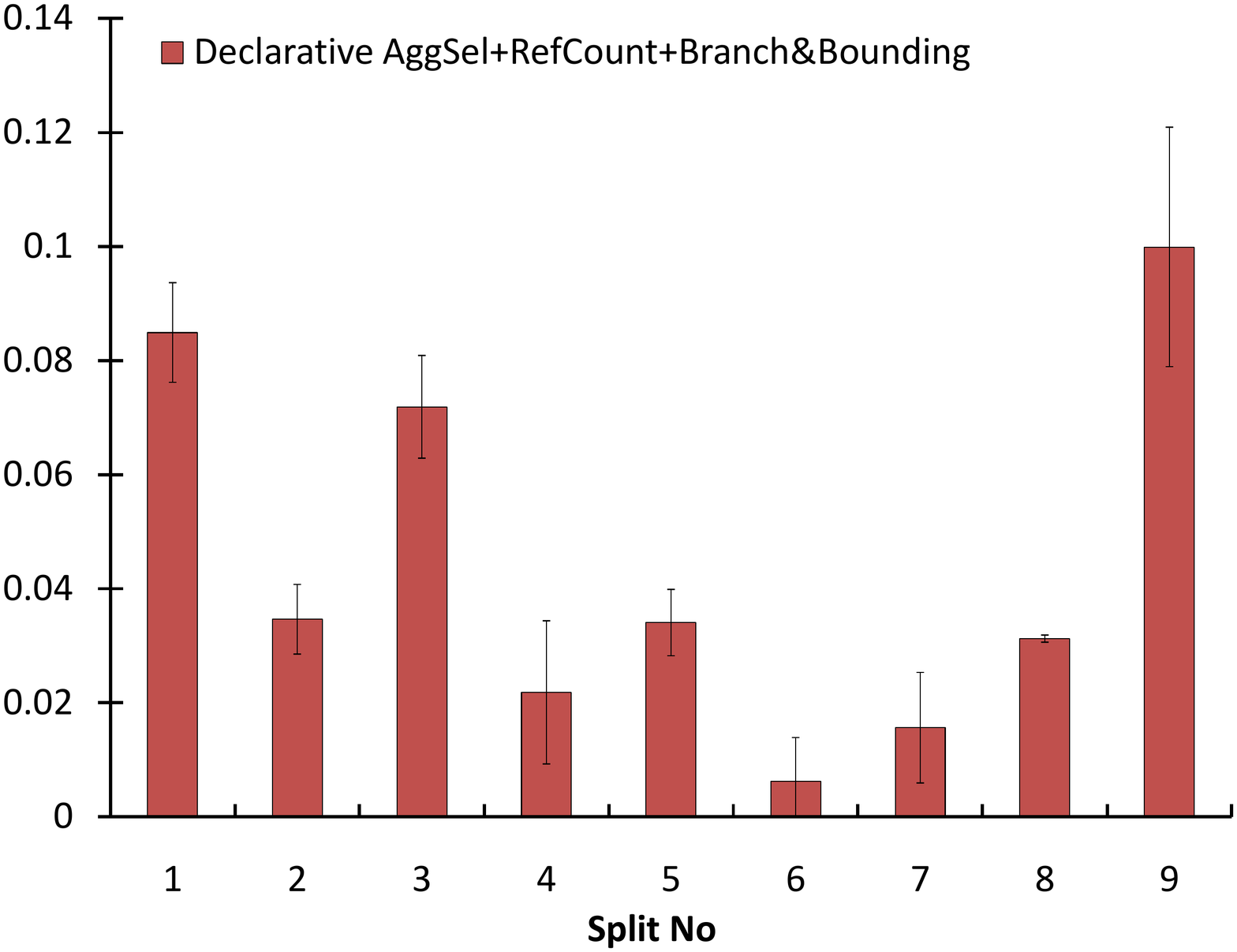}
\vspace{-2mm}
\small (a) Execution time (normalized to Volcano)
\end{center}
\end{minipage}
\hfill
\begin{minipage}[t]{2.25in}
\begin{center}
\includegraphics[width=2.25in,height=1.25in]{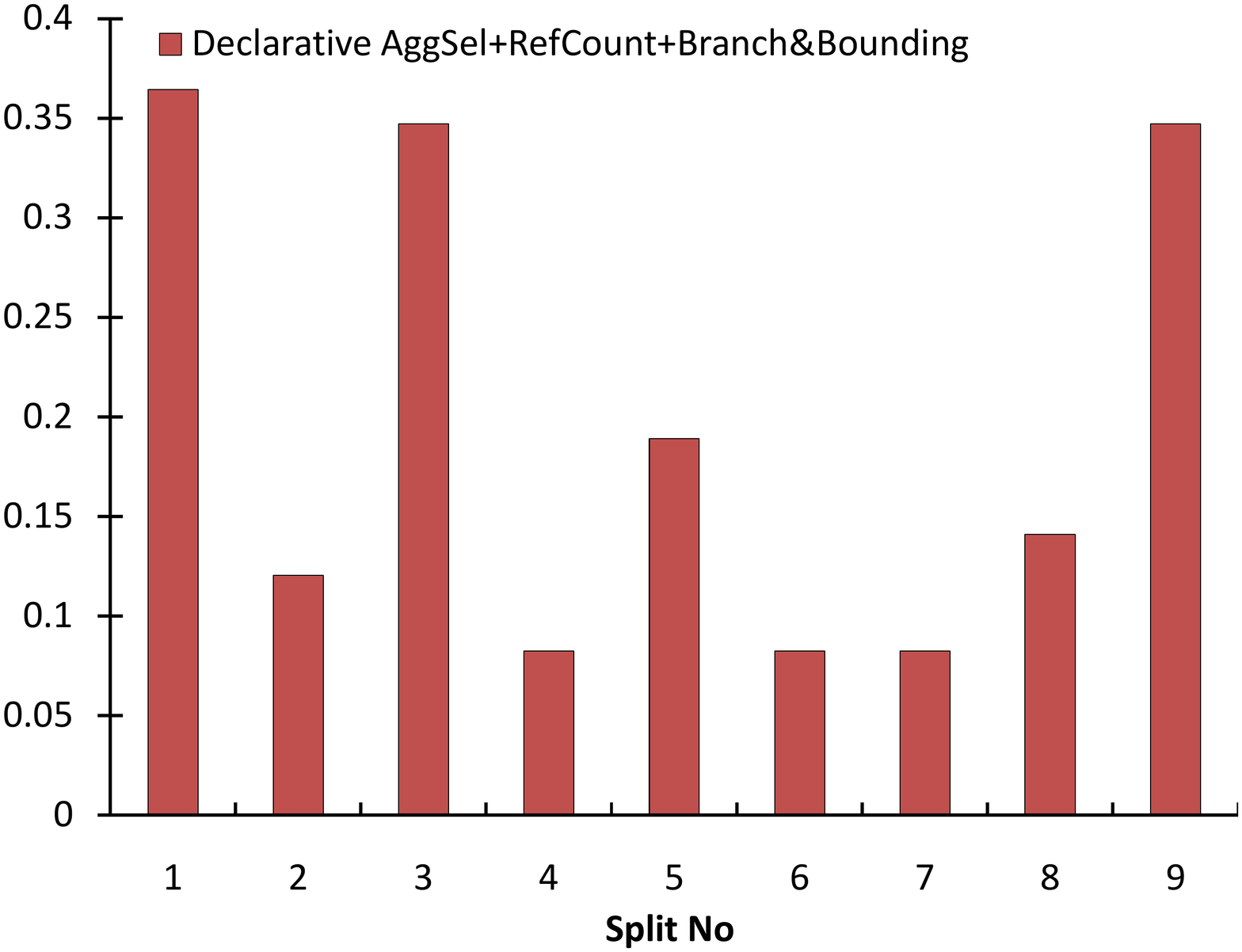}
\vspace{-2mm}
\small (b) Update ratio: plan table entries
\end{center}
\end{minipage}
\hfill
\begin{minipage}[t]{2.25in}
\begin{center}
\includegraphics[width=2.25in,height=1.25in]{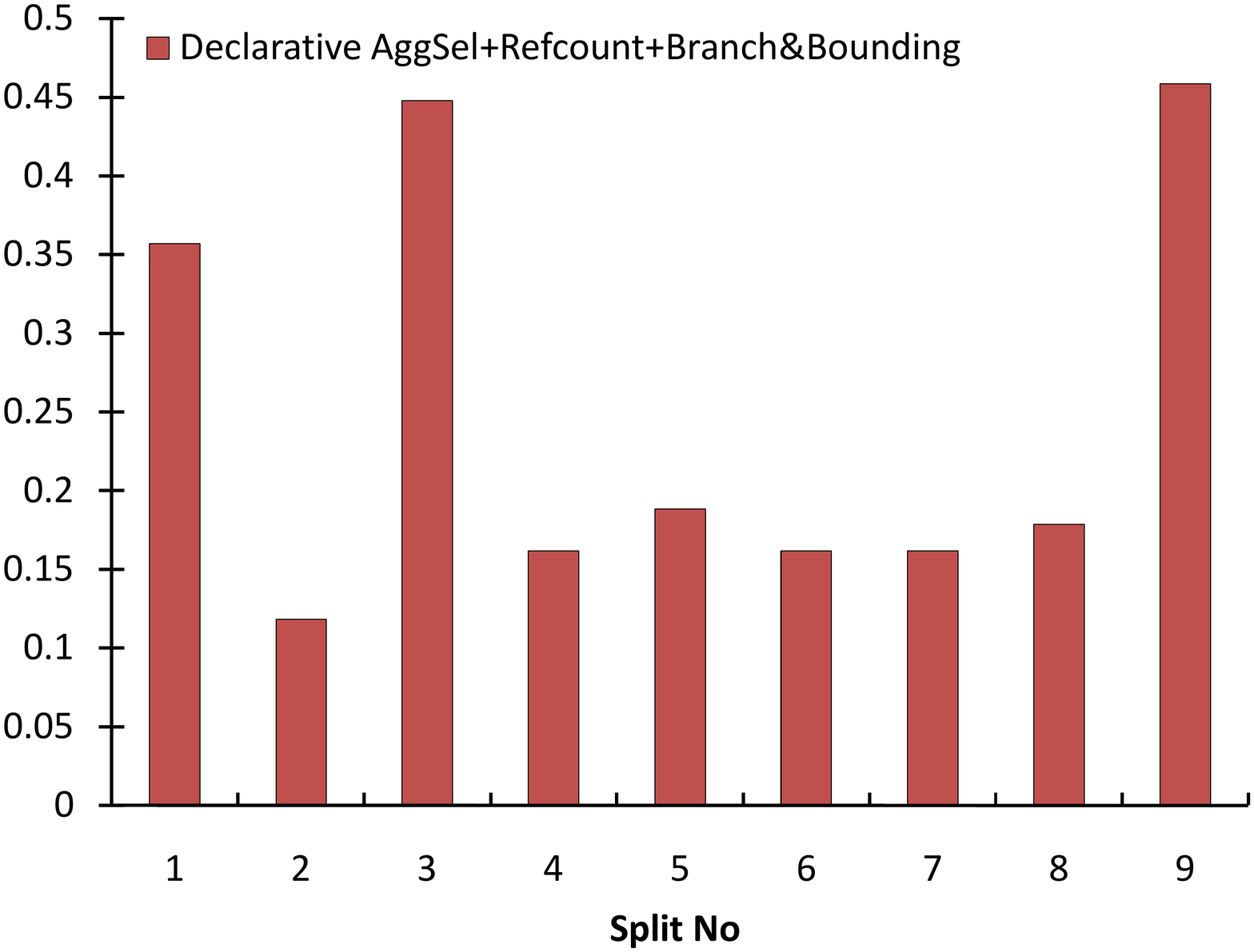}
\vspace{-2mm}
\small (c) Update ratio: plan alternatives
\end{center}
\end{minipage}
\caption{\small Performance during incremental re-optimization of TPC-H Q5 --- updates to costs based on real execution over skewed data \label{fig:incremental-Q5-real-workload}}
\end{figure*}

\begin{figure*}[t]
\begin{minipage}[t]{2.25in}
\begin{center}
\includegraphics[width=2.25in,height=1.25in]{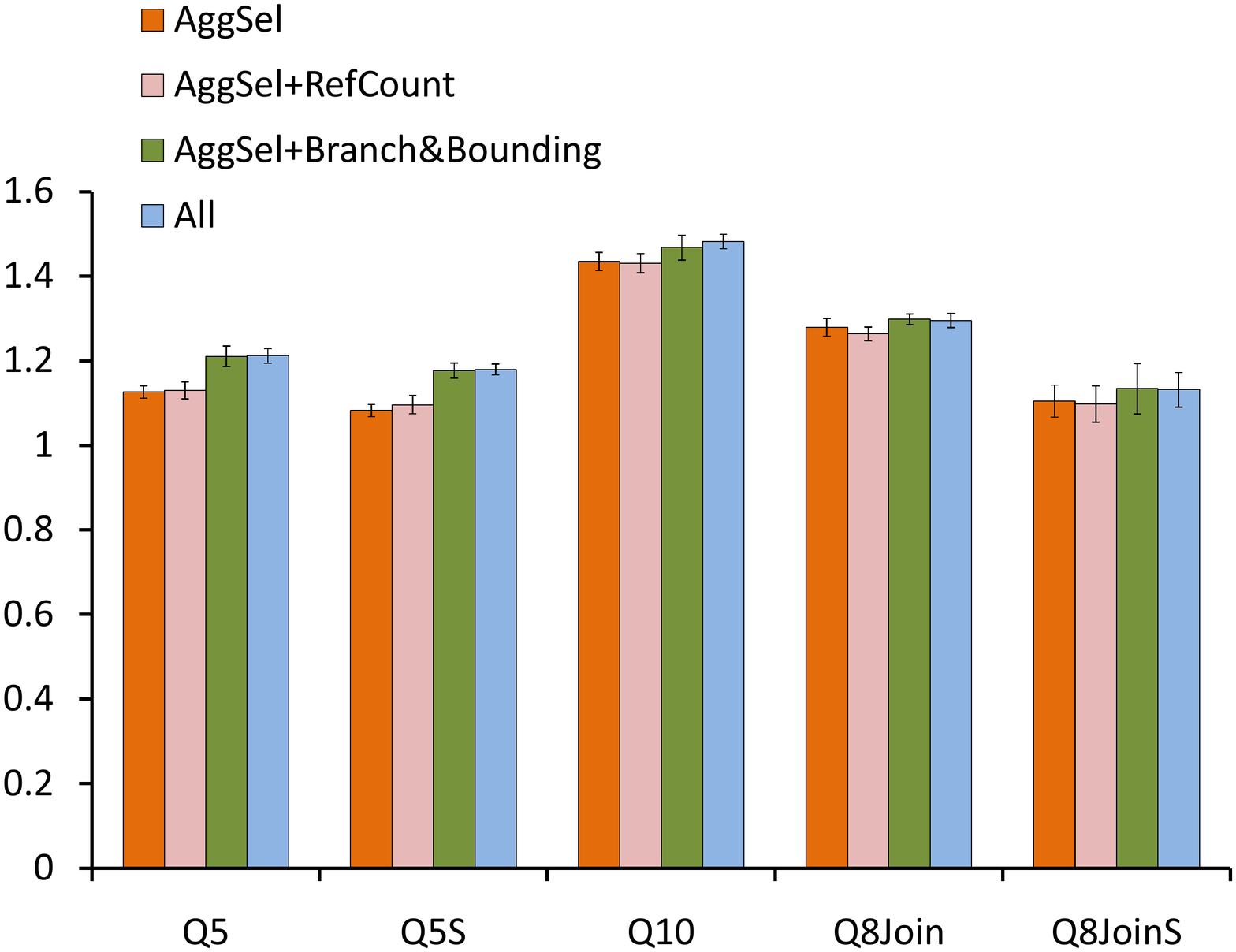}
\vspace{-2mm}
\small (a) Execution time (normalized to Volcano)
\end{center}
\end{minipage}
\hfill
\begin{minipage}[t]{2.25in}
\begin{center}
\includegraphics[width=2.25in,height=1.25in]{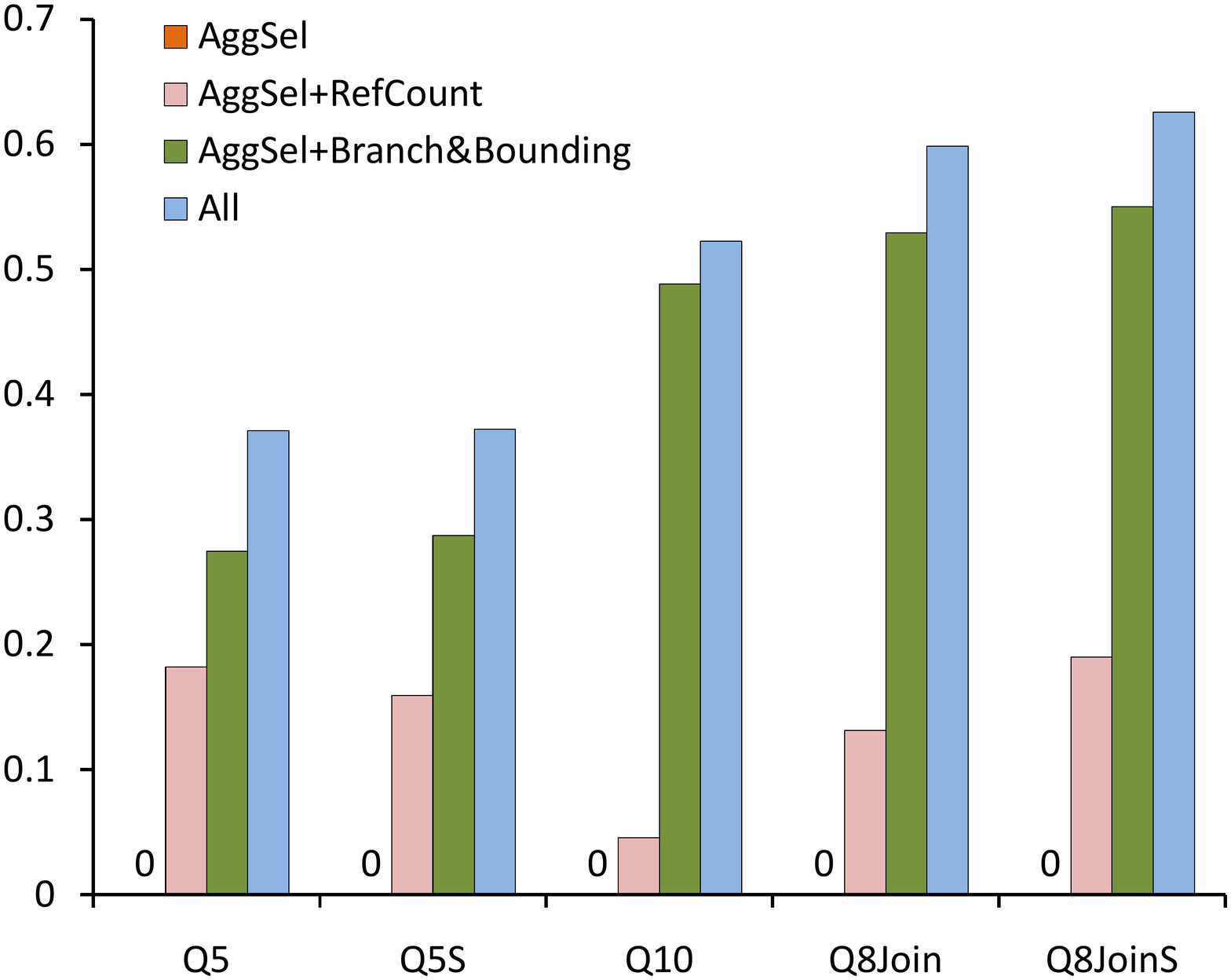}
\vspace{-2mm}
\small (b) Pruning ratio: plan table entries
\end{center}
\end{minipage}
\hfill
\begin{minipage}[t]{2.25in}
\begin{center}
\includegraphics[width=2.25in,height=1.25in]{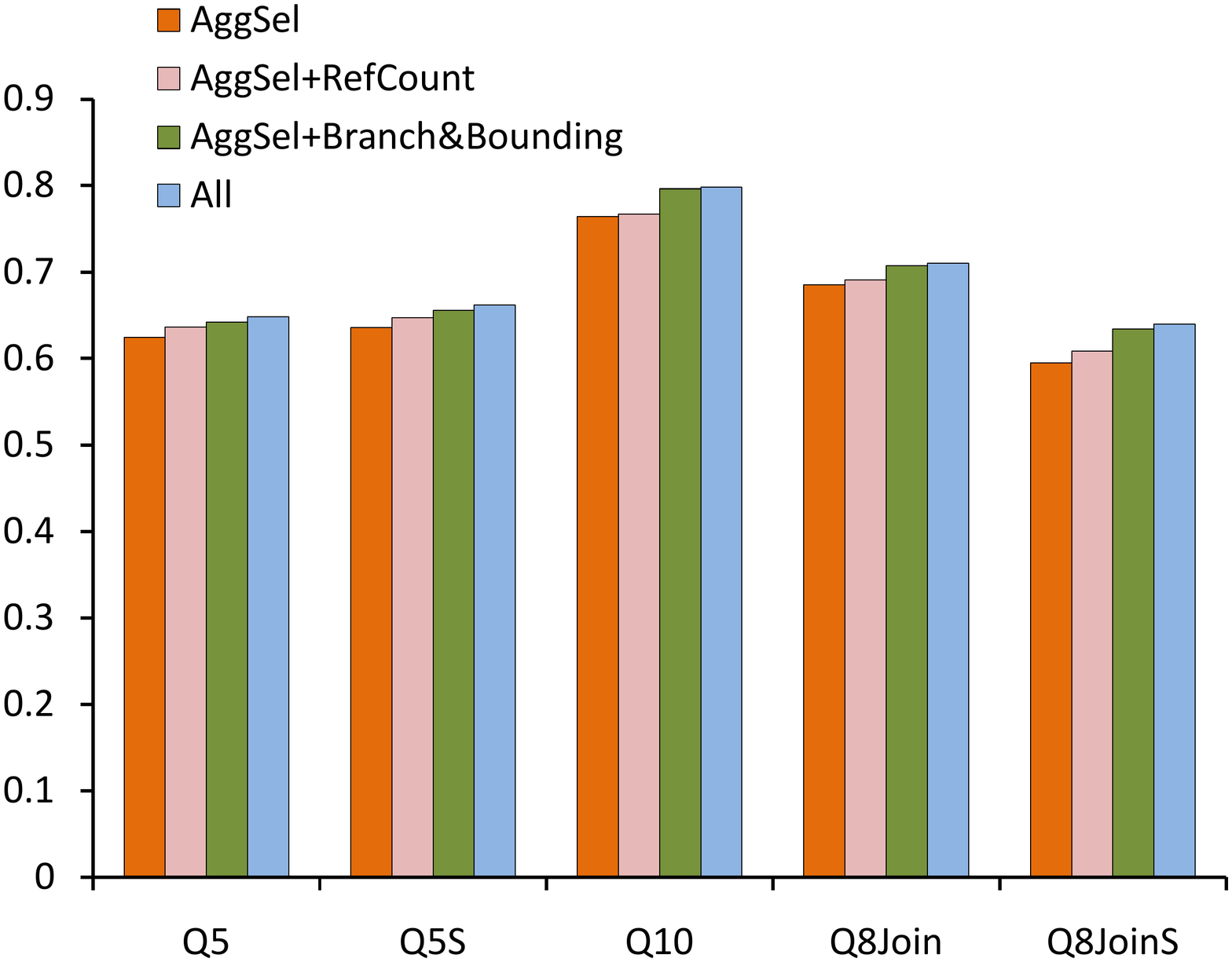}
\vspace{-2mm}
\small (c) Pruning ratio: plan alternatives
\end{center}
\end{minipage}
\caption{\small Performance breakdown of pruning techniques for initial optimization, across full query workload
\label{fig:stategies-preoptimization}}
\end{figure*}

\eat{
\reminder{ML: We should address concerns that we ignore many tricks and customization to the traditional approaches and their variants. We need to explain why we compare with Volcano/System R, because their variants are currently used in Tukwila/Cape for re-opt, and these are state-of-the-art adp systems. Mention we are comparing against stream versions of non-incremental re-optimization, maybe not Volcano/System R but its variants in Tukwila/Cape. We are not comparing against heuristics-based approaches because we generate totally different solutions: we generate plans that are provenly optimal in our cost model, but theirs are heuristics-based rather than cost-based. The results are different, the performance comparison thus has little meaning.}}

In this section, we discuss the implementation and evaluate the performance of our declarative optimizer: both 
versus other strategies, and as a primitive for adaptive query processing.  (Note that we reuse existing
adaptive techniques from~\cite{tukwila-04,cape-04}; our focus is on showing that incremental re-optimization \emph{improves} these.)  


We implemented the optimizer as 10
datalog rules (see Appendix) plus 8 external functions (involving histograms, cost estimation,
and expression decomposition).  Our goal was to implement as a proof of concept the \textbf{common core} of optimizer techniques --- not an exhaustive set.  We executed the optimizer in a modified version of the ASPEN system's query engine~\cite{recursive-views}, as obtained from its
authors.  To support the pruning and incremental update propagation features in this paper,
we added approximately 10K lines of code to the query engine.  In addition, we developed a plan generator 
to translate the declarative optimizer into a dataflow graph as in
Figure~\ref{fig:execution-flow}.  Our experiments were performed on a single local node.


For comparison, we implemented in Java a Volcano-style top-down query
optimizer and a System-R-style dynamic programming optimizer, which
reuse the histogram, cost estimation, and other core components as our declarative optimizer.  We
also built a variant of our declarative optimizer that only uses the
pruning strategies of the Evita Raced declarative
optimizer~\cite{evita-raced}.  Wherever possible we used common
code across the implementations.

\vspace{-1mm}
\Paragraph{Experimental Workload.} For \textbf{repeated optimization} scenarios we use TPC-H queries, with data from
the TPC-H and skewed TPC-D data generators~\cite{tpcd} (Scale Factor
1, with Zipfian skew factor 0 for the latter). We focused on the single-block
SQL queries: Q1, Q3, Q5, Q6 and Q10. (Q1 and Q6 are
aggregation-only queries; Q3 joins 3 relations; Q10 joins 4; and Q5
joins 6 relations). Our experiments showed that Q1, Q3, and Q6 are all
simple enough to optimize that (1) there is not a compelling need to
adapt, since there are few plan alternatives; (2) they completed in
under 80msec on all implementations.  (The declarative approach tended
to add 10-50msec to these settings, as it has higher initialization
costs.)
Thus we focus our presentation on join queries with more than 3-way
joins.  To create greater query diversity, we modified the 4-way and
larger join queries by removing aggregation --- we constructed a 
simplified query Q5S.  Finally, to test scale-up to larger queries, we
manually constructed an eight-way join query, Q8Join, and its simplified version (removing aggregates),
Q8JoinS. For \textbf{adaptive stream processing} we used the Linear Road benchmark~\cite{linearRoad}: We modified the largest query, called SegToll.
\eat{The standard TPC-H queries can be found in
Appendix~\ref{sec:appendix-queries};} We show our new queries Q8Join and SegTollS in Table~\ref{table:queries}.


\begin{table}
\footnotesize
\begin{tabular}{|p{8cm}|}
\hline
{\bf Q8Join:}  SELECT c\_name, p\_name, ps\_availqty, s\_name, o\_custkey, r\_name, n\_name, sum(l\_extendedprice * (1 -l\_discount)) FROM orders, lineitem, customer, part, partsupp, supplier, nation, region WHERE o\_orderkey = l\_orderkey and c\_custkey = o\_custkey and p\_partkey = l\_partkey and ps\_partkey = p\_partkey and s\_suppkey = ps\_suppkey and r\_regionkey = n\_regionkey and s\_nationkey = n\_nationkey GROUPBY c\_name, p\_name, ps\_availqty, s\_name, o\_custkey, r\_name, n\_name; \\
\hline
{\bf SegTollS:} SELECT r1\_expway, r1\_dir, r1\_seg, COUNT(distinct r5\_xpos) FROM CarLocStr [size 300 time] as r1, CarLocStr [size 1 tuple partition by expway, dir, seg] as r2, CarLocStr [size 1 tuple partition by caid] as r3, CarLocStr [size 30 time] as r4, CarLocStr [size 4 tuple partition by carid] as r5 WHERE r2\_expway = r3\_expway and r2\_dir = 0 and r3\_dir = 0 and r2\_seg < r3\_seg and r2\_seg > r3\_seg - 10 and r3\_carid = r4\_carid and r3\_carid = r5\_carid and r1\_expway = r2\_expway and r1\_dir = r2\_dir and r1\_seg = r2\_seg GROUP BY r5\_carid, r2\_expway, r2\_dir, r2\_seg; \\
\hline
\end{tabular}
\vspace{-3mm}
\caption{\label{table:queries} Queries modified based on TPC-H and LinearRoad benchmark queries used in our experiments}
\end{table}

\vspace{-1mm}
\Paragraph{Experimental Methodology.}
We aim to answer four questions:
\begin{compactList}

\vspace{-2mm}

\item Can a declarative query optimizer perform at a rate competitive
 with procedural optimizers, for 4-way-join queries and larger?

\vspace{-2mm}
\item Does incremental query re-optimization show running time and
  search space benefits versus non-incremental re-optimization, for repeated query execution-over-static-data scenarios?
\vspace{-2mm}
\item How does each of our three pruning strategies 
  (aggregate selection, reference counting, and recursive bounding)
  contribute to the performance?
\vspace{-2mm}
\item Does incremental re-optimization improve the performance of cost-based adaptive query processing techniques for streaming?
\vspace{-2mm}
\end{compactList}
 
The TPC-H benchmark experiments are conducted on a single local desktop machine: a dual-core
Intel Core 2 2.40GHz with 2GB memory running 32-bit Windows XP Professional,
and Java JDK 1.6. The Linear Road benchmark experiments are conducted on a single server machine: a dual-core Intel Xeon  2.83GHz with 8 GB memory running 64-bit Windows Server Standard. Performance results are averaged across 10 runs, and 95\% confidence intervals are shown. We mark as 0 any results that are exactly zero.

\eat{
We measure the following parameters:
\begin{enumerate}
\vspace{-2mm}
\item {\bf Execution time (normalized to Volcano):} the execution time of the optimizer divided by the execution time of the existing procedural-based Volcano-style query optimizer.
\vspace{-2mm}
\item {\bf Ratio of Pruned And Node}: the pruned amount of ``AND'' node state ($PlanCost$ explorations) divided by the total amount of ``AND'' node state, measuring the amount of explorations saved.  
\vspace{-2mm}
\item {\bf Ratio of Pruned Or Node}: the pruned amount of ``OR'' node state ($BestCost$ propagations) divided by the total amount of ``OR'' node state, measuring the amount of propagations saved.
\vspace{-2mm}
\item {\bf Ratio of Updated And Node}: the amount of updated ``AND'' node state in incremental query re-optimization divided by the total amount of ``AND'' node state in the original query pre-optimization, measuring the amount of explorations done.
\vspace{-2mm}
\item {\bf Ratio of Updated Or Node}: the amount of updated ``OR'' node state in incremental query re-optimization divided by the total amount of ``OR'' node state in the original query pre-optimization, measuring the amount of propagations done.
\end{enumerate}
}

\Subsection{Declarative Optimization Performance}
\label{subsec:declarative-procedural}

\eat{
Question: I don not believe declarative query optimizer can possibly match a customized optimizer. 
Answer: True if you only optimize once; however, the performance overhead is within factor of 1.5 for queries of optimization time > 0.05s. As shown in Figure~\ref{fig:declarative-procedural-execution-time}.

Question: Does the declarative query optimizer get effective pruning?
Answer: Yes. Compared to Volcano, Evita-Raced in Figure~\ref{fig:declarative-state-size}.
}

Our initial experiments focus on the question of \textbf{whether our
declarative query optimizer can be competitive with procedural
optimizers} based on the System-R (bottom-up enumeration through
dynamic programming) and Volcano (top-down enumeration with
memoization and branch-and-bound pruning) models.  To show the
value of the pruning techniques developed in this paper, we also measure
the performance when our engine is limited to the pruning
techniques developed in Evita Raced~\cite{evita-raced}
(where pruning is only done against logically equivalent plans for the
same output properties).  Recall that all of our implementations share
the same procedural logic (e.g., histogram derivation); 
their differences are in search strategy, dataflow, and
pruning techniques.

We begin with a running time comparison among Volcano-style,
System-R-style, and declarative implementations (one using our
sideways information passing strategies, and one based on the Evita
Raced pruning heuristics) --- shown in
Figure~\ref{fig:declarative-procedural} (a).  This graph is normalized
against the running times for our Volcano-style implementation (which
is also included as a bar for visual comparison of the running times).
Actual Volcano running times are shown directly above the bar.
Observe from the graph that the Volcano strategy is always the
fastest, though System-R-style enumeration often approaches its
performance due to simpler (thus, slightly faster) exploration logic.
Our declarative implementation is not quite as fast as the dedicated
procedural optimizers, with an overhead of 10-50\%, but this is
natural given the extra overhead of using a general-purpose engine and
supporting incremental update processing.  The Evita Raced-style
declarative implementation is marginally faster in this setting, as it
does less pruning.  We shall see in later experiments that there are
significant benefits to our more aggressive strategies during
re-optimization --- which is our focus in this work.

\eat{
In the first set of experiments, we compare the performance of our
declarative query optimizer against a procedural-based query optimizer
for query pre-optimization (a normal non-incremental optimization).
The goal is to understand the performance overhead of a declarative
query optimizer and the effectiveness of its pruning algorithms. The
four schemes compared in this set of experiments include: Volcano
(top-down plan enumeration with branch\&bounding and memoization);
System R (bottom-up plan enumeration and dynamic programming);
Evita-Raced~\cite{evita-raced} (declarative query optimization with
pruning from the sibling nodes, implemented in our declarative engine
to test the performance); and Declarative (our fully optimized
declarative query optimization with aggregate selection, reference
count and branch\&bounding ). To ensure fair comparison, we maintain
the same user-defined functions for both the declarative end and the
procedural end. The main difference is the enumeration order for
$SearchSpace$ and for $PlanCost$. Hence, we mimic Volcano and System
R's search order using procedural code; and implement Evita-Raced's
strategies in our declarative engine.

The results are shown in Figure~\ref{fig:declarative-procedural}. (a)
demonstrates the normalized execution time of query pre-optimization.
We pick 5 queries from our workloads that have the running time over 0.05
seconds, and compare Volcano, System R and Declarative approaches here.
Volcano is shown as the baseline which is normalized to 1, with its actual
running times shown on the labels. The figure shows that System R
always performs slightly slower than Volcano, because it does not
exploit branch\&bounding. Our declarative optimizer has at most 1.5x
of the execution time of a Volcano optimizer for all queries. This is reasonable
as a declarative optimizer has some unavoidable overhead to support the
stream-based data model even during pre-optimization, e.g., maintaining the 
internal state when joining left subplans and right subplans; inserting, deleting 
or updating a $PlanCost$ tuple incrementally, etc. Nevertheless, our results 
show that one can extend an existing procedural-based query optimizer 
into a declarative one with tolerable performance overhead for pre-optimization.
We will demonstrate a huge benefit on the re-optimization performance in the 
later sections. }

To better understand the performance of the different options, we next
study their effectiveness in \emph{pruning} the search space.  We
divide this into two parts: (1) pruning of expression-property entries
in the plan table, such that we do not need to compute and maintain
\emph{any} plans for a particular expression yielding a particular
property; (2) pruning of \emph{plan alternatives} for a particular
expression-property pair.  In terms of the and-or graph formulation of
Figure~\ref{fig:example-and-or-graph-Q3S}, the first case prunes
or-nodes and the second prunes and-nodes.  We show these two cases in
Figure~\ref{fig:declarative-procedural} parts (b) and (c).  We omit
System R from this discussion, as it uses a dynamic programming-based
pruning model that is difficult to directly compare.

Part (b) shows that our declarative implementation achieves pruning of
approximately 35-80\% of the plan table entries, resulting in large
reductions in state (and, in many cases, reduced computation).  We
compare with the strategies used by Evita Raced, which we can see
never prunes plan table entries, and with our Volcano-style
implementation.  Observe that our pruning strategies --- 
which are quite flexible with respect to order of
processing --- are often more effective than the Volcano strategy,
which is limited to top-down enumeration with
branch-and-bound pruning.  (All pruning strategies' effectiveness depends on the specific order in
which nodes are explored: better pruning is achieved when inexpensive
options are considered early.  However, in the common case, high
levels of pruning are observed.)

Part (c) looks in more detail at the number of alternative query plans
that are pruned: here our declarative implementation
prunes approximately 55-75\% of the space of plans.  It exceeds the
pruning ratios obtained by the Evita Raced strategies by around 4-8\%,
and often results in significantly greater pruning than Volcano.

Our conclusions from these experiments are that, even for initial
query optimization ``from scratch,'' a declarative optimizer can be
performance-competitive with a procedural one --- both in terms of
running time and pruning performance.  Moreover, given that our plan
enumeration and pruning strategies are completely decoupled, we plan
to further study whether there are effective heuristics for exploring
the search space in our model.

\eat{
Figure~\ref{fig:declarative-procedural}(b)(c) measures the effectiveness of pruning the state of ``OR'' node and ``AND'' node respectively. Our declarative approach always achieves better pruning compared to Evita-Raced, because our optimization strategies makes the bound of one plan derive from all its ancestor plans whereas Evita-Raced just maintains the bounds deriving from immediate sibling plans. On the other hand, for some queries such as Q5/Q5S and Q10/Q10S, we pruned more versus the Volcano baseline; whereas for others such as Q8Join/Q8JoinS we prune less. Since pruning largely depends on the search order: if an algorithm can reach the min cost plan early, it can use this cost as the bound to prune others more effectively. Our declarative optimizer's search order is dependent on our push-based query processor's default execution flow, which actually enumerates plans in the top-down iterative deepening order and estimates costs in the bottom-up iterative deepening order. On the other hand, Volcano enumerates plans in the depth-first pre-order and estimates costs in the depth-first post-order. There is no order that is optimal for every workload. In the future work, we plan to mimic Volcano-style enumeration and System R-style enumeration in our declarative optimizer to enable fully customizable optimization. }

\Subsection{Incremental Re-optimization}
\label{subsec:incremental}

Now that we understand the relative performance of our declarative
optimizer in a conventional setting, we move on to study it how it handles
incremental changes to costs.  A typical setting in a non-streaming context
would be to improve performance during the repeated execution of a 
query, such as for a prepared statement where only a binding changes.  The
question we ask is how expensive --- given a typical update
--- it is to re-optimize the query and produce the new, predicted-optimal plan.

Note that there exists no comparable techniques for incremental
cost-based re-optimization, so we compare the gains versus those of
re-running a complete optimization (as is done
in~\cite{tukwila-04,cape-04}).  In these
experiments, we consider running time --- versus the running times for
the best-performing initial optimization strategy, namely that of our
Volcano-style implementation --- as well as how much of the total
search space gets re-enumerated.  We consider re-optimization under
``microbenchmark''-style simulated changes to costs, for synthetic
updates as well as observed execution conditions over skewed
data.  We measured performance across the \textbf{full suite of queries} in our
workload.  However, due to space constraints, and since the results are
representative, we focus our presentation on query Q5. \eat{Reviewers think Q5 is not enough. Maybe show one more graph about its representativeness of the other queries.}

\eat{
\Subsubsection{Changes to Leaf Operator Costs}

Figure~\ref{fig:incremental-Q5-scan-cost} shows the results of
incremental re-optimization for TPC-H Q5, focusing on the same
parameters as the previous section, when the cost of scanning a single
input relation gets updated (e.g., as might occur if there were
synergistic or adversarial interactions with other queries running
concurrently).  We generate cost updates synthetically, re-estimating
the costs (as low as 1/8 of the original estimates,
and as high as 8 times the original estimates) and re-optimizing, for
each of the source relations to Q5.  As we see in
Figure~\ref{fig:incremental-Q5-scan-cost} (a), changes to some
relations have greater impact than others, but re-optimization is
between 4.5 and 10 times as fast as re-running a complete optimization
under the Volcano strategy.  Perhaps not surprisingly, changes to the
\reln{lineitem} table, which is by far the largest in the TPC-H
dataset, have the largest re-optimization cost: changes to this table
are likely to have the most far-reaching effects on the costs of
potential query plans.

The differences in running time are explained in
Figures~\ref{fig:incremental-Q5-scan-cost} (b) and (c), where we see
that changes to \reln{lineitem} require us to re-evaluate
approximately 62\% of the alternative plans, and nearly 90\% of 
expression-property pairs.  Changes to the other relations
generally had much less effect on the plan alternatives (25-50\%) and
especially plan table entries (34-55\%).  In general we see that,
across a fairly broad set of possible changes, re-optimization
typically only revisits around 1/4 to 3/5 of the set of possible
plans.  As we shall see later, much of this savings is due to our
pruning and incremental maintenance techniques.  Observe that reductions in
search space, especially on the plan table, have a greater-than-linear
effect on running time.  This is in part because we cache state (e.g.,
derived histograms) between initial and subsequent optimizations, and
scale and reuse it between re-optimizations.
}

\begin{figure*}[t]
\begin{minipage}[t]{5.25in}
\begin{minipage}[t]{1.72in}
\begin{center}
\includegraphics[width=\textwidth,height=1.25in]{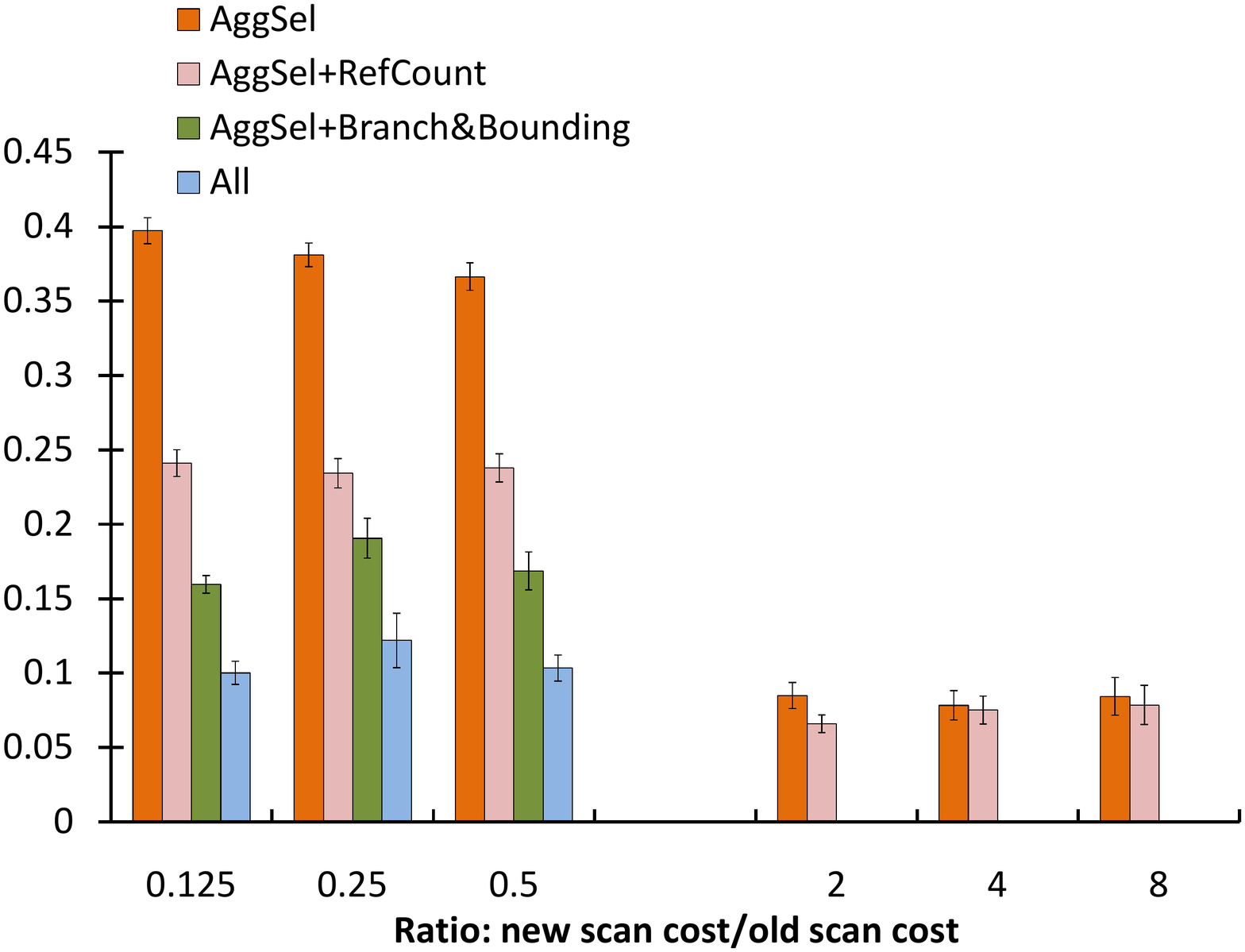}
\vspace{-2mm}
\small (a) Execution time (vs Volcano)
\end{center}
\end{minipage}
\hfill
\begin{minipage}[t]{1.72in}
\begin{center}
\includegraphics[width=\textwidth,height=1.25in]{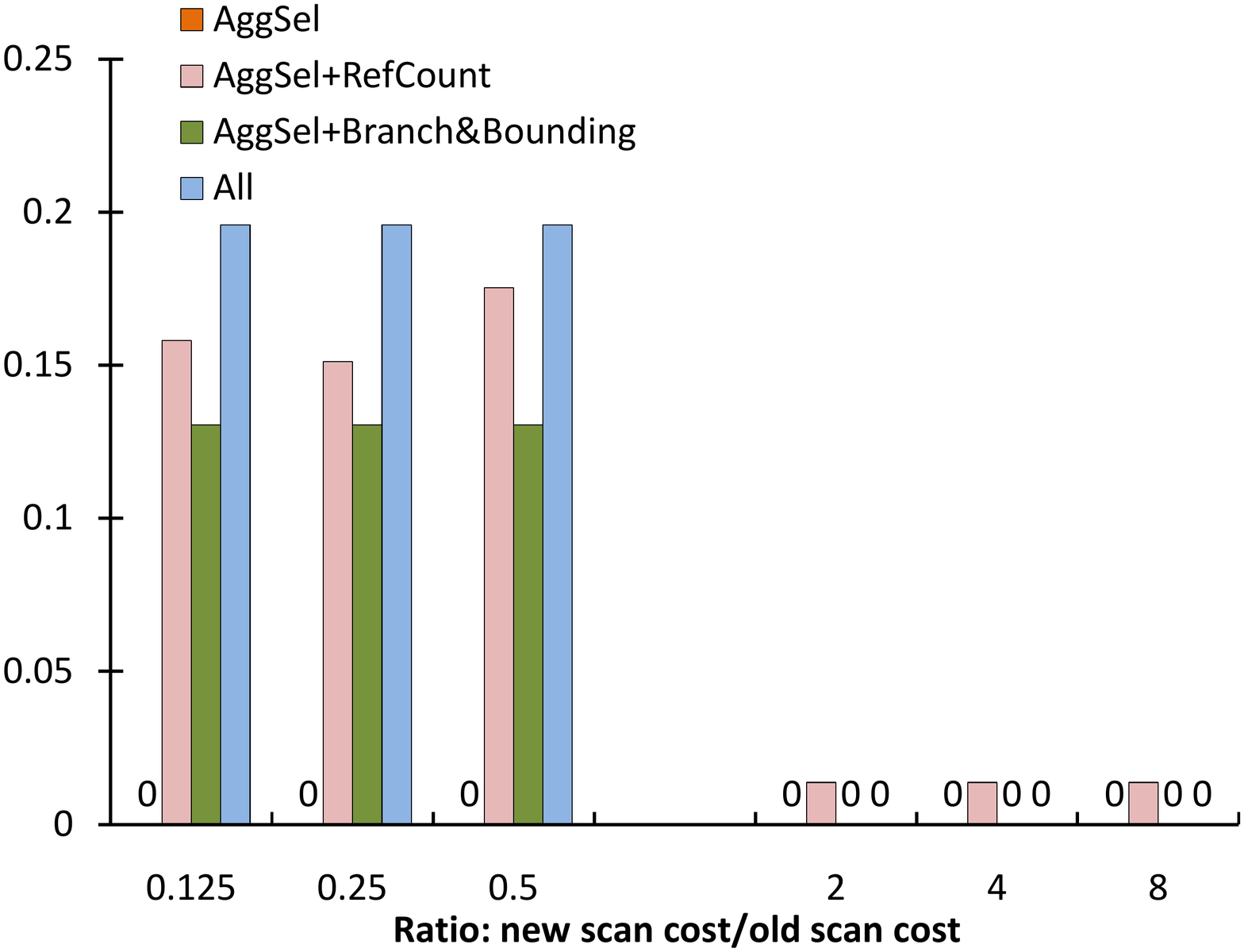}
\vspace{-2mm}
\small (b) Pruning ratio: plan table entries
\end{center}
\end{minipage}
\hfill
\begin{minipage}[t]{1.72in}
\begin{center}
\includegraphics[width=\textwidth,height=1.25in]{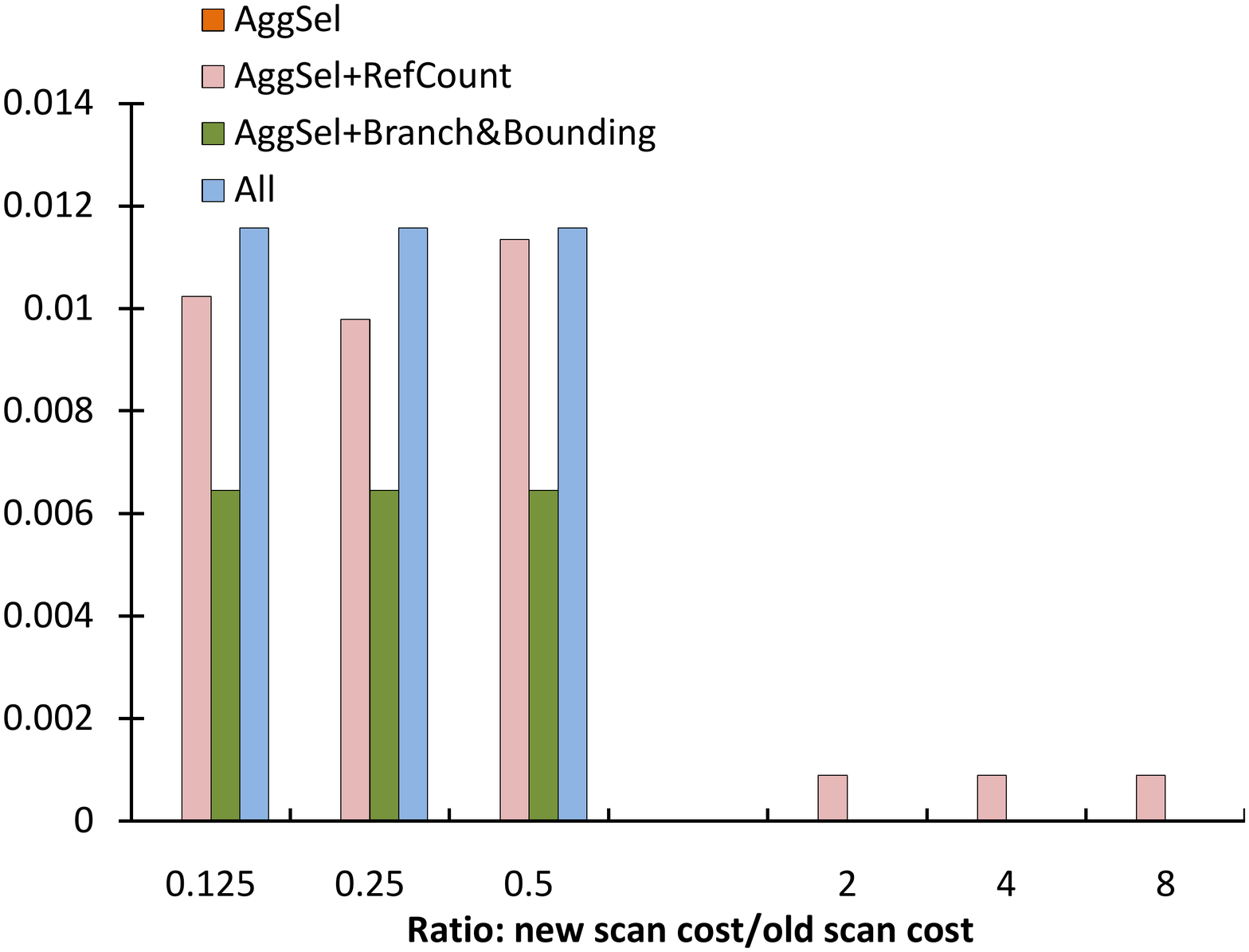}
\vspace{-2mm}
\small (c) Pruning ratio: plan alternatives
\end{center}
\end{minipage}
\caption{\small Performance breakdown of pruning techniques during incremental re-optimization of Q5 when $Orders$ has updated scan cost
\label{fig:stategies-reoptimization-Q5-Orders}}
\end{minipage}
\hfill
\begin{minipage}[t]{1.25in}
\begin{center} \includegraphics[height=1.15in,width=\textwidth]{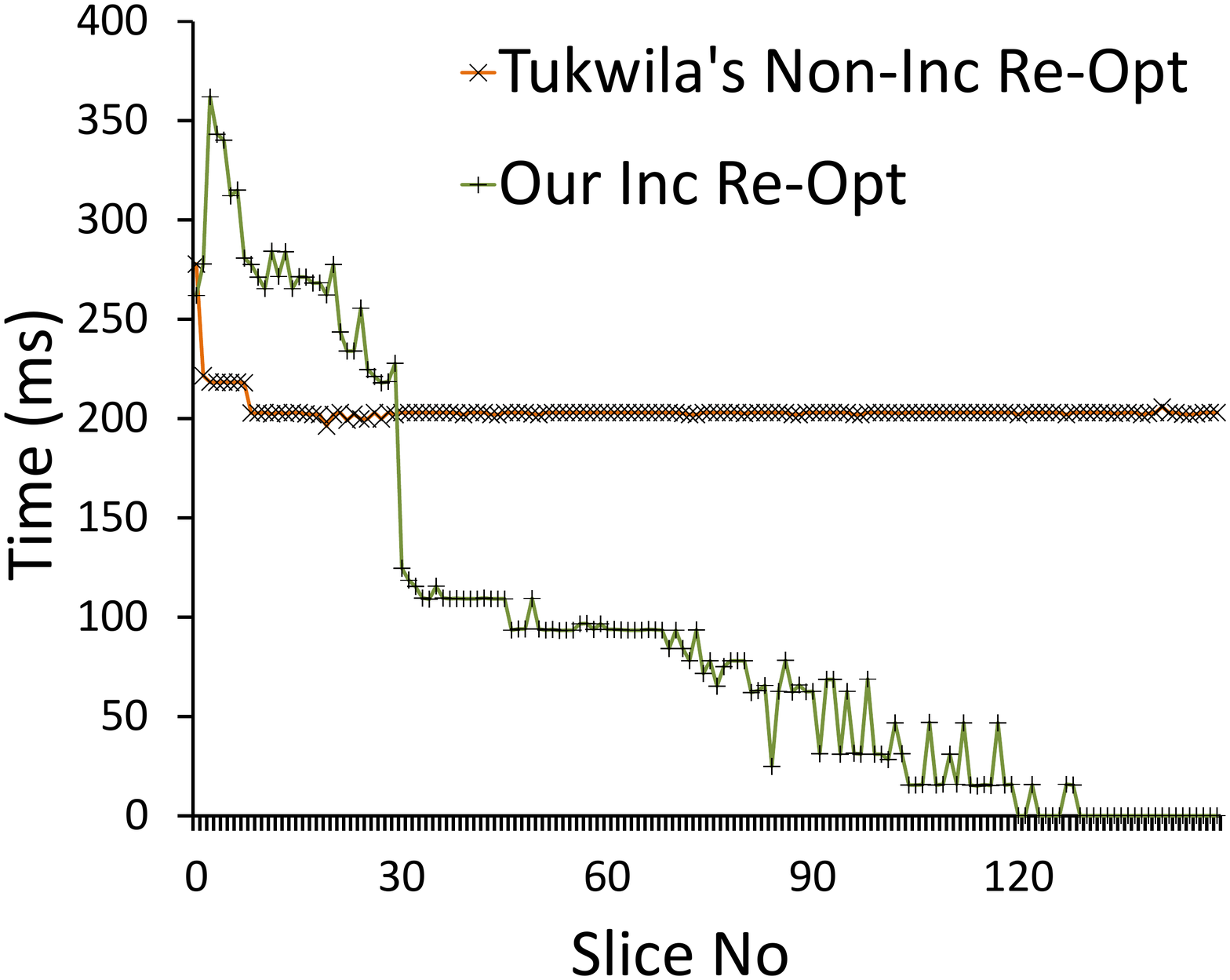} 
\vspace{-4mm}
\caption{\small AQP re-opt\label{fig:aqp-opt}}
\end{center}
\vspace{-2mm}
\end{minipage}
\vspace{-3mm}
\end{figure*}

\eat{
We synthetically update the scan costs of
the relation $Customer$, $Orders$, $Lineitem$, $Supplier$, $Nation$
and $Region$, where the scan costs are associated with the leaf plans
contributing to the previous best plan. These changes reflect the most
amount of work for incremental re-optimization that may originate from
a leaf update on the $PlanCost$.
Figure~\ref{fig:incremental-Q5-scan-cost} (a) shows the normalized
execution time of query re-optimization compared to the Volcano
baseline(0.50s). The X-axis is the ratio of the newly updated scan
cost to the old scan cost: if it is 1/8, it means that the scan cost
has decreased 7/8 of the original cost, etc. All the execution times
resulting from a single input scan relation are within 0.2x of the
Volcano baseline, with some cases nearly zero.  An interesting
observation from the figure is that the largest re-optimization time
comes from changing the input cost of $Lineitem$. This is probably
because $Lineitem$ is the most sensitive base relation to the best
plan.  Figure~\ref{fig:incremental-Q5-scan-cost} (b) and (c) show the
ratio of incrementally updated ``OR'' node state and ``AND'' node
state to the non-incremental total state. They represent the amount of
work done incrementally, and the trend is very similar to
Figure~\ref{fig:incremental-Q5-scan-cost} (a). This explains that the
execution time is a result of the amount of state being incremental
updated during re-optimization.}

\eat{
Similar explanations of  Figure~\ref{fig:incremental-Q10-scan-cost}(a)(b)(c) and Figure~\ref{fig:incremental-Q8Join-scan-cost}(a)(b)(c) for TPCH Q10 and Q8Join.  Volcano baseline of the execution time is 0.34s, and 2.46s respectively. }

\Subsubsection{Synthetic Changes to Subplan Costs}

\eat{Reviewer thinks 1/8x to 8x is not enough to cover order of magnitude errors. We should provide a wider range.}

We first simulate what
happens if we discover that an operator's output is not in accordance with our original selectivity
estimates.  
Figures~\ref{fig:incremental-Q5-join-selectivity} (a)-(c) show the
impact of synthetically injecting changes for each join expression's selectivity, and therefore the \reln{PlanCost} of the related plans and their
super-plans.  For conciseness
in the graph captions, we assign a symbol with each expression, e.g.,
the first join $Region\Join Nation$ is expression $A$, and the second
join expression combines the output of $A$ with data from the
$Customer$ table, yielding $B=Customer\Join A$.  We expect that changes to smaller subplans will take longer to re-optimize, and changes to larger subplans will take less time (due to the number of recursive propagation steps involved).  We separately plot
the results of changing each expression's selectivity value, as we change it
along a range from 1/8 the predicted size through 8 times larger than
the predicted size.  Running times in part (a) are plotted
relative to the Volcano implementation's performance: we see that the speedups are
at least a factor of 12, when the lowest-level join cost is updated;
going up to over 300, when the topmost join operator's selectivity is
changed.  In general the speedups confirm that larger
expressions are cheaper to update.  \eat{However, there are some exceptions
due to interactions with pruning: changes to the size of Join $C$ are
often more expensive to process than those to Join $B$.  We see in
parts (b) and (c) that this is because some of the changes to Join $B$
are almost immediately pruned from the search space. } We
can observe from these last two figures that we recompute only a small
portion of the search space.

\eat{
results of incremental re-optimization for TPCH Q5, when a single
intermediate ``AND'' node has changed its statistics such as its join
selectivity, thereby changing its $PlanCost$. As the previous set of
experiments, we synthetically update the join selectivities of
intermediate nodes along the path of the previous best plan, e.g.,
``AND'' node $A=Region\Join Nation$, ``AND'' node $B=Customer\Join A$,
and so forth.  This results in the most amount of work for query
re-optimization that is originated from an intermediate node.
Figure~\ref{fig:incremental-Q5-join-selectivity}(a) shows the
normalized execution time of query re-optimization compared to the
Volcano baseline, changing the newly updated scan cost to the old scan
cost. All the execution times of incremental re-optimization in this
figure are within 0.1x of the Volcano baseline. Another important
observation is that the closer the updated ``AND'' node to the root
node, the less amount of state that should be incrementally updated.
Hence, the execution time of updating $A$ is bigger than $B$, and so
forth.  The only exception is when increasing the cost of
$B=Customer\Join A$, because even increasing its cost to 8x of the
original, it is still the best cost at its parent ``OR'' node.
Figure~\ref{fig:incremental-Q5-join-selectivity}(b) and (c) also
reflect the similar trend of (a), however, compared to
Figure~\ref{fig:incremental-Q5-scan-cost}(b) and (c), it has decreased
the amount of updated state from 90\% to 20\%, and 60\% to 10\%
respectively. In general, changing the join selectivity is not as
expensive as changing the scan cost.  This validates the conclusion
that any incremental re-optimization from a single update has an order
of magnitude gains over the Volcano baseline.}

\Subsubsection{Changes based on Real Execution}

We now look at what happens when costs are updated according to an actual query execution.  We took
TPC-H Q5 and to gain better generality, we divided its input into 10 partitions (each having uniform distribution and independent variables) that would result in equal-sized output.
We optimized the query over one such partition, using histograms from the TPC-H dataset.  Then we ran the
resulting query over different partitions of \textbf{skewed data} (Zipf skew factor 0.5, from the Microsoft Research skewed
TPC-D generator~\cite{tpcd}); each of which exhibits different properties.  At the end we re-optimized the given the cumulatively
observed statistics from the partition.  We performed re-optimization on each of such interval, given the current plan and the revised statistics.

\eat{In addition to the synthetic workload evaluations, we also conduct a
study on the real workloads. We take a skewed data set generated by
the TPC-D generator (~\cite{tpcd}) with Zipfian skewed factor 0.5,
divide the dataset into 10 splits for all the relations except for
$Region$ and $Nation$, and perform query re-optimization incrementally
for each split. Since statistics may be wrongly estimated on a skewed
data set, especially the join selectivities, we expect a series of
updates on the intermediate subplans that occur simultaneously.}

Figure~\ref{fig:incremental-Q5-real-workload} (a) shows the execution
times for each round of incremental re-optimization, normalized
against the running time of Volcano.  We see that, as with the join
re-estimation experiments of
Figure~\ref{fig:incremental-Q5-join-selectivity}, there are speedups
of a factor of 10 or greater.  In terms of throughput, the Volcano
model takes 500msec to perform one optimization, meaning it can
perform 2 re-optimizations per second; whereas our declarative
incremental re-optimizer can achieve 20-60 optimizations per second,
and it can respond to changing conditions in 10-100msec.  Again,
Figure~\ref{fig:incremental-Q5-real-workload} (b) and~(c) show that the
speedup is due to significant reductions in the amount of state that
must be recomputed.

\Subsection{Contributions of Pruning Strategies}
\label{subsec:strategies}

\eat{
Question: How does each of the optimization strategy contribute, in isolation and together?
Answer: We present four results: AggSel, RefCount plus AggSel, BranchAndBounding (which implies AggSel, but without RefCount) and BranchAndBounding plus RefCount (which also implies AggSel). We show results of pre-optimization of TPCH Q5,Q5S,Q10,Q8Join, Q8JoinS in Figure~\ref{fig:stategies-preoptimization}(a)(b)(c); and results of re-optimization of TPCH Q5 changing two different relations $Supplier$ and $Orders$ in Figure~\ref{fig:stategies-reoptimization-Q5-Supplier}(a)(b)(c) and Figure~\ref{fig:stategies-reoptimization-Q5-Orders}(a)(b)(c);}

Here we investigate how each of our pruning and incremental
strategies from Sections~\ref{sec:sip} and~\ref{sec:incremental}
contribute to the overall performance of our declarative optimizer.
We systematically considered all techniques individually and in
combination, unless they did not make sense (e.g., reference counting
must be combined with one of the other techniques, and
branch-and-bound requires aggregate selection to perform
pruning of the search space).  See
Figure~\ref{fig:stategies-preoptimization}, where \emph{AggSel} refers
to aggregate selection with source tuple suppression; \emph{RefCount} refers to
reference counting; and \emph{Branch\&\-Bounding} refers to recursive
bounding. We consider aggregate selection in isolation and
in combination with the other techniques individually and together.
(We also considered the case where none of the pruning techniques are enabled: 
here running times were over 2 minutes,
due to a complete lack of pruning.  We omit this from the graphs.)

\eat{
In this section, we evaluation how our each optimization strategy
contribute to the performance, in isolation and together. In
particular, we compare four schemes here: AggSel, RefCount+AggSel
(with both AggSel and RefCount), Branch\&Bounding (without Refcount)
and Branch\&Bounding (with RefCount). Since Branch\&Bounding strictly
contains AggSel, the last one is when we enable all the three
optimizations, and what we use to compare against other in the
previous experiments. We should note here that every scheme we compare
here exploits sideways information passing: the pruned state is
propagated back to $SearchSpace$ state and some tuples are
pre-inserted and then re-deleted. We study the performance of both
pre-optimization and incremental re-optimization below.}

Figures~\ref{fig:stategies-preoptimization} (a)-(c) compare the three
pruning strategies when performing initial optimization on 
various TPC-H queries.  It can be observed that each of the pruning 
techniques adds a small bit of runtime overhead (never
more than 10\%) in this setting, as each requires greater computation
and data propagation.  Parts (b) and (c) show that each
technique adds greater pruning capability, however.

Once we move to the incremental setting --- shown in
Figures~\ref{fig:stategies-reoptimization-Q5-Orders} (a)-(c) for query
Q5 and changes to the \reln{orders} table, over different cost
estimate changes --- we see significant benefits in running time as
well as pruned search space.  Note that in contrast to our other
graphs for incremental re-optimization, plots (b) and (c) isolate the
amount of pruning performed, rather than showing the total state
updated.  We see here that our different techniques work best in
combination, and that each increases the amount of pruning.


\eat{
branch\&bounding scheme compared to the AggSel one is generally within
0.1x of the original time. Here the overhead mainly comes from the
computation overhead of bounds and propagations, and maintenance of
ref counts when applicable. In the next two set of figures, we can see
a great performance improvement in incremental re-optimization. Note
that in (b) and (c) we measure the amount of pruned state, rather than
the updated state, hence the larger the more effective in pruning. We
can see that schemes with Branch\&Bounding is most effective in
pruning. For example, it prunes 40\% more ``OR'' nodes and 5\% more
``AND'' nodes for TPCH Q10S.
}

\eat{
Figure~\ref{fig:stategies-reoptimization-Q5-Supplier}(a)(b)(c) show the performance for TPCH Q5 when the scan cost of $Supplier$ is updated. }

\eat{
In the next set of experiments, we compare the three optimization strategies when \emph{re-optimizing} TPCH queries. Here each strategy is applied to the pre-optimization phase, and then to incremental re-optimization to measure the performance. Figure~\ref{fig:stategies-reoptimization-Q5-Orders}(a)(b)(c) show the performance for TPCH Q5 when the scan cost of $Orders$ is updated. We can see from (a) that each strategy has result in better performance. For example, when the ratio is 0.125, AggSel takes about 0.4x of the baseline; RefCount+AggSel about 0.23x; Branch\&Bounding about 0.16x; and Branch\&Bounding about 0.09x. When the ratio of input update is bigger than 1, Branch\&Bounding schemes have nearly zero execution time. (b) and (c) measure the incrementally pruned state. Since Branch\&Bounding schemes already prune a lot during query pre-optimization, the incrementally pruned state is sometimes near zero; however, in other cases that are non-zero, Branch\&Bounding with RefCount always prune the most among all the schemes. }

Our pruning strategies enable incremental updates to be supported with relatively minor space overhead:  even for the largest query (Q8Join), the total optimizer state was under 100MB.

\Subsection{Incremental Reoptimization for AQP}
\label{subsec:linearRoad}

\begin{figure}[ht]
 \begin{center}
 \includegraphics[width=1.85in]{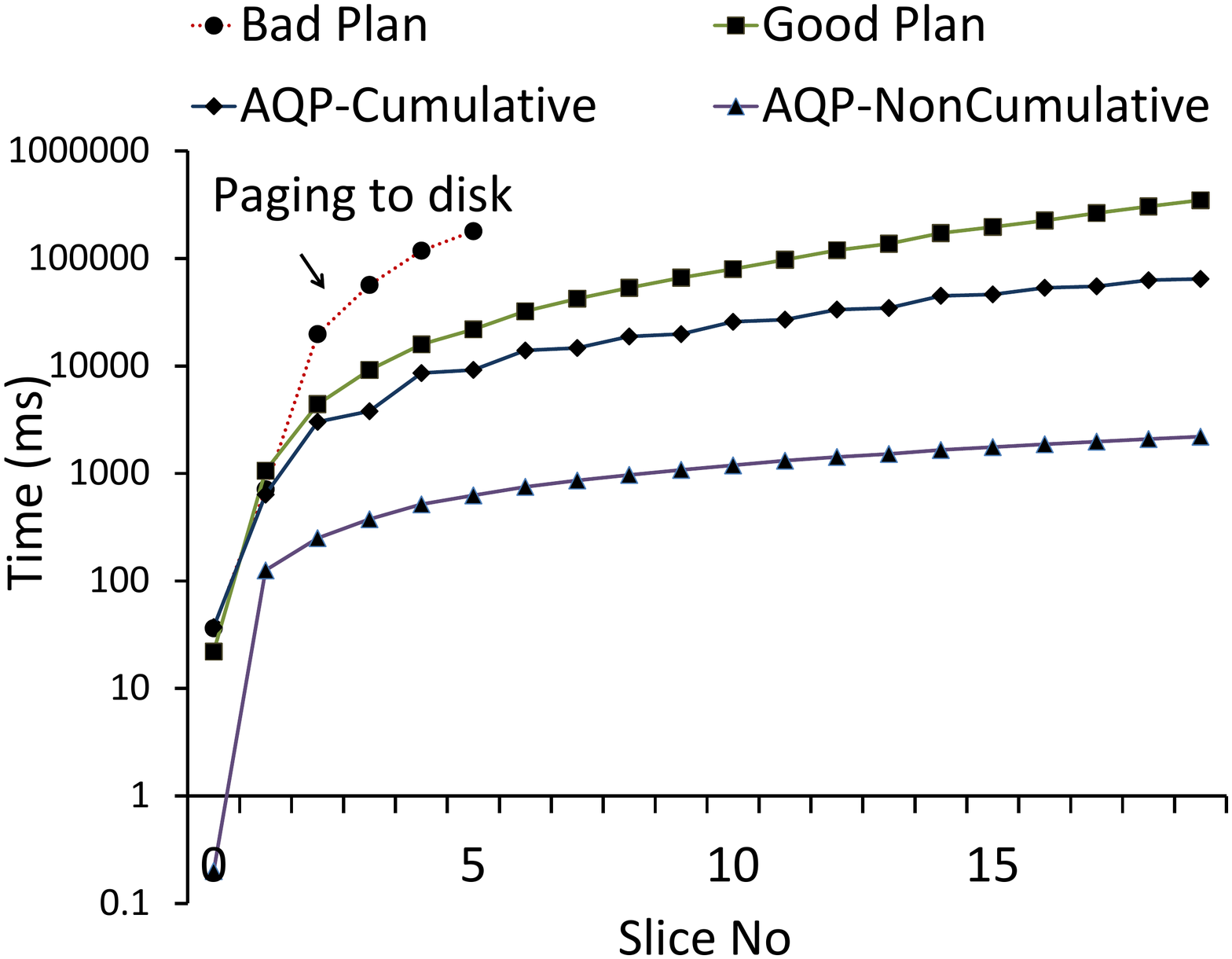} 
\vspace{-6mm}
\caption{\small AQP execution time\label{fig:aqp-exe}}
\vspace{-4mm}
\end{center}
\end{figure}

\eat{
\begin{figure}[ht]
\begin{center}
  \includegraphics[width=2.25in]
  {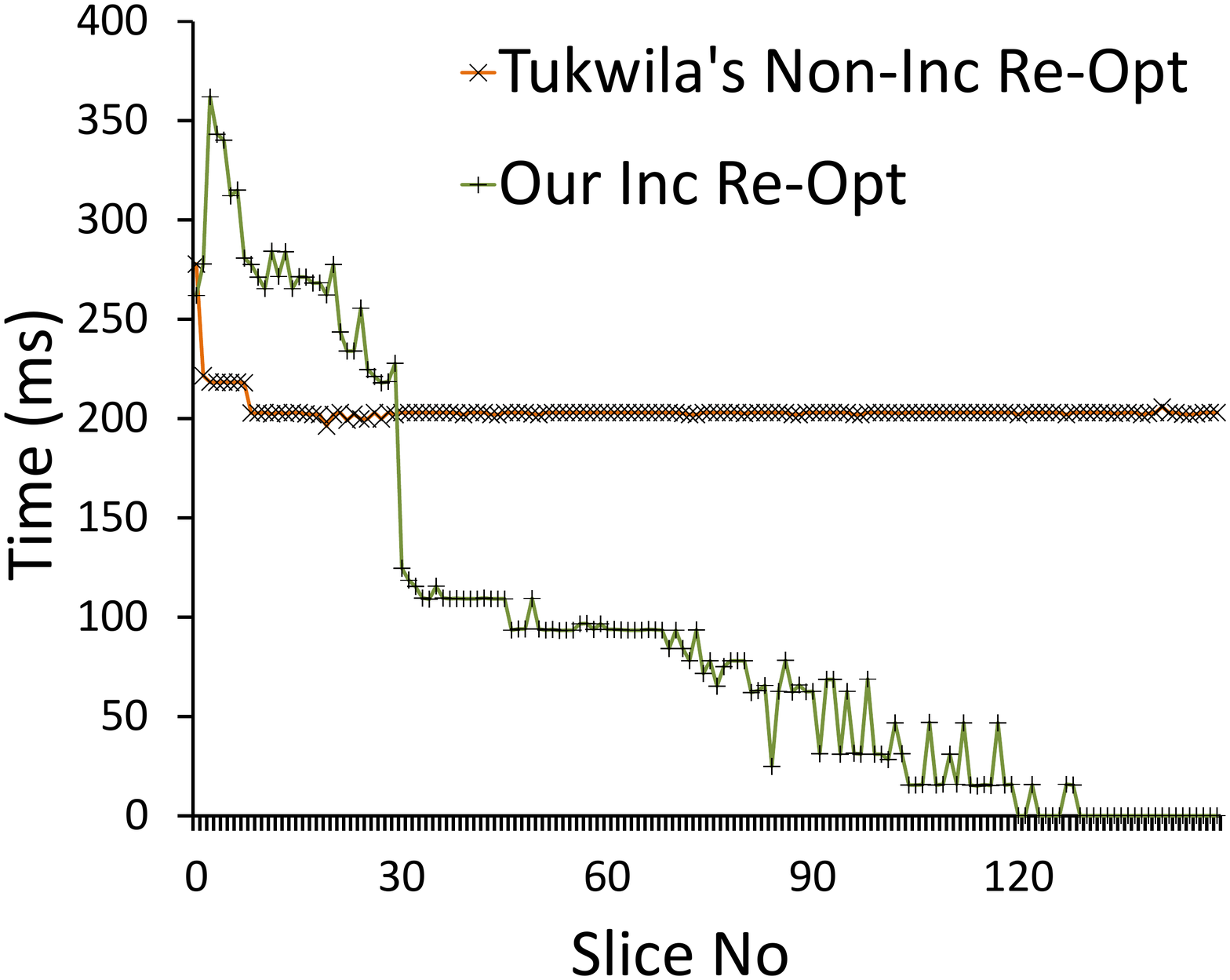}
\small (a) Adapting every 1s for non-cumulative data
\end{center}
\end{figure}
}


\eat{
\begin{figure}[ht]
\begin{center}
\includegraphics[width=2.25in]{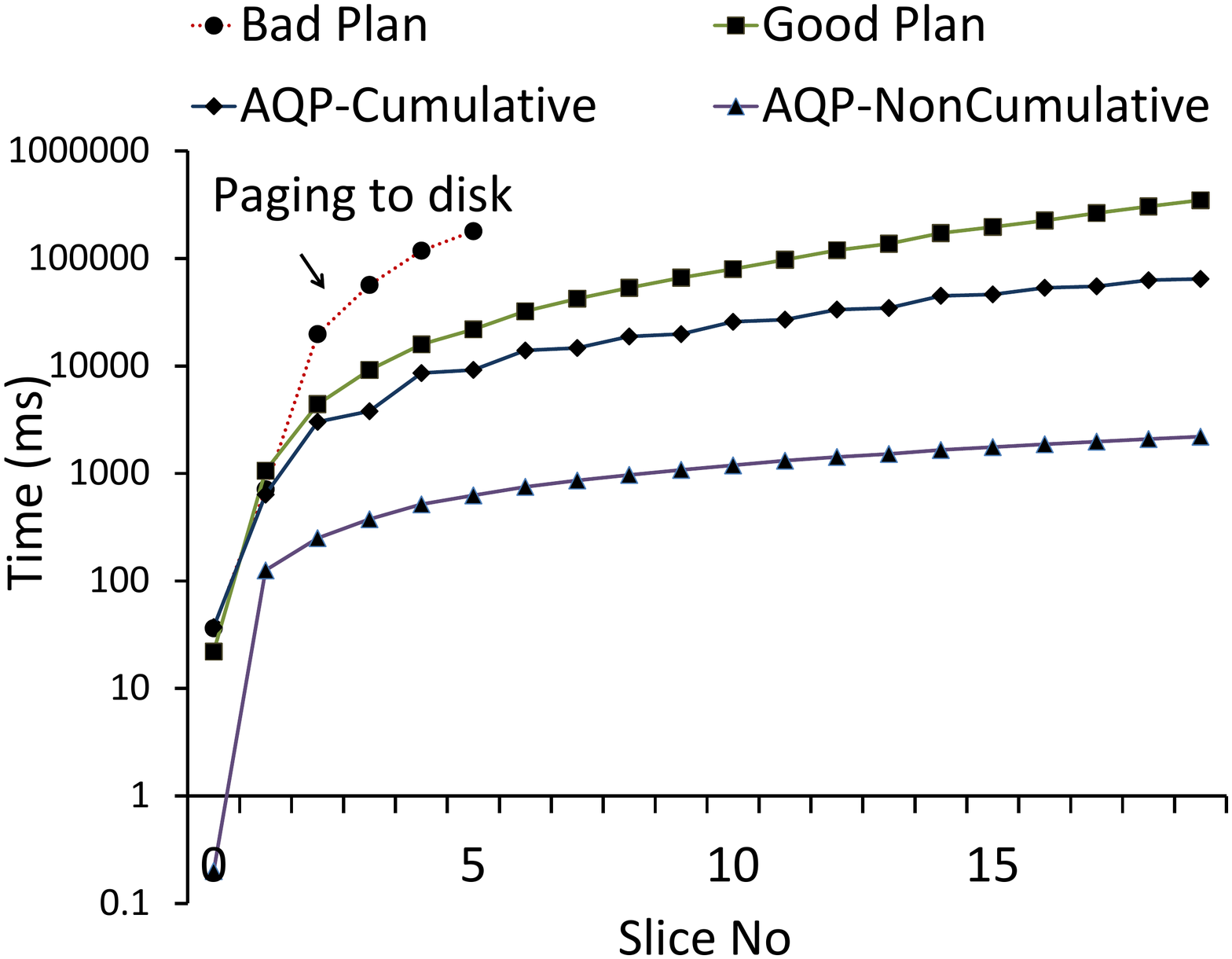}
\small (c) Adapting every 1s for cumulative data
\end{center}
\caption{\label{fig:linearRoad-SegToll}\small Incremental re-optimization and adaptive execution performance of SegTollS query over Linear Road Benchmark data~\cite{linearRoad}.}
\vspace{-1mm}
\end{figure}

}

A major motivation for our work was to facilitate better \emph{cost-based adaptive query processing}, especially for
continuous optimization of stream queries.  Our goal is to show the benefits of incremental reoptimization; we leave as future work a broader comparison of adaptive query processing techniques.  Our final set of experiments shows how our techniques can be used within
a standard cost-based adaptive framework, one based on the \emph{data-partitioned} model of~\cite{tukwila-04} where the optimizer periodically pauses plan execution, forming a ``split'' point from which it may choose a new plan and continue execution.  In general, if we change plans at a split point, there is a challenge of determining how to combine state across the split.  In contrast to~\cite{tukwila-04} we chose \emph{not} to defer the cross-split join execution until the end: rather, we used CAPS's \emph{state migration} strategy~\cite{cape-04} to transfer all existing state from the prior plan into the current one.  As necessary, new intermediate state is computed.  Note that our data-partitioned model could be combined with other cost-based adaptive schemes such as~\cite{db2-mqreopt,DBLP:conf/sigmod/KabraD98}.

To evaluate this setting, we combine unfold the various views comprising the SegToll query from the Linear Road benchmark~\cite{linearRoad} --- resulting in a 5-way join plan SegTollS with multiple windowed aggregates.  We use the standard Linear Road data generator to synthesize data whose characteristics frequently change.

\begin{table}
\begin{center}
\begin{tabular}{|c|c|c|c|}
\hline
Per Slice & Re-Opt Time & N.C. Exec Time & Total Time \\
\hline
1s & 5.75s & 2.20s & 7.95s \\
\hline
5s & 1.23s & 6.82s & 8.05s\\
\hline
10s & 0.63s & 13.35s & 13.98s \\
\hline
\end{tabular}
\end{center}
\vspace{-5mm}
\caption{\label{table:aqp} Frequency of Adaptation (20 sec stream)}
\end{table}

\eat{
Our results are shown in Figure~\ref{fig:linearRoad-SegToll}. In (a) and (b), we compare the performance of our incremental re-optimizer against a non-incremental Tukwila-like approach~\cite{tukwila-04}, and measure the execution time of the adapted plans chosen by our re-optimizer, in comparison with static non-adaptive plans. For comparison, we obtain a ``good'' single static plan by treating the entire stream as a relation and let our optimizer picks the best for the workload. This good plan is important for us to understand how well our re-optimizer behaves, but may not be achievable by a stream optimizer. We also obtain a ``bad'' plan computed by an optimizer assuming it only has partial knowledge about the stream, e.g., the first several seconds of the stream information. This plan is still optimal for that period of time, but in the long run it is likely not going to perform well.

In (a), we measure the performance for adapting for every 1 second slice of the input stream. We focus on the execution times of \emph{non-cumulative} data computation within each slice to emphasize the impact of the plan picked by an optimizer on the performance of new data computation. We measure our incremental re-optimization overhead versus a non-incremental Tukwila-like optimizer~\cite{tukwila-04}. Our re-optimization time starts off from around 0.4s and drops towards 0 ever since, whereas their performance holds onto 0.2s throughout the stream. In total we take 14.6s for 300s of input stream, whereas they need 60.8s which is roughly 4x better. Ironically, as plan execution time holds well around 0.1s, full optimization wastes resources on exploration rather than exploitation. As expected, our incremental re-optimizer spends more time in the first half of the stream because data characteristics fluctuate more wildly before they have been stabilized inside windows. After around slice 130, our re-optimizer only changes the plan once near 190 and retain the same plan till the end of the stream --- as we trigger delta cost updates only when input changes more than 1\%, this shows that  mild changes of data characteristics make our re-optimizer extremely cheap. 

We also measure the execution time of the adapted plans picked by our re-optimizer. It behaves roughly well as the good single plan (with no adaptation), which is quite encouraging because a good plan assumes it knows the complete stream statistics but we make decisions based on the information of the stream received up till to the adaptation point. On the other hand, the bad single plan with just the knowledge of the first 1s of data statistics takes about 300ms-400ms in executing each slice. Overall our approach spends 36.9s in executing queries, which is 1/3x of the bad plan, and 0.2s better than the good plan.  Even adding optimization time and execution time, our performance is still better than the bad plan, although not as ideal as the good plan. This is because the optimization to execution ratio here is relatively high that the optimization time cannot be amortized as well as in bigger chunks.

In (b) we experiment the same metrics as in (a), the only difference is the granularity of adaptation, here we re-optimize every 10s rather than every 1s, which is a much coarser granularity. We pick the same good single plan and bad single plan as in (a) for comparison. This time, the bad plan makes the execution engine go out of memory (8GB) quickly after 30s, and since it pages to disk, the performance is not representative enough to be shown on the figure. Overall our adaptive execution performance encouragingly matches the good plan again around 9s. In optimizing, our incremental approach shows another cumulative gain over non-incremental one, totaling 2.9s against 5.5s. The fewer the adaptations, the smaller the optimization overhead.
}

In the adaptive setting, our goal is to have the optimizer start with zero statistical information on the data, and find a sequence of plans whose running time equals or betters the \emph{single best static plan} that it would pick given complete information (e.g., histograms).  

Figure~\ref{fig:aqp-exe} shows that in fact our incremental AQP scheme provides \textbf{superior performance to the single best plan} (``good single plan''), if we re-optimize every 1 second. This is because the adaptive scheme has a chance to ``fit'' its plan to the local characteristics of whatever data is currently in the window, rather than having to choose one plan for all data. 

A natural question is where the ``sweet spot'' is between query execution overhead and query optimizer overhead, such that we can achieve best performance.  We measured, for a variety of slice sizes, the total running time of query re-optimization and execution \emph{over each new slice of data} (not considering the additional overhead of state migration, which depends on how similar the plans are).  We can see the results in Table~\ref{table:aqp}:  there are significant gains in shrinking the interval from 10sec to 5sec, but little more gain to be had in going down to 1sec.  Figure 9 helps explain this:  as we execute and re-optimize, the overhead of a non-incremental re-optimizer remains constant (about 200 msec each time), whereas the incremental re-optimization time drops off rapidly, going to nearly zero.  This means that the system has essentially \emph{converged on a plan} and that new executions do not affect the final plan.




\eat{
We then measure the more general (and genuinely adaptive) cumulative data execution performance in (c), where data can be computed (e.g., joined) across slices. In order to correctly measure cumulative execution time in our adaptive query engine, we implemented the state-of-the-art state migration techniques as described in Cape~\cite{cape-04}. Basically as plan changes during execution, the old operator state in the old plan needs to be \emph{migrated} to the new plan, and certain new state needs to be properly recomputed, in order to ensure complete and non-redundant results. We take into account of this migration cost when measuring the execution time, and compare our performance with non-adaptive plans. In general, we see that the bad plan pages to disk again, and the good plan linearly grows towards 10s per slice (for an 1s input stream this is certainly not ideal) and our engine piggy-backs around 1s and 5s during execution. We observe that our adaptive engine switches the plan after every slice, and perhaps not surprisingly, when it alternates among two plans it has a better performance than merely sticking with either one. Efficient state matching and state sharing mechanisms enable our query execution engine to spend only the essential portion of efforts on recomputing state, and the plan picked by our re-optimizer shows its competitiveness compared with even the good plan knowing the entire stream information. }

\eat{We obtain the following observations from these experiments. First, by adapting plans we are able to avoid extremely bad plans like those that run out of memory like in b) and c) as early as possible. In most stream-based workloads, as arrival rates and data characteristics were hard to predict, we can not avoid choosing those bad plans. Adaptive query processing is shown to be of paramount importance in this scenario. Second, the performance of query re-optimization is extremely important in enabling fine-grained adaptations. For example, in (a), data arrive every 1s whereas a non-incremental re-optimizer takes a constant 0.2s to merely choosing a plan. It is even more than the average 0.1s of execution time for non-cumulative data. Resources would be wasted on repeated work of optimizing, more importantly, for bursty streams the current Tukwila-like stream optimizer is unable to provide satisfactory quick turnaround of optimization. The more time spent on re-optimizing, the more time is going to be spent on the bad plans. Third, Our incremental schemes work best when data characteristics remain mostly the same at re-optimization, as we see in the later part of (a). When data are still accumulating in a window, we expect cardinalities to change dramatically over slices, as shown in the earlier part of (a). }

\eat{
Left figure of 1s per chunk experiment: 1) It shows that our execution time picked by the re-optimization (red line) is similar to the golden standard: the best plan that would have been picked by an optimizer if it knows the entire 300s of stream information before optimizing (drawn as the best plan for the entire stream, blue line). In reality one cannot know the stream statistics beforehand, so our re-optimizer does online refinement and picks nearly optimal plans. 2) If an optimizer does one-time optimization, based information on the first 1s chunk, and does no re-optimization in the future, the execution time is shown in green line, which is approximately 3x of our execution time. This is because first 1s of data characteristics cannot represent the whole data stream, which is rather typical in data stream scenarios. 3) Incremental re-optimization time is acceptable. In the first half, the optimization time is more frequent because data characteristics tend to fluctuate before they reach a relatively stable window size. When it is more stable, the re-optimizer spent zero time in re-optimizing, and leaving the optimal plan running for the rest of the data. The total optimization time is 14.6s, <400s per thunk, and the total execution time of new data joining new data is 36.9s, as a whole it is 51.6s. the first plan based on 1s of data takes 102s, which is 3x of our execution time, and 2x of our optimization+execution time. We save 100\% here! An ideal golden standard plan takes 37.1s, although we perform even better in execution time, our re-optimization overhead is big here because as this is 1s chunk, we re-optimize very frequently (the re-optimization time amortizes better for the larger 10s chunk). 

In the 10s chunk experiment, our optimization time totals 2.9s, our execution time totals 229s. The worst plan takes more than 150s each split * 30 splits, so we are at least 20x better. The golden standard plan takes 232.5s, so we are also slightly better than it even counting optimization time. 
}

\Subsection{Experimental Conclusions}
\label{subsec:experiment-conclusion}

We summarize our evaluation results by providing the answers to the
questions raised earlier in this section. First, our declarative query
optimizer performs respectably when compared to a procedural query
optimizer, for initial optimization: despite the overhead of starting
up a full query processor, it gets within 10-50\% of the running times
of a dedicated optimizer.  It more than recovers this overhead during
incremental re-optimization, where --- across a variety of queries and
changes --- it typically shows an order-of-magnitude speedup, or
better.  These gains are largely due to having to re-enumerate a much
smaller space of plans than a full re-optimization.  In addition, we find
that our pruning techniques developed in Section~\ref{sec:sip} and
Section~\ref{sec:incremental} each contribute in a meaningful way to
the overall performance of incremental re-optimization. Finally, we show that our incremental re-optimization techniques introduced in this paper help cost-based adaptation techniques provide \emph{finer-grained} adaptivity and hence better overall performance.  Overhead decreases as the system \emph{converges on a single plan}.

\eat{
It also results in
much less amount of work in terms of ``AND'' and ``OR'' node state
updates. Finally, our three optimization strategies described in
Section~\ref{sec:sip} and Section~\ref{sec:incremental} have each
contribute their savings on the performance. In query
pre-optimization, their performance overhead is less than 0.1x of the
original; however, in incremental re-optimization, the fully
Branch\&Bounding with RefCount scheme can save 0.3x of the original
compared to the one with just AggSel.}


\vspace{-3mm}
\Section{Related Work}
\label{sec:related}

\eat{
Eddies, StreaMon, SteMs: we are cost-based re-opt but they are heuristics-based; we guarantee optimal results if the cost model is perfect; our declarative flavor is hugely different from theirs. We are plan-based and they are tuple-based, when a change has cumulative effect on many downstream operators, we need a cost model with statistics. 

Tukwila and Cape: their re-optimizer is essentially a tweak of Volcano or System-R. They did stream workloads. Our comparison to Volcano or System-R in the stream case can be regarded as comparison to these state-of-the-art ADP systems.

Run-time heuristics such as choose-plan, mid-query re-opt, progressive re-optimization: they are local heuristics, decided at run-time rather than by a full-bloom optimizer. They only work on partitioned plans, not on partitioned data.
}


Our work takes a first step towards supporting continuous adaptivity in a distributed (e.g., cloud)
setting where correlations and runtime costs may be unpredictable at
each node.  Fine-grained adaptivity has previously only been addressed
in the query processing literature via heuristics, such as flow
rates~\cite{DBLP:conf/sigmod/AvnurH00,viglas-rate}, that continuously
``explore'' alternative plans rather than using long-term cost estimates.  
Exploration adds overhead even when a good plan has been found; moreover, 
for joins and other stateful operators, the flow heuristics has been shown to result in state that later
incurs significant costs~\cite{stairs}.  Other strategies based on
filter reordering~\cite{stanford-aqp} are provably optimal, but only work for selection-like predicates.
Full-blown cost-based
re-optimization can avoid these future costs but was previously only
possible on a coarse-grained (high multiple seconds)
interval~\cite{tukwila-04,cape-04}.


Our use of declarative techniques to specify the optimizer was
inspired in part by the Evita Raced~\cite{evita-raced} system.
However, their work aims to construct an entire optimizer using
reprogrammable datalog rules, whereas our goal is to effectively perform
incremental maintenance of the output query plan.  We seek to
fully match the pruning techniques of conventional optimizers
following the System R~\cite{systemR} and Volcano~\cite{volcano}
models.  Our results show for the first time that a declarative
optimizer \emph{can} be competitive with a procedural one, even for
one-time ``static'' optimizations, and produce large benefits for
future optimizations.

\eat{
Note that our approach differs from System R~\cite{systemR} and
Volcano~\cite{volcano}, neither of which can easily support
incremental reoptimization. System R enumerates plans and estimates
costs in the same order of \emph{bottom-up iterative deepening}. This
order is effective for dynamic programming as best sub-plans are
always computed before the parent plan; however, it has no pruning at
all, because every child node is already cost-estimated when a parent
node is decided to have been pruned. On the other hand, Volcano
enumerates plans in depth-first pre-order traversal of the search
space, and estimates costs in depth-first post-order traversal of the
search space. Here cost estimation is a bottom-up traversal, however,
it is called via recursive functions hence it must have the same
depth-first order as plan enumeration. Volcano's search order needs to
achieve dynamic programming via memoization; on the other hand, it is
effective in exploiting branch-and-bounding because depth-first
traversal ensures that a complete plan from the root is computed early
enough that it can be used as a bound to potentially prune others.
}

\Section{Conclusions and Future Work}
\label{sec:conclusion}

\eat{ML: Instead of repeating intro and enumerating contributions, I would suggest adding some retrospects of our experience. Declarative framework buys us many interesting features, like customizable search orders, potential for parallelization, etc, that traditional DP approaches did not have. It can do more stuff than merely incremental re-opt. This is a framework that has lots of potentials, in this paper we show its benefits in abstracting the problem of re-optimization by exploiting the optimization techniques in logic programming, in Evita-Raced it was introduced for expressing different optimization strategies, maybe it has more potentials for other goals. If an optimizer can be re-architected in this way, it would be really amazing to achieve something that were not thought about before because of the lack of such abstraction point of view. Our abstraction approach might not be the only declarative model, but this framework is highly extensible to various other tasks, like A* searching, etc. Indeed, although it sounds like a drastic change from the original engine, most of the functions can be reused, best of all, we exploit query engines to not only process queries but perform optimizer computations as well! If we think about query optimization as more of a data-intensive task, this is really the way to go! Imagine that the computation of a query optimizer is complex enough that it should be run on 1000 machines using Hadoop. With more than 20 joins or so, or multi-query optimization, as it is an exponential problem, this might be essential. In this paper we are merely using a database stream query processing engine to replace Hadoop here, and that's why we use Datalog because Datalog has the same expressive power as SQL + recursion, hence a Datalog plan can be executed by a database engine. Other mild ideas like expressing query optimization in Map and Reduce tasks is certainly interesting and possible as well.}

To build large-scale, pipelined query processors that are reactive to
conditions across a cluster, we must develop new adaptive query
processing techniques.  This paper represents the first step 
towards that goal: namely, a fully cost-based architecture for
incrementally re-optimizing queries.  We have made the following
contributions:

\begin{compactList}
\vspace{-1mm}
\item A rule-based, declarative approach to query (re)optimization in
  adaptive query processing systems.
\vspace{-2mm}

\item Novel optimization techniques to
  prune the optimizer state: \emph{aggregate
  selection}, \emph{reference counting}, and \emph{recursive bounding}.
\vspace{-2mm}

\item A formulation of query re-optimization as an incremental view
  maintenance problem, for which we develop novel incremental
  algorithms to deal with insertions, deletions and updates over
  runtime cost parameters.
\vspace{-2mm}

\item An implementation over the ASPEN query
  engine~\cite{recursive-views}, with a comprehensive
  evaluation of performance against alternative approaches, over a
  diverse workload, showing order-of-magnitude speedups for
  incremental re-optimization.
\vspace{-2mm}
\end{compactList}

We believe this basic architecture leaves a great deal of room for
future exploration. We plan to study how our declarative execution
model parallelizes across multi-core hardware and clusters, and how it
can be extended to consider the cost of \emph{changing} a plan given
existing query execution state.

\vspace{-3mm}

\balance

{\scriptsize
\setlength{\bibsep}{0.011in}
\bibliographystyle{abbrv}
\bibliography{zives-short,boon,mengmeng}

\begin{thebibliography}{10}

\bibitem{aurora}
D.~J. Abadi, D.~Carney, U.~Cetintemel, M.~Cherniack, C.~Convey, S.~Lee,
  M.~Stonebraker, N.~Tatbul, and S.~Zdonik.
\newblock Aurora: a new model and architecture for data stream management.
\newblock {\em VLDB J.}, 12(2), August 2003.

\bibitem{boom-analytics}
P.~Alvaro, T.~Condie, N.~Conway, K.~Elmeleegy, J.~M. Hellerstein, and R.~Sears.
\newblock Boom analytics: exploring data-centric, declarative programming for
  the cloud.
\newblock In {\em EuroSys}, 2010.

\bibitem{linearRoad}
A.~Arasu, M.~Cherniack, E.~Galvez, D.~Maier, A.~S. Maskey, E.~Ryvkina,
  M.~Stonebraker, and R.~Tibbetts.
\newblock Linear road: a stream data management benchmark.
\newblock In {\em VLDB}, VLDB, pages 480--491. VLDB Endowment, 2004.

\bibitem{DBLP:conf/sigmod/AvnurH00}
R.~Avnur and J.~M. Hellerstein.
\newblock Eddies: Continuously adaptive query processing.
\newblock In {\em SIGMOD}, 2000.

\bibitem{stanford-aqp}
S.~Babu, R.~Motwani, K.~Munagala, I.~Nishizawa, and J.~Widom.
\newblock Adaptive ordering of pipelined stream filters.
\newblock In {\em SIGMOD}, 2004.

\bibitem{asterix}
A.~Behm, V.~R. Borkar, M.~J. Carey, R.~Grover, C.~Li, N.~Onose, R.~Vernica,
  A.~Deutsch, Y.~Papakonstantinou, and V.~J. Tsotras.
\newblock Asterix: towards a scalable, semistructured data platform for
  evolving-world models.
\newblock {\em Distributed and Parallel Databases}, 29(3), 2011.

\bibitem{telegraphcq}
S.~Chandrasekaran, O.~Cooper, A.~Deshpande, M.~J. Franklin, J.~M. Hellerstein,
  W.~Hong, S.~Krishnamurthy, S.~Madden, V.~Raman, F.~Reiss, and M.~A. Shah.
\newblock {TelegraphCQ}: Continuous dataflow processing for an uncertain world.
\newblock In {\em CIDR}, 2003.

\bibitem{evita-raced}
T.~Condie, D.~Chu, J.~M. Hellerstein, and P.~Maniatis.
\newblock Evita raced: metacompilation for declarative networks.
\newblock {\em PVLDB}, 1(1), 2008.

\bibitem{stairs}
A.~Deshpande and J.~M. Hellerstein.
\newblock Lifting the burden of history from adaptive query processing.
\newblock In {\em VLDB}, 2004.

\bibitem{aqp-survey}
A.~Deshpande, Z.~Ives, and V.~Raman.
\newblock Adaptive query processing.
\newblock {\em Foundations and Trends in Databases}, 2007.

\bibitem{DBLP:journals/debu/Graefe95a}
G.~Graefe.
\newblock The {Cascades} framework for query optimization.
\newblock {\em IEEE Data Eng. Bull.}, 18(3), 1995.

\bibitem{volcano}
G.~Graefe and W.~J. McKenna.
\newblock The {Volcano} optimizer generator: Extensibility and efficient
  search.
\newblock In {\em ICDE}, 1993.

\bibitem{springerlink:10.1007/BFb0014149}
A.~Gupta, H.~Jagadish, and I.~S. Mumick.
\newblock In P.~Apers, M.~Bouzeghoub, and G.~Gardarin, editors, {\em EDBT}.
  Berlin / Heidelberg, 1996.

\bibitem{gms93-dred}
A.~Gupta, I.~S. Mumick, and V.~S. Subrahmanian.
\newblock Maintaining views incrementally.
\newblock In {\em SIGMOD}, 1993.

\bibitem{tukwila-04}
Z.~G. Ives, A.~Y. Halevy, and D.~S. Weld.
\newblock Adapting to source properties in processing data integration queries.
\newblock In {\em SIGMOD}, pages 395--406, 2004.

\bibitem{DBLP:conf/sigmod/KabraD98}
N.~Kabra and D.~J. DeWitt.
\newblock Efficient mid-query re-optimization of sub-optimal query execution
  plans.
\newblock In {\em SIGMOD}, 1998.

\bibitem{liu_diff_eval}
L.~Liu, C.~Pu, R.~Barga, and T.~Zhou.
\newblock Differential evaluation of continual queries.
\newblock Technical Report TR95-17, University of Alberta, June 1995.

\bibitem{recursive-views}
M.~Liu, N.~E. Taylor, W.~Zhou, Z.~G. Ives, and B.~T. Loo.
\newblock Recursive computation of regions and connectivity in networks.
\newblock In {\em ICDE}, 2009.

\bibitem{db2-mqreopt}
V.~Markl, V.~Raman, G.~Lohman, H.~Pirahesh, D.~Simmen, and M.~Cilimdzic.
\newblock Robust query processing through progressive optimization.
\newblock In {\em SIGMOD}, 2004.

\bibitem{DBLP:journals/pvldb/MelnikGLRSTV10}
S.~Melnik, A.~Gubarev, J.~J. Long, G.~Romer, S.~Shivakumar, M.~Tolton, and
  T.~Vassilakis.
\newblock Dremel: Interactive analysis of web-scale datasets.
\newblock {\em PVLDB}, 3(1), 2010.

\bibitem{stanford-stream}
R.~Motwani, J.~Widom, A.~Arasu, B.~Babcock, S.~Babu, M.~Datar, G.~Manku,
  C.~Olston, J.~Rosenstein, and R.~Varma.
\newblock Query processing, resource management, and approximation in a data
  stream management system.
\newblock In {\em CIDR}, 2003.

\bibitem{tpcd}
V.~Narasayya.
\newblock {TPC-D} skewed data generator.
\newblock 1999.

\bibitem{systemR}
P.~G. Selinger, M.~M. Astrahan, D.~D. Chamberlin, R.~A. Lorie, and T.~G. Price.
\newblock Access path selection in a relational database management system.
\newblock In {\em SIGMOD}, pages 23--34, 1979.

\bibitem{sudarshan91aggregation}
S.~Sudarshan and R.~Ramakrishnan.
\newblock Aggregation and relevance in deductive databases.
\newblock In {\em VLDB}, 1991.

\bibitem{viglas-rate}
S.~D. Viglas and J.~F. Naughton.
\newblock Rate-based query optimization for streaming information sources.
\newblock In {\em SIGMOD}, 2002.

\bibitem{cape-04}
Y.~Zhu, E.~A. Rundensteiner, and G.~T. Heineman.
\newblock Dynamic plan migration for continuous queries over data streams.
\newblock In {\em SIGMOD}, 2004.

\end{thebibliography}
}





\vspace{-4mm}
\begin{appendix}
\vspace{-2mm}
\Section{Datalog Rules for Optimizer}
\vspace{-2mm}
\label{sec:opt-rules}


\scriptsize
\linespread{0.9} {
\begin{small}
\begin{datalog}
\heading{R1:}\>\dlrel{SearchSpace}\dlvar{(expr, prop, index, logOp, phyOp, lExpr,} \\
\>\>\hspace{+.15in}\dlvar{lProp, rExpr, rProp)} \dlog \\
\>\>\atom{Expr}{expr, prop}, \atom{Fn\_isleaf}{expr, {false}}, \\
\>\>\dlrel{Fn\_split}\dlvar{(expr, prop, \underline{index}, \underline{logOp}, \underline{phyOp},	\underline{lExpr}}, \\
\>\>\hspace{+.15in}\dlvar{\underline{lProp}, \underline{rExpr}, \underline{rProp})}; \\
\heading{R2:}\>\dlrel{SearchSpace}\dlvar{(expr, prop, index, logOp, phyOp,}\\
\>\>\hspace{+.15in}\dlvar{lExpr, lProp, rExpr, rProp)} \dlog \\
\>\>\atom{SearchSpace}{-, -, -, -, -, expr, prop, -, -},\\
\>\>\atom{Fn\_isleaf}{expr, {false}}, \\
\>\>\dlrel{Fn\_split}\dlvar{(expr, prop, \underline{index}, \underline{logOp}, \underline{phyOp},}\\
\>\>\hspace{+.15in}\dlvar{\underline{lExpr}, \underline{lProp}, \underline{rExpr}, \underline{rProp})};\\
\heading{R3:}\>\dlrel{SearchSpace}\dlvar{(expr, prop, index, logOp, phyOp,}\\
\>\>\hspace{+.15in}\dlvar{lExpr, lProp, rExpr, rProp)} \dlog \\
\>\>\atom{SearchSpace}{-, -, -, -, -, -, -, expr, prop}, \\
\>\>\atom{Fn\_isleaf}{expr, {false})}, \\
\>\>\dlrel{Fn\_split}\dlvar{(expr, prop, \underline{index}, \underline{logOp}, \underline{phyOp},}\\
\>\>\hspace{+.15in}\dlvar{\underline{lExpr}, \underline{lProp}, \underline{rExpr}, \underline{rProp})};\\
\heading{R4:}\>\head{SearchSpace}{expr, prop, -, 'scan', phyOp, -, -, -, -)} \dlog \\
\>\>\atom{SearchSpace}{-, -, -, -, -, expr, prop, -, -},  \\
\>\>\atom{Fn\_isleaf}{expr, {true}}, \atom{Fn\_phyOp}{prop, \underline{phyOp}}; \\
\heading{R5:}\>\head{SearchSpace}{expr, prop, -, 'scan', phyOp, -, -, -, -} \dlog \\
\>\>\atom{SearchSpace}{-, -, -, -, -, -, -, expr, prop},\\
\>\>\atom{Fn\_isleaf}{expr, {true}}, \atom{Fn\_phyOp}{prop, \underline{phyOp}};\\
\heading{R6:}\>\dlrel{PlanCost}\dlvar{(expr, prop, index, logOp, phyOp, -, -, -, -,} \\
\>\>\hspace{+.15in}\dlvar{md, cost)} \dlog \\
\>\>\atom{SearchSpace}{expr, prop, index, logOp, phyOp, -, -, -, -},\\
\>\>\atom{Fn\_scansummary}{expr, prop, \underline{md}}, \\
\>\>\atom{Fn\_scancost}{expr, prop, md, \underline{cost}};\\
\heading{R7:}\>\dlrel{PlanCost}\dlvar{(expr, prop, index, logOp, phyOp,} \\
\>\>\hspace{+.15in}\dlvar{lExpr, lProp, -, -, md, cost)} \dlog \\
\>\>\dlrel{SearchSpace}\dlvar{(expr, prop, index, logOp, phyOp,} \\
\>\>\hspace{+.15in}\dlvar{lExpr, lProp, -, -}, \atom{Fn\_isleaf}{lExpr, {false}},\\
\>\>\atom{PlanCost}{lExpr, lProp, -, -, -, -, -, -, -, lMd, lCost}, \\
\>\>\atom{Fn\_nonscansummary}{expr, prop, index, logOp, lMd, -, \underline{md}},  \\
\>\>\dlrel{Fn\_nonscancost}\dlvar{(expr, prop, index, logOp, phyOp,}\\
\>\>\hspace{+.15in}\dlvar{lExpr, lProp, -, -, md, \underline{localCost})}, \\
\>\>\atom{Fn\_sum}{lCost, null, localCost, \underline{cost}}; \\
\heading{R8:}\>\dlrel{PlanCost}\dlvar{(expr, prop, index, logOp, phyOp,} \\
\>\>\hspace{+.15in}\dlvar{lExpr, lProp, rExpr, rProp, md, cost)} \dlog \\
\>\>\dlrel{SearchSpace}\dlvar{(expr, prop, index, logOp, phyOp,}\\
\>\>\hspace{+.15in}\dlvar{lExpr, lProp, rExpr, rProp)}, \\
\>\>\atom{Fn\_isleaf}{lExpr, {false}), Fn\_isleaf(rExpr, {false}},\\
\>\>\atom{PlanCost}{lExpr, lProp, -, -, -, -, -, -, -, lMd, lCost}, \\
\>\>\atom{PlanCost}{rExpr, rProp, -, -, -, -, -, -, -, rMd, rCost}, \\
\>\>\atom{Fn\_nonscansummary}{expr, prop, index, logOp, lMd, rMd, \underline{md}}, \\
\>\>\dlrel{Fn\_nonscancost}\dlvar{(expr, prop, index, logOp, phyOp,} \\
\>\>\hspace{+.15in}\dlvar{lExpr, lProp, rExpr, rProp, md, \underline{localCost})}, \\
\>\>\atom{Fn\_sum}{lCost, rCost, localCost, \underline{cost}};\\
\heading{R9:}\> \head{BestCost}{expr, prop, min<cost>} \dlog \\
\>\>\dlrel{PlanCost}\dlvar{(expr, prop, index, logOp, phyOp,}\\
\>\>\hspace{+.15in}\dlvar{lExpr, lProp, rExpr, rProp, md, cost)};\\
\heading{R10:}\> \dlrel{BestPlan}\dlvar{(expr, prop, index, logOp, phyOp,}\\
\>\>\hspace{+.15in}\dlvar{lExpr, lProp, rExpr, rProp, md, cost)} \dlog \\
\>\>\atom{BestCost}{expr, prop, cost}, \\
\>\>\dlrel{PlanCost}\dlvar{(expr, prop, index, logOp, phyOp,} \\
\>\>\hspace{+.15in}\dlvar{lExpr, lProp, rExpr, rProp, md, cost)};
\end{datalog}
\end{small}
}
\vspace{-3mm}

\end{appendix}
\eat{
\begin{appendix}
\section{Declarative Optimizer}
\label{sec:append-declarative}



Here we present the datalog specification of our declarative
optimizer.  Attributes prefixed with an ampersand are free variables
in function calls (i.e., they are returned by the function after the
other --- bound --- variables are fed as inputs to the function).

\begin{Dlog}
R1: SearchSpace(expr, prop, index, logOp, phyOp, 
							lExpr, lProp, rExpr, rProp) :- 
		Expr(expr, prop), Fn_isleaf(expr, &false), 
		Fn_split(expr, prop, &index, &logOp, &phyOp,
							&lExpr, &lProp, &rExpr, &rProp)

R2: SearchSpace(expr, prop, index, logOp, phyOp,
							lExpr, lProp, rExpr, rProp) :- 
		SearchSpace(-, -, -, -, -, expr, prop, -, -), 
		Fn_isleaf(expr, &false), 
		Fn_split(expr, prop, &index, &logOp, &phyOp,
								&lExpr, &lProp, &rExpr, &rProp)

R3: SearchSpace(expr, prop, index, logOp, phyOp,
							lExpr, lProp, rExpr, rProp) :- 
		SearchSpace(-, -, -, -, -, -, -, expr, prop), 
		Fn_isleaf(expr, &false), 
		Fn_split(expr, prop, &index, &logOp, &phyOp,
								&lExpr, &lProp, &rExpr, &rProp)
								
R4: SearchSpace(expr, prop, index, logOp, phyOp,
							null, null, null, null) :-
		SearchSpace(-, -, -, -, -, expr, prop, -, -),
		Fn_isleaf(expr, &true)		
			
R5: SearchSpace(expr, prop, index, logOp, phyOp,
							null, null, null, null) :-
		SearchSpace(-, -, -, -, -, -, -, expr, prop),
		Fn_isleaf(expr, &true)		
					

R6: PlanCost(expr, prop, index, logOp, phyOp,
							null, null, null, null, metadata, cost) :- 
		SearchSpace(expr, prop, index, logOp, phyOp, 
							null, null, null, null),  			
		Fn_scansummary(expr, prop, &metadata), 
		Fn_scancost(expr, prop, metadata, &cost)


R7: PlanCost(expr, prop, index, logOp, phyOp,
							lExpr, lProp, null, null, metadata, cost) :- 
		SearchSpace(expr, prop, index, logOp, phyOp,
							lExpr, lProp, null, null), lExpr != null, lProp != null, 
		PlanCost(lExpr, lProp, -, -, -, -, -, -, -, lMetadata, lCost), 
		Fn_nonccansummary(expr, prop, index, logOp, lMetadata, 
							null, &metadata), 
		Fn_nonscancost(expr, prop, index, logOp, phyOp,
							lExpr, lProp, null, null, metadata, &localCost), 
		Fn_sum(lCost, null, localCost, &cost)


R8: PlanCost(expr, prop, index, logOp, phyOp, 
							lExpr, lProp, rExpr, rProp, metadata, cost) :- 
		SearchSpace(expr, prop, index, logOp, phyOp, 
							lExpr, lProp, rExpr, rProp), 
							lExpr != null, lProp != null, rExpr != null, rProp != null,
		PlanCost(lExpr, lProp, -, -, -, -, -, -, -, lMetadata, lCost), 
		PlanCost(rExpr, rProp, -, -, -, -, -, -, -, rMetadata, rCost), 
		Fn_nonscansummary(expr, prop, index, logOp, lMetadata, 
							rMetadata, &metadata),  
		Fn_nonscancost(expr, prop, index, logOp, phyOp, 
							lExpr, lProp, rExpr, rProp, metadata, &localCost), 
		Fn_sum(lCost, rCost, localCost, &cost)

R9: BestCost(expr, prop, min<cost>) :- 
		PlanCost(expr, prop, index, logOp, phyOp, 
							lExpr, lProp, rExpr, rProp, metadata, cost)

R10: BestPlan(expr, prop, index, logOp, phyOp,
							lExpr, lProp, rExpr, rProp, metadata, cost) :- 
		BestCost(expr, prop, cost), 
		PlanCost(expr, prop, index, logOp, phyOp,
							lExpr, lProp, rExpr, rProp, metadata, cost)


\end{Dlog}

\eat{
\subsection{Q3 Example}

\reminder{We can drop this subsection if space is an issue.}
\reminder{Put this in table form if possible.}

We revisit our earlier Q3 example, and show example output tuples,
based on the and-or-graph in
Figure~\ref{fig:example-and-or-graph-Q3S}. The following two
$SearchSpace$ tuples correspond to the level 3 "AND" nodes:

\begin{Dlog}
SearchSpace('COL', 'None', 1, 'Join', 'Sort-Merge Join', 
'C', 'Sort on C_custkey', 'OL', 'Sort on O_custkey')

SearchSpace('COL', 'None', 2, 'Join', 'Indexed Nested-Loop Join', 
'L', 'Index on L_orderkey', 'CO', null)
\end{Dlog}

The two $SearchSpace$ tuples represent the alternative ways of
splitting the expression $COL$ into $C\Join (OL)$ and $L\Join (CO)$
respectively. Considering the first $SearchSpace$ tuple, the left
expression is $C$ and its right expression is $OL$. The join
implementation is based on Sort-Merge, as a result, the left and right
physical property requires a sort order based on $C\_Custkey$ and
$O\_Custkey$ respectively. The second tuple uses an Indexed
Nested-Loop Join as its physical operator. The left expression refers
to the inner join relation indexed on $L\_Orderkey$, while there are
no ordering restrictions on the right expression.


Each $PlanCost$ tuple maintains a cost for each physical plan
corresponding to a logical ``AND'' node. Below are two example tuples
of the $PlanCost$ relation:

\begin{Dlog}
PlanCost('COL', 'None', 1, 'Join', 'Sort-Merge Join', 
'C', 'Sort on C_custkey', 'OL', 'Sort on O_custkey', 1.00)

PlanCost('COL', 'None', 2, 'Join', 'Indexed Nested-Loop Join', 
'L', 'Index on L_orderkey', 'CO', 'None', 1.01)
\end{Dlog}

Finally, we show all the $BestCost$ and $BestPlan$ tuples generated
for the example query.

\begin{Dlog}
BestCost('COL', 'None', 1.00)
BestCost('OL', 'Sort on O_custkey', 0.93)
BestCost('CO', 'None', 0.30)
BestCost('C', 'Sort on C_custkey', 0.04)
BestCost('L', 'None', 0.68)
BestCost('O', 'Sort on O_custkey', 0.19)
\end{Dlog}


\begin{Dlog}
BestPlan('COL', 'None', 1, 'Join', 'Sort-Merge Join', 
'C', 'Sort on C_custkey', 'OL', 'Sort on O_custkey', 1.00)
BestPlan('OL', 'Sort on O_custkey', 1, 'Join',
'Pipelined-Hash Join', 'O', 'Sort on O_custkey', 
'L', 'None', 0.93)
BestPlan('C', 'Sort on C_custkey', 1, 'Scan', 'Index-Scan', 
null, null, null, null, 0.04)
BestPlan('L', 'None', 1, 'Scan', 'Local-Scan', 
null, null, null, null, 0.68)
BestPlan('O', 'Sort on O_custkey', 1, 'Scan', 'Index-Scan', 
null, null, null, null, 0.19)
\end{Dlog}
}


\eat{
\begin{table}
\begin{tabular}{|p{8cm}|}
\hline
Relations (* represents the key attributes)\\
expr : expression signature \\
prop : physical property (e.g., interesting order) \\
index : the $i$th logical way of splitting an expression \\
logOp : the logical operator \\
phyOp : the physical operator \\
lExpr : left expression signature \\
lProp : left physical property \\
rExpr : right expression signature) \\
rProp : right physical property \\
metadata : summaries (e.g., histograms) \\
cost : plan cost \\
\hline
\hline
Expr(*expr, *prop) \\
\hline
SearchSpace(*expr, *prop, *index, logOp, *phyOp, lExpr, lProp, rExpr, rProp) \\
\hline
PlanCost(*expr, *prop, *index, logOp, *phyOp, lExpr, lProp, rExpr, rProp, metadata, cost) \\
\hline
BestCost(*expr, *prop, minCost) \\
\hline
BestPlan(*expr, *prop, *index, logOp, *phyOp, lExpr, lProp, rExpr, rProp, metadata, cost) \\
\hline
\end{tabular}
\caption{\label{table:schemas} Relation schemas used in our declarative program. * represents the key attributes of a
relation.}
\end{table}

Table~\ref{table:schemas} shows a list of relations in our declarative
optimizer. We focus primarily on $Expression$, and $SearchSpace$, and
defer discussing other relations which are used in subsequent
components.
}

\eat{
While we could have used two relations
separately for logical and physical search spaces, merging into one
simplifies our datalog program\footnote{In the implementation given
  the one-to-many mappings between a logical plan and its
  corresponding physical plan, we carefully propagated the tuples
  between operators in our execution to ensure that duplicate logical
  plans are not propagated multiple times}.
}

\eat{Rules R1-R5 are used to generate both the logical and physical search
spaces. Our rules merges logical plan enumeration and physical plan
enumeration through a single $F\_Split$ function. This splitting
process can be down recursively hence can be executed with a Fixpoint
operator - it splits any expression until this expression is a
leaf-level scan expression.}


\eat{
\subsection{Cost Enumeration}
\label{sec:append-costenumeration}

Rule R6 computes the $PlanCost$ for those ``AND'' nodes with zero
input, for example, the node associated with the Scan operator; Rule 7
computes the $PlanCost$ for those $AND$ nodes with one input, for
example, the logical operators of Functions and Aggregates (this case
does not exist in this example); Rule 8 computes the $PlanCost$ for
those ``AND'' nodes with two inputs, for example, Level 3 ``AND'' node
$(C, OL)$. The cost is computed via two functions in each case. In the
zero-input case, since the cost of a scan operation is the leaf level
cost, it is computed via function $F\_Scan\_Summary$ to get the base
scan histogram summaries, and $F\_Scan\_Cost$ is to compute an
estimated scan cost based on the scan summaries. In the case of
one-input and two-input ``AND'' nodes, since $PlanCost$ is a cost
associated with an intermediate ``AND'' node in the graph, it should
sum up its local cost, its left subtree cost and its right subtree
cost. In particular, in the case of one-input node, the left cost is
set by default as the input cost, hence the cost of the parent a
summation of its local cost and its left cost, with no right cost. In
the case of two-input node, the plan cost is a summation of its local
cost, its left cost and its right cost. The local cost is computed two
functions: $F\_Non\_Scan\_Summary$ computes a histogram summary of an
intermediate ``AND'' node based on its inputs, e.g., a histogram of
the output of an join operator is an intersection of the histograms of
its two inputs; then based on the non-scan histogram summaries,
$F\_Non\_Scan\_Cost$ computes an estimated cost. The main reason to
differentiate scan cases and non-scan cases is because their different
inputs to the functions and different mechanisms to compute histogram
summaries and cost estimations.

}


\section{TPC-H Query Workload}
\label{sec:appendix-queries}

See Table~\ref{tab:tpch} for the SQL queries used from the TPC-H benchmark.

\eat{
\section{Four-levels of query optimization}

The logical plan search space contains all the logical plans that
correspond to subexpressions of the original query expression up to
any logically equivalent transformations such as commutativity and
associativity of join operations. In traditional query optimizers, a
data structure called and-or-graph is maintained to enumerate the
logical plan search space. As an example, the and-or-graph of a
four-way join query is shown in
Figure~\ref{fig:example-and-or-graph}. For brevity, in the figure, we
use $AB$ to denote $A\Join B$, $(expr)$ to denote an ``OR'' node
corresponding to the expression $expr$, and $(expr, idx)$ to denote an
``AND'' node corresponding to the pair of expression $expr$ and child
index number $idx$. For example, $(ABC)$ is an ``OR'' node of
expression $ABC$ while $(ABC, 1)$ is an ``AND'' node which is the
first child of its parent ``OR'' node $ABC$. The edges in the graph
can be classified into two categories: the edges that go from an
``OR'' node to its ``AND'' children represent different ways of
splitting that ``OR'' expression, hence they are mutually ``OR''ed;
the edges that go from an ``AND'' node to its ``OR'' children
represent the expression-subexpression relationship, and these edges
are mutually ``AND''ed. Generally, an ``AND'' node has at most two
children, but an ``OR'' node may have up to $2^{n-1}-1$ children where
$n$ is the maximum number of relations to be joined at the node. In
contrast, here we use the relation $LogicalPlanSearchSpace$ to
represent the entire logical plan search space. In addition, the
tuples in this relation are generated via multiple recursive datalog
rules rather than a procedural function.

\paragraph{Level 2: Enumerate the Physical Plan Search Space}
A physical plan search space extends a logical plan search space in that it enumerates all the physical operator trees which implement a certain algebraic expression. For example, Join is a logical operator, but it may have multiple physical implementations such as Hash-Join, Index-Join, and Merge-Join. A physical plan should not only identify its physical operator, but also the physical properties over inputs, such as sorted input for a merge-join operator; sometimes a direction is also needed to distinguish the left child and right child of a join if the physical implementation is not a symmetric join operator, such as Index-Join. In most cases, the physical plan search space is a linear extension to the logical plan search space. 

\paragraph{Level 3: Estimate the Cost}
A cost estimator assigns an estimated cost to any partial or complete plan in the physical plan search space. The first step is to maintain a set of statistics on the input relations and indexes, e.g., number of tuples in a relation, number of tuples in an index, number of distinct values in a column, etc. We use $RelMetadata$ to store the statistics maintained on the input relations, and $LogMetadata$ to store the statistics maintained on the intermediate subexpressions. Note that $LogMetadata$ is a logical property, because logically equivalent expressions should have the same output cardinality, and $LogMetadata$ depends on $LogicalPlanSearchSpace$ rather than $PhysicalPlanSearchSpace$.  

Based on all the statistics in $RelMetadata$ and $LogMetadata$, the cost of a plan can be computed. We compute the combined factors such as CPU, I/O, bandwidth and energy into a single cost metric, and store the cost of each physical operator into the relation $LocalCost$. $LocalCost$ is a physical plan property because each $LocalCost$ tuple corresponds to a particular physical plan. The triple of $logId$, $dir$, $phyOp$ uniquely identifies a physical plan search space tuple, and the purpose of $LocalCost$ is to compute a {\em cost} to this physical plan. 

Given $LocalCost$, the next step is to compute an aggregate sum of the cost the root operator. We still the cost into $PlanCost$. $PlanCost$ has the same schema as $LocalCost$, the only difference is the definition of the cost: $LocalCost$ computes of the cost of the operator, whereas $PlanCost$ computes the cost of the sum of its subtree. Indeed, the cost in $PlanCost$ of every operator is the sum of the $PlanCost$ of its left and right children, and its local operator cost in $LocalCost$.  

\paragraph{Level 4: Find the Optimal Plan}
The final stage of our query optimizer is to compute the optimal plan based on the estimated costs available in $PlanCost$. The lowest cost value of each plan is maintained in $BestCost$. However, sometimes $BestCost$ is not enough, because we still want to compute the best plan itself, namely, the physical operator tree. Hence, we use $BestPlan$ to store the tree structure of the optimal plan.

\paragraph{Putting all together}
Figure~\ref{fig:plan_dependency} shows the relation dependency graph
that combines these four levels. Due to the limit of pages, we omit
the rules here. From the figure, we can see that our declarative
program is in nature recursive; and there are multiple fixpoint
operators. One can imagine that general techniques that can improve
this high-level logic program may benefit the performance of
declarative query optimizer to a large extent. We are currently
exploring optimization techniques and priority schemes in this space.

\section{Order of plan enumerations}

Below listed three types query optimizers: Volcano-style, System R-style and logic-based semi-naive evaluation. Each is compared with their order of enumerating the plan search space, the order of computing the cost of plans, and the order of branch-and-bounding. 

\paragraph{Volcano-style optimizer:}
\begin{itemize}
\item{The order of enumerating the plan search space (depth-first pre-order with memoization):}
\begin{verbatim}
(ABCD,1), A, (BCD,1), B, (CD,1), C, D, (BCD,2), 
(BD,1), (BCD,3), (BC,1), (ABCD,2), (ACD,1), 
(ACD,2), (AD,1), (ACD,3), (AC,1), (ABCD,3), 
(ABD,1), (ABD,2), (ABD,3), (AB,1), (ABCD,4), 
(ABC,1), (ABC,2), (ABC,3), (ABCD,5), (ABCD,6), 
(ABCD,7)
\end{verbatim}

\item{The order of computing the cost of plans (depth-first post-order with memoization):}
\begin{verbatim}
A, B, C, D, (CD,1), (BCD,1), (ABCD,1), (BD,1), 
(BCD,2), (BC,1), (BCD,3), (ACD,1), (AD,1), 
(ACD,2), (AC,1), (ACD,3), (ABCD,2), (ABD,1), 
(ABD,2), (AB,1), (ABD,3), (ABCD,3), (ABC,1), 
(ABC,2), (ABC,3), (ABCD,4), (ABCD,5), (ABCD,6), 
(ABCD,7)
\end{verbatim}

\item{The order of branch-and-bounding:}
\begin{verbatim}
A, B, C, D, (CD,1), (BCD,1), (ABCD,1), (BD,1), 
(BCD,2), (BC,1), (BCD,3), (ACD,1), (AD,1), 
(ACD,2), (AC,1), (ACD,3), (ABCD,2), (ABD,1), 
(ABD,2), (AB,1), (ABD,3), (ABCD,3), (ABC,1), 
(ABC,2), (ABC,3), (ABCD,4), (ABCD,5), (ABCD,6), 
(ABCD,7)
\end{verbatim}

If any expression is pruned, its ancestors should be pruned. For example, if $BD$ is pruned, then its ancestors that have not been visited $(BCD,2),(ABD,1),(ABCD,6),(ABCD,3)$ should be pruned. 

\item{Pruning Conditions:}
\begin{verbatim}
A, B, C, D, CD, BCD, ABCD, BD(<BCD-C), BCD
(<BCD,<ABCD-A), BC(<BCD-D), BCD(<BCD,
<ABCD-A), ACD(<ABCD-B), AD(<ACD-C,<ABCD-BC), 
ACD(<ACD,<ABCD-B), AC(<ACD-D,<ABCD-BD), 
ACD(<ACD,<ABCD-B), ABCD(<ABCD), ABD(<ABCD-C), 
ABD(<ABD, <ABCD-C), AB(<ABD-D,<ABCD-CD), 
ABD(<ABD,<ABCD-C), ABCD(<ABCD), ABC(<ABCD-D), 
ABC(<ABC,<ABCD-D), ABC(<ABC,<ABCD-D), 
ABCD(<ABCD), ABCD(<ABCD), ABCD(<ABCD), 
ABCD(<ABCD)
\end{verbatim}
\end{itemize}

\paragraph{System R-style optimizer:}

\begin{itemize}
\item{The order of enumerating the plan search space (bottom-up iterative deepening):}
\begin{verbatim}
A, B, C, D, (AB,1), (AC,1), (AD,1), (BC,1), 
(BD,1), (CD,1), (ABC,1), (ABC,2), (ABC,3), 
(ABD,1), (ABD,2), (ABD,3), (ACD,1), (ACD,2), 
(ACD,3, (BCD,1), (BCD,2), (BCD,3), (ABCD,1), 
(ABCD,2), (ABCD,3), (ABCD,4), (ABCD,5), 
(ABCD,6), (ABCD,7)
\end{verbatim}

\item{The order of enumerating the cost of plans (bottom-up iterative deepening):}
\begin{verbatim}
A, B, C, D, (AB,1), (AC,1), (AD,1), (BC,1), 
(BD,1), (CD,1), (ABC,1), (ABC,2), (ABC,3), 
(ABD,1), (ABD,2), (ABD,3), (ACD,1), (ACD,2), 
(ACD,3), (BCD,1), (BCD,2), (BCD,3), (ABCD,1), 
(ABCD,2), (ABCD,3), (ABCD,4), (ABCD,5), 
(ABCD,6), (ABCD,7)
\end{verbatim}

The order of branch-and-bounding:
N/A. Because it is totally bottom-up. 
\end{itemize}

\paragraph{Semi-naive Fixpoint implementation:}
\begin{itemize}
\item{The order of enumerating the plan search space (top-down iterative deepening):}
\begin{verbatim}
(ABCD,1), (ABCD,2), (ABCD,3), (ABCD,4), 
(ABCD,5), (ABCD,6), (ABCD,7), A, (BCD,1), 
(BCD,2), (BCD,3), B, (ACD,1), (ACD,2), 
(ACD,3), C, (ABD,1), (ABD,2), (ABD,3), D, 
(ABC,1), (ABC,2), (ABC,3), (AB,1), (CD,1), 
(AC,1), (BD,1), (AD,1), (BC,1)
\end{verbatim}

\item{The order of enumerating the cost of plans: (bottom-up iterative deepening):}
\begin{verbatim}
A, B, C, D, (AB,1), (CD,1), (AC,1), (BD,1), 
(AD,1), (BC,1), (BCD,1), (BCD,2), (BCD,3), 
(ACD,1), (ACD,2), (ACD,3), (ABD,1), (ABD,2), 
(ABD,3), (ABC,1), (ABC,2), (ABC,3), (ABCD,1), 
(ABCD,2), (ABCD,3), (ABCD,4), (ABCD,5), 
(ABCD,6), (ABCD,7)
\end{verbatim}

\item{The order of branch-and-bounding:}
\begin{verbatim}
A, B, C, D, (AB,1), (CD,1), (ABCD,5), (AC,1), 
(BD,1), (ABCD,6), (AD,1), (BC,1), (ABCD,7), 
(BCD,1), (BCD,2), (BCD,3), (ABCD,1), (ACD,1), 
(ACD,2), (ACD,3), (ABCD,2), (ABD,1), (ABD,2), 
(ABD,3), (ABCD,3), (ABC,1), (ABC,2), (ABC,3), 
(ABCD,4)
\end{verbatim}

\item{Pruning Conditions:}
\begin{verbatim}
A, B, C, D, AB, CD, ABCD, AC, BD(<ABCD-AC), 
ABCD(<ABCD), AD, BC(<ABCD-AD), ABCD(<ABCD), 
BCD(<ABCD-A), BCD(<ABCD,<ABCD-A), BCD(<ABCD,
<ABCD-A), ABCD(<ABCD), ACD(<ABCD-B), ACD
(<ABCD-B,<ABCD), ACD(<ABCD-B,<ABCD), 
ABCD(<ABCD), ABD, ABCD(<ABCD), ABC(<ABCD), 
ABCD(<ABCD)
\end{verbatim}
\end{itemize}
}





\begin{table}[t]
\footnotesize
\begin{tabular}{|l|p{7cm}|}
\hline
{\bf Q1} & SELECT l\_returnflag, l\_linestatus, sum(l\_quantity) as sum\_qty, sum(l\_extendedprice) as sum\_base\_price, sum(l\_extendedprice*(1-l\_discount)) as sum\_disc\_price, sum(l\_extendedprice*(1-l\_discount)*(1+l\_tax)) as sum\_charge, avg(l\_quantity) as avg\_qty, avg(l\_extendedprice) as avg\_price, avg(l\_discount) as avg\_disc, count(*) as count\_order FROM lineitem WHERE l\_shipdate $\leq$ '1998-09-01' GROUPBY l\_returnflag, l\_linestatus; \\
\hline
{\bf Q3} & SELECT l\_orderkey, sum(l\_extendedprice*(1-l\_discount)) as revenue, o\_orderdate, o\_shippriority FROM customer, orders, lineitem WHERE c\_mktsegment = 'MACHINERY' and c\_custkey = o\_custkey and l\_orderkey = o\_orderkey and o\_orderdate $<$ '1995-03-15' and l\_shipdate $>$ '1995-03-15' GROUPBY l\_orderkey, o\_orderdate, o\_shippriority; \\
\hline
{\bf Q5} & SELECT n\_name, sum(l\_extendedprice * (1 - l\_discount)) as revenue FROM customer, orders, lineitem, supplier, nation, region WHERE c\_custkey = o\_custkey and l\_orderkey = o\_orderkey and l\_suppkey = s\_suppkey and c\_nationkey = s\_nationkey and s\_nationkey = n\_nationkey and n\_regionkey = r\_regionkey and r\_name = 'AMERICA' " and o\_orderdate $\geq$ CAST( '1993-01-01' and o\_orderdate $<$ '1994-01-01' GROUPBY n\_name;\\
\hline
{\bf Q6} & SELECT sum(l\_extendedprice*l\_discount) as revenue FROM lineitem WHERE l\_shipdate $\geq$ '1994-01-01' and l\_shipdate $<$ '1995-01-01' and l\_discount $\geq$ 0.06 - 0.01 and l\_discount $\leq$ 0.06 + 0.01 and l\_quantity $<$ 24.2; \\
\hline
{\bf Q10} & SELECT c\_custkey, c\_name, sum(l\_extendedprice * (1 - l\_discount)) as revenue, c\_acctbal, n\_name, c\_address, c\_phone, c\_comment FROM customer, orders, lineitem, nation WHERE c\_custkey = o\_custkey and l\_orderkey = o\_orderkey and o\_orderdate $\geq$ '1993-06-01' and o\_orderdate $<$ '1993-09-01' and l\_returnflag = 'R' and c\_nationkey = n\_nationkey GROUPBY c\_custkey, c\_name, c\_acctbal, c\_phone, n\_name, c\_address, c\_comment; \\
\hline
\end{tabular}
\caption{\label{tab:tpch} TPC-H queries, reproduced}
\end{table}

\eat{
\section{Proofs}

\begin{proof}[of Proposition~\ref{prop:viability}]
Prove by contradiction. If a plan tree $T^s$ of the subexpression $E^s$ is suboptimal, then there must exist an optimal plan tree other than $P(E^S)$, suppose it is $P_{OPT}{(E^S)}$, we have $PlanCost(P(E^S))$ $>$ $PlanCost(P_{OPT}{(E^S)})$. Since $P(E^S)$ is a subtree of $P_{OPT}{(E)}$, if we substitute subtree $P(E)$ with subtree $P_{OPT}{(E)}$, then we get a new plan of the original expression $E$, which has a smaller cost than $P_{OPT}{(E)}$. However, by definition, $P_{OPT}{(E)}$ has the smallest cost of every plan to expression $E$. Contradiction.
\end{proof}

\begin{proof}[of Proposition~\ref{prop:refcount}]
Prove by contradiction. If a plan $P(E^S)$ is the sub-plan of $P_{OPT}{(E)}$, then it must have at least parent in the plan tree of $P_{OPT}{(E)}$. This contradicts the definition of reference counts. 
\end{proof}

\begin{proof}[of Proposition~\ref{prop:bound}]
Prove by contradiction. Assume the plan $P(E^S)$ which has $PlanCost$ $(P(E^S))$ $>$ $Bound$ $(E^S)$ is the sub-tree of the optimal plan tree $P_{OPT}{(E)}$. Suppose the parent of $E^S$ on the optimal plan tree is $PE^S$, the sibling of $E^S$ is $SE^S$. According to the definition of $Bound$,  we have $Bound$ $(E^S)$ $\geq$ $Bound$ $(PE^S)$ $-$ $LocalCost$ $(PE^S)$ $-$ $BestCost$ $(SE^S)$ or $Bound$ $(E^S)$ $=$ $BestCost$ $(E^S)$. Hence either condition 1, $PlanCost$ $(P(E^S))$ $>$ $Bound$ $(PE^S)$ $-$ $LocalCost$ $(PE^S)$ $-$ $BestCost$ $(SE^S)$ or condition 2, $PlanCost$ $(P(E^S))$ $>$ $BestCost$ $(E^S)$ applies. Since $P(E^S)$ is the sub-tree of the optimal plan tree $P_{OPT}{(E)}$, $P(E^S)$ must have the optimal cost for expression $E^S$, that is, $PlanCost$ $(P(E^S))$ $=$ $BestCost$ $(E^S)$. Therefore, condition 1 is valid but condition 2 is invalid. On the other hand, according to the discussion in Section~\ref{subsec:branch-and-bounding}, the optimal substructure ensures that $BestCost$ $(PE^S)$ $=$ $BestCost$ $(E^S)$ $+$ $BestCost$ $(SE^S)$ + $LocalCost$ $(PE^S)$. Hence we get $BestCost$ $(PE^S)$ $>$ $Bound$ $(PE^S)$. 

If we deduct this process continuously until we reach the root in the optimal plan tree, we get $BestCost(E)$ $>$ $Bound(E)$. Since we set $Bound(E)$ $=$ $BestCost(E)$ to start the branch-and-bounding process. Contradiction. 
\end{proof}
}

\section{Additional Experimental Results}

For greater completeness, we show representative results for a variety
of additional queries in
Figure~\ref{fig:stategies-reoptimization-Q5-Supplier} (for incremental
re-optimization with changes to a different table); Figure
~\ref{fig:incremental-Q10-scan-cost} (for incremental re-optimization
of a different query, Q10);
Figure~\ref{fig:incremental-Q8Join-scan-cost} (for incremental
re-optimization of query Q8Join); and finally
Figure~\ref{fig:incremental-Q8Join-real-workload} (for incremental
re-optimization of query Q8Join using observed conditions from real
query execution).

\begin{figure*}[t]
\begin{minipage}[t]{2.25in}
\begin{center}
\includegraphics[width=2.25in]{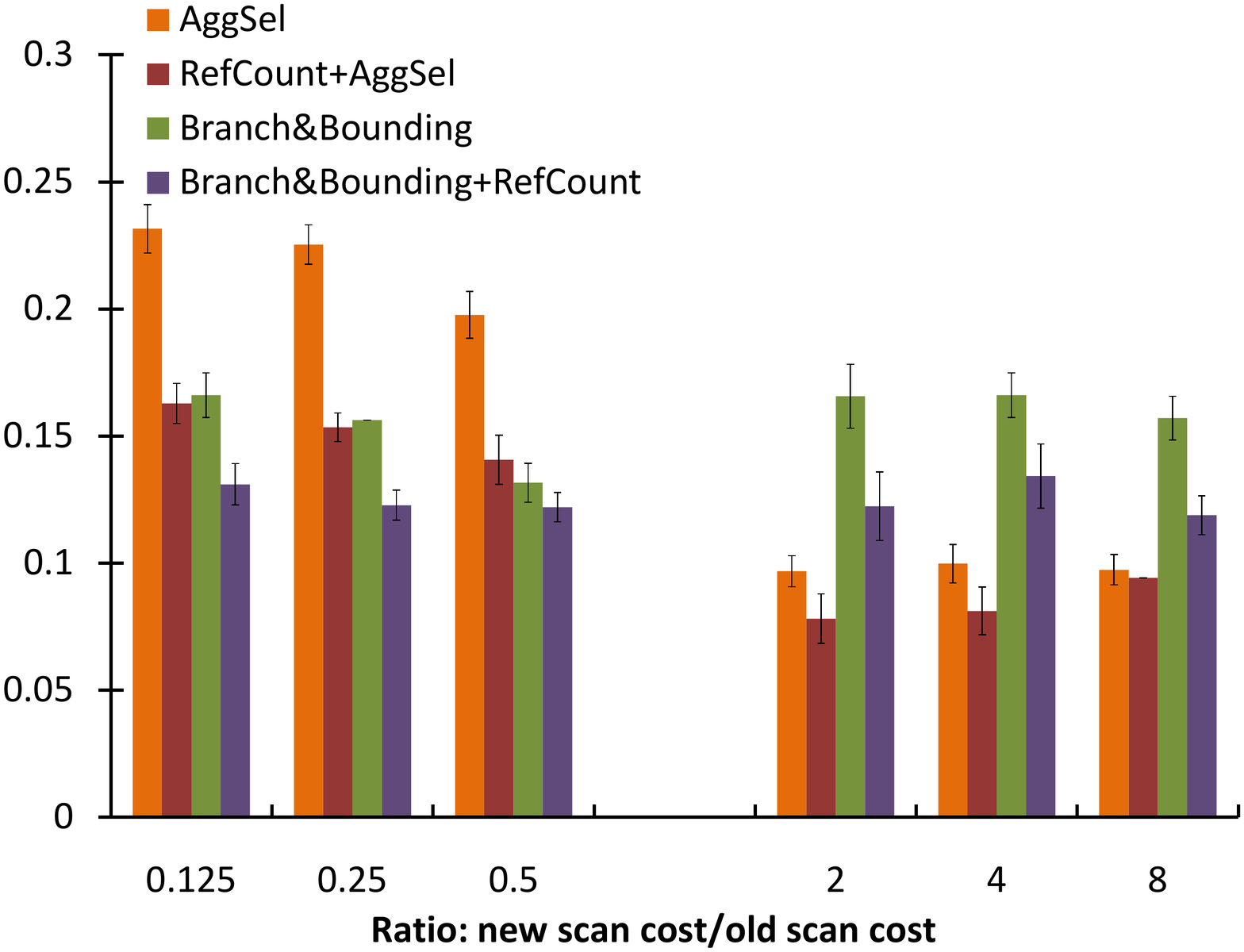}
\vspace{-3mm}
\small (a) Execution time (normalized to Volcano)
\end{center}
\end{minipage}
\hfill
\begin{minipage}[t]{2.25in}
\begin{center}
\includegraphics[width=2.25in]{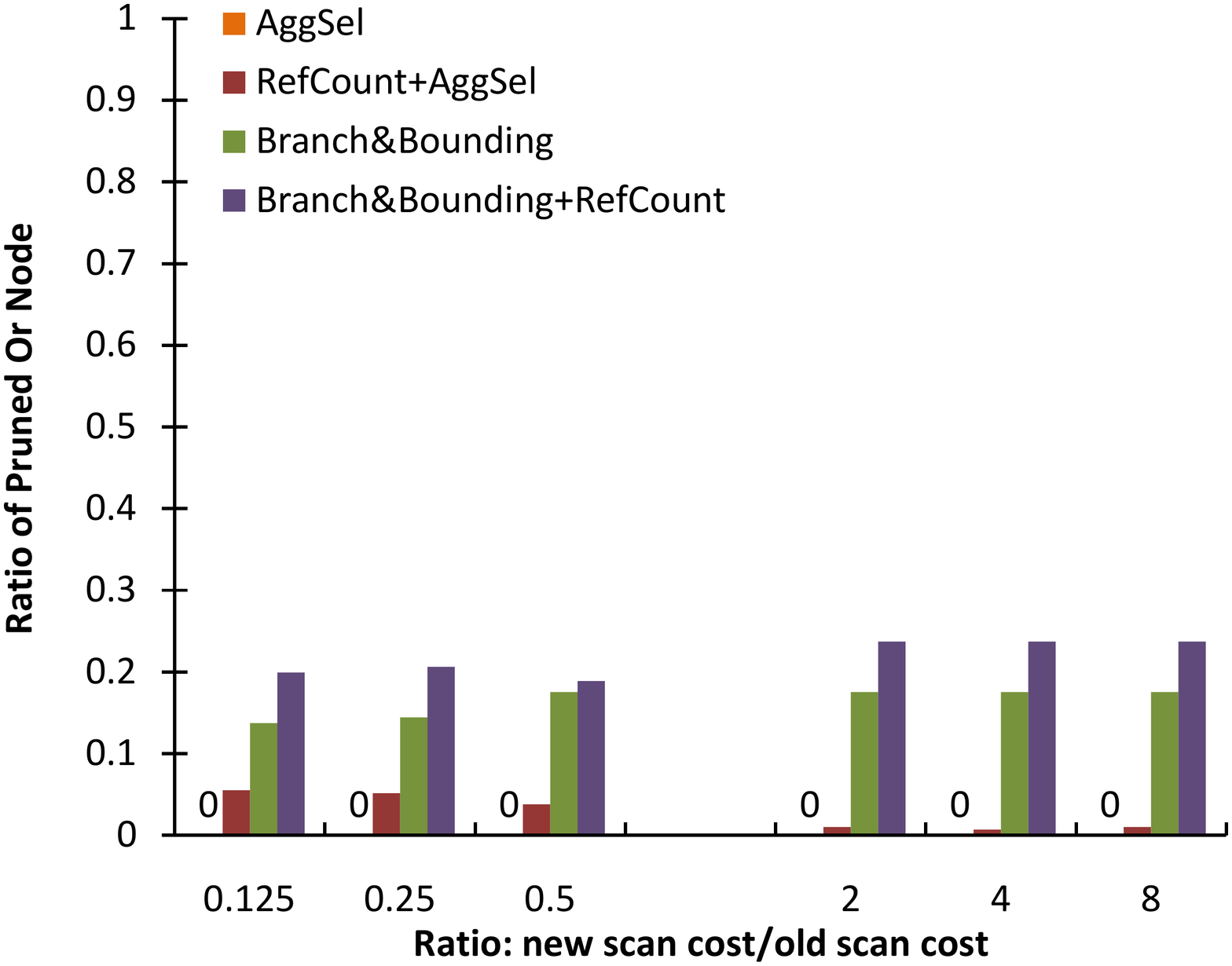}
\vspace{-3mm}
\small (b) Pruning ratio: plan table entries
\end{center}
\end{minipage}
\hfill
\begin{minipage}[t]{2.25in}
\begin{center}
\includegraphics[width=2.25in]{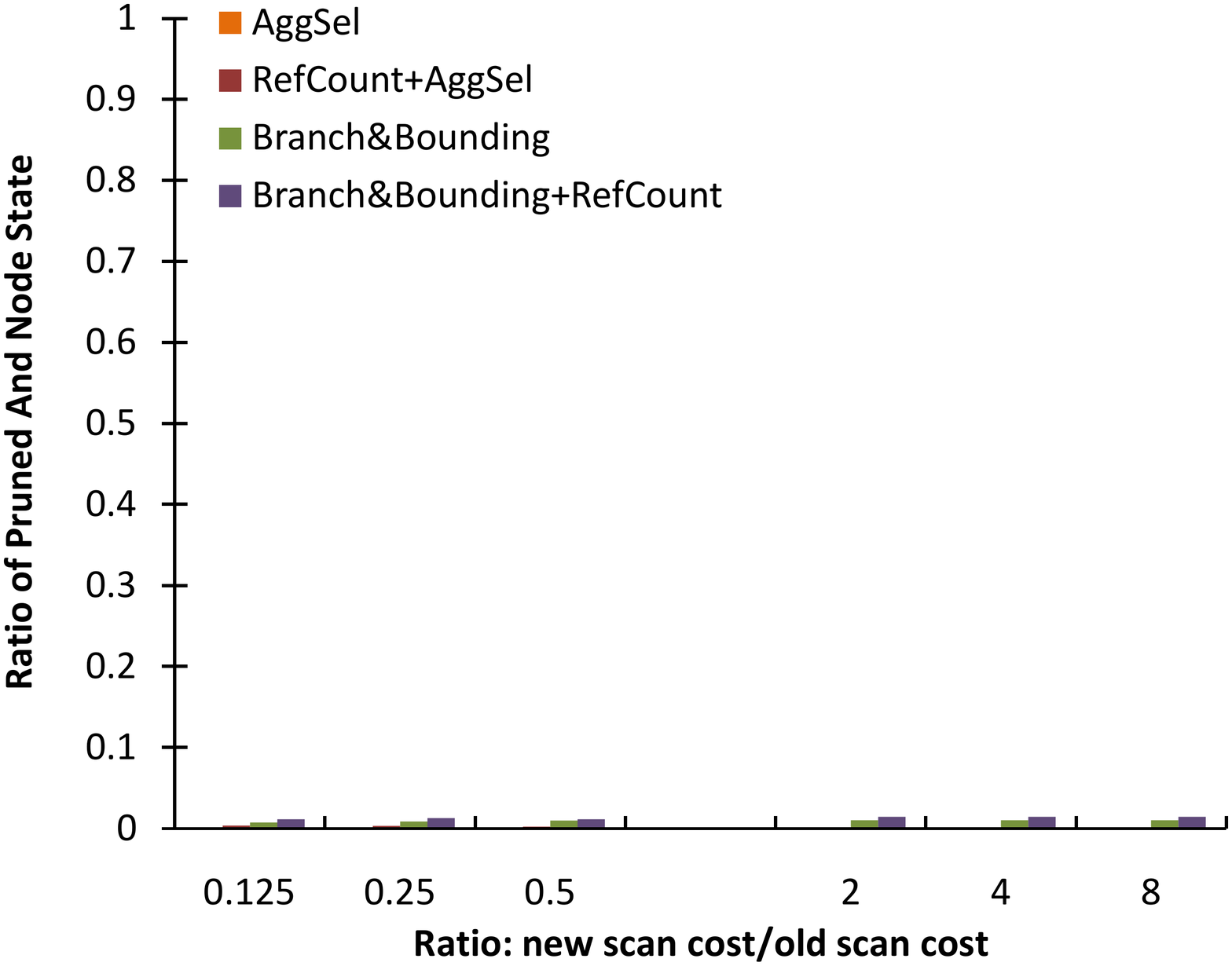}
\vspace{-3mm}
\small (c) Pruning ratio: plan alternatives
\end{center}
\end{minipage}
\caption{\small Incremental Re-optimization of TPCH Q5 when $Supplier$ has updated scan cost
\label{fig:stategies-reoptimization-Q5-Supplier}}
\end{figure*}

\begin{figure*}[t]
\begin{minipage}[t]{2.25in}
\begin{center}
\includegraphics[width=2.25in]{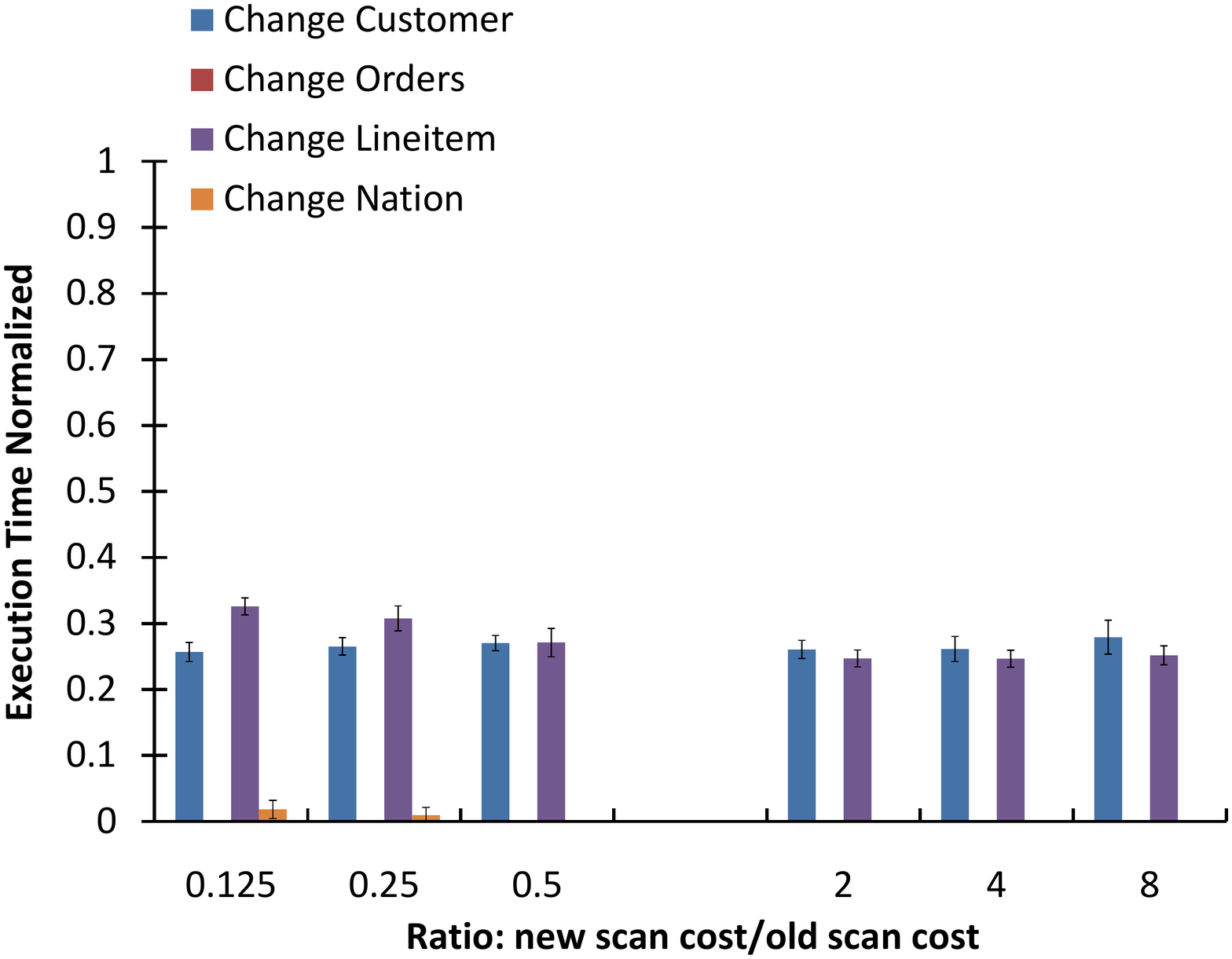}
\vspace{-3mm}
\small (a) Execution time (normalized to Volcano)
\end{center}
\end{minipage}
\hfill
\begin{minipage}[t]{2.25in}
\begin{center}
\includegraphics[width=2.25in]{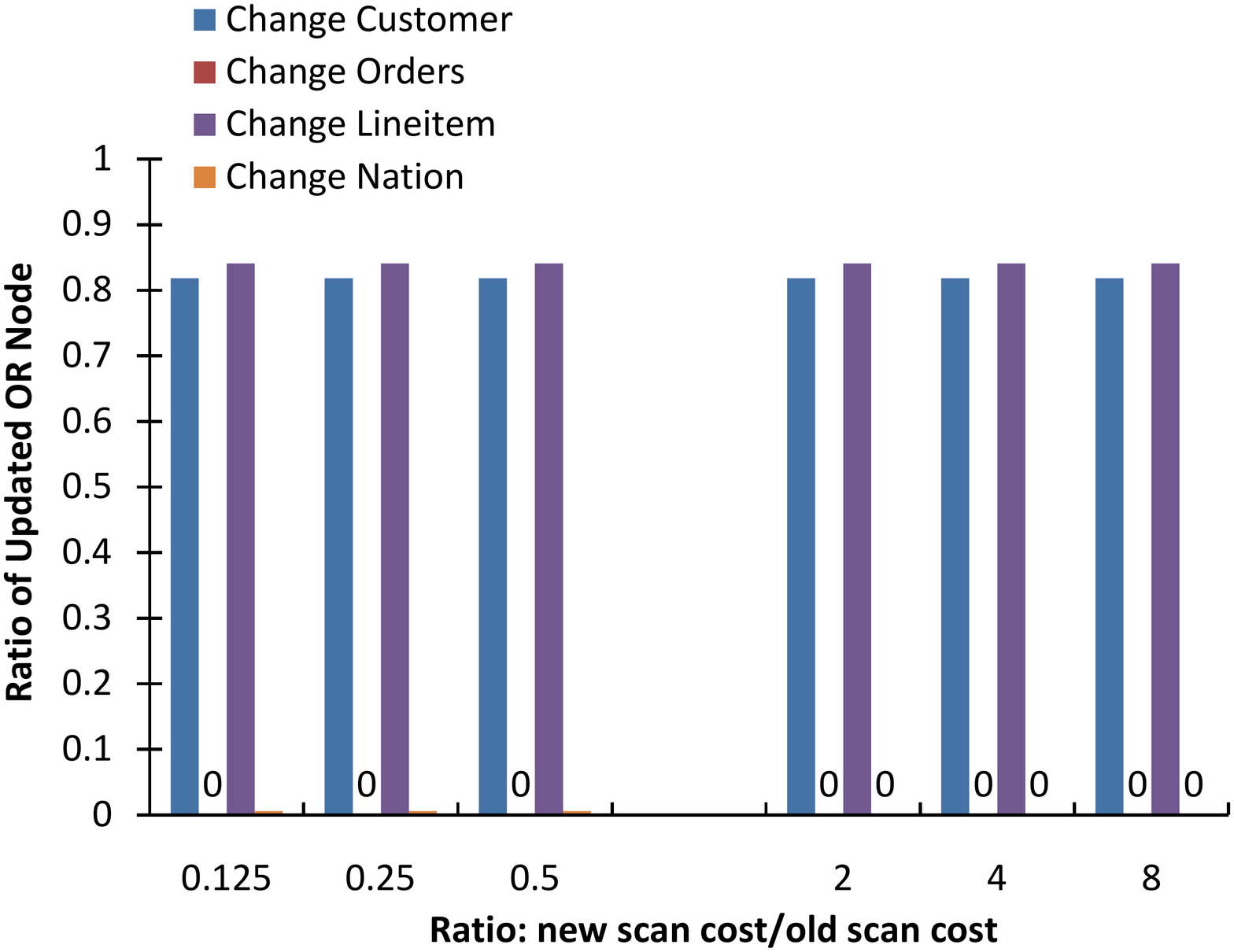}
\vspace{-3mm}
\small (b) Update ratio: plan table entries
\end{center}
\end{minipage}
\hfill
\begin{minipage}[t]{2.25in}
\begin{center}
\includegraphics[width=2.25in]{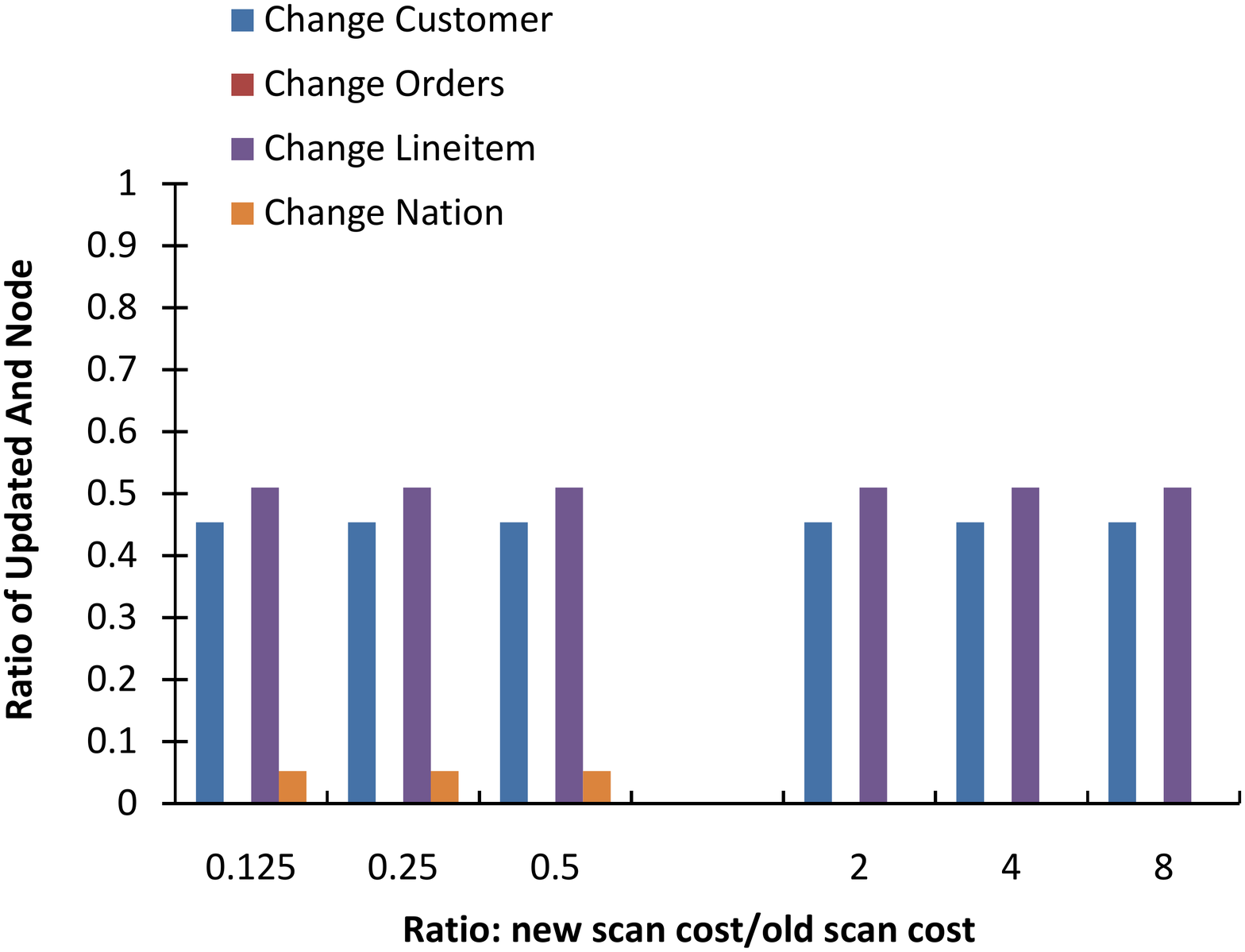}
\vspace{-3mm}
\small (c) Update ratio: plan alternatives
\end{center}
\end{minipage}
\caption{\small Declarative incremental re-optimization for TPCH Q10 when a base relation scan cost is changed \label{fig:incremental-Q10-scan-cost}}
\end{figure*}

\begin{figure*}[t]
\begin{minipage}[t]{2.25in}
\begin{center}
\includegraphics[width=2.25in]{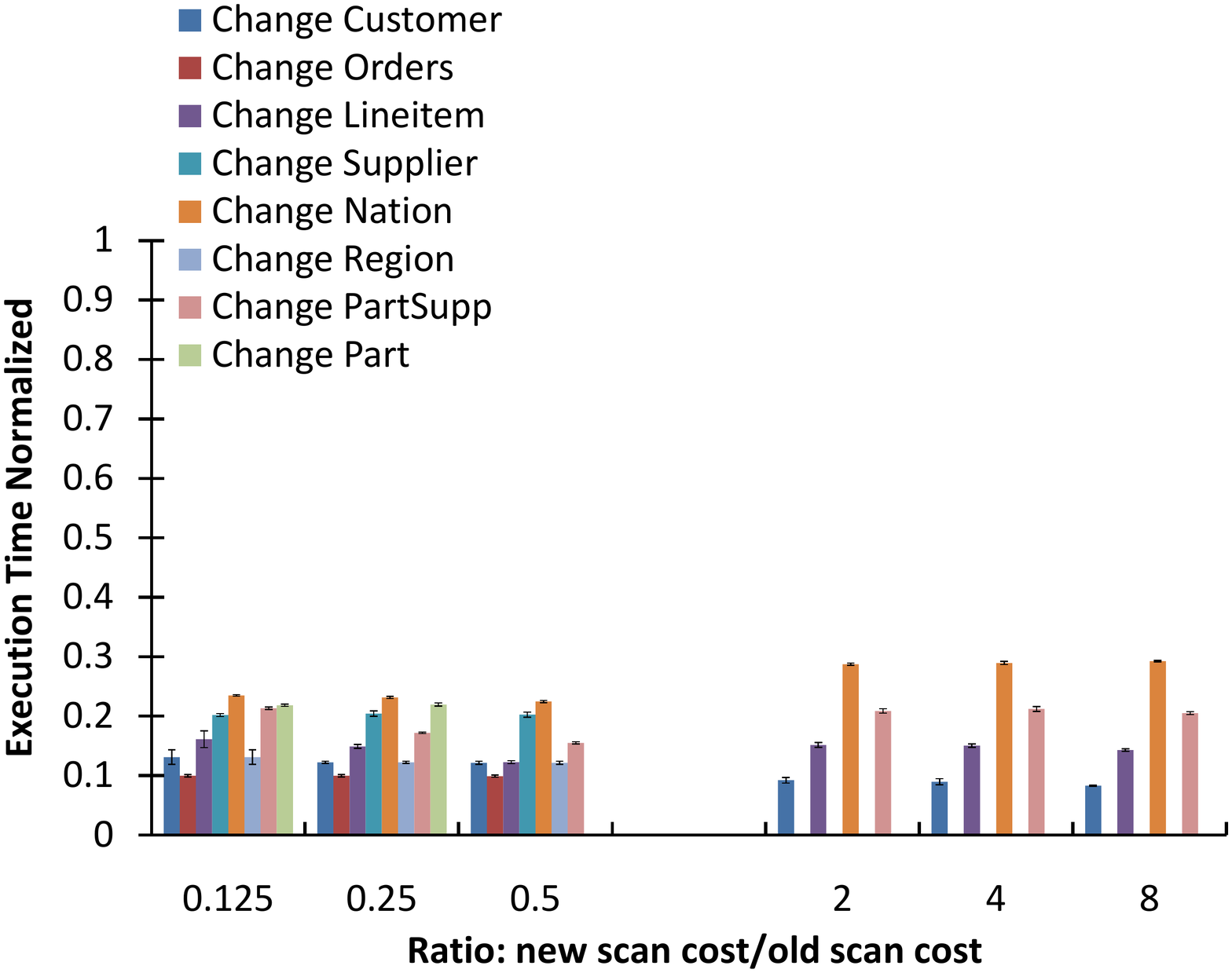}
\vspace{-3mm}
\small (a) Execution time (normalized to Volcano)
\end{center}
\end{minipage}
\hfill
\begin{minipage}[t]{2.25in}
\begin{center}
\includegraphics[width=2.25in]{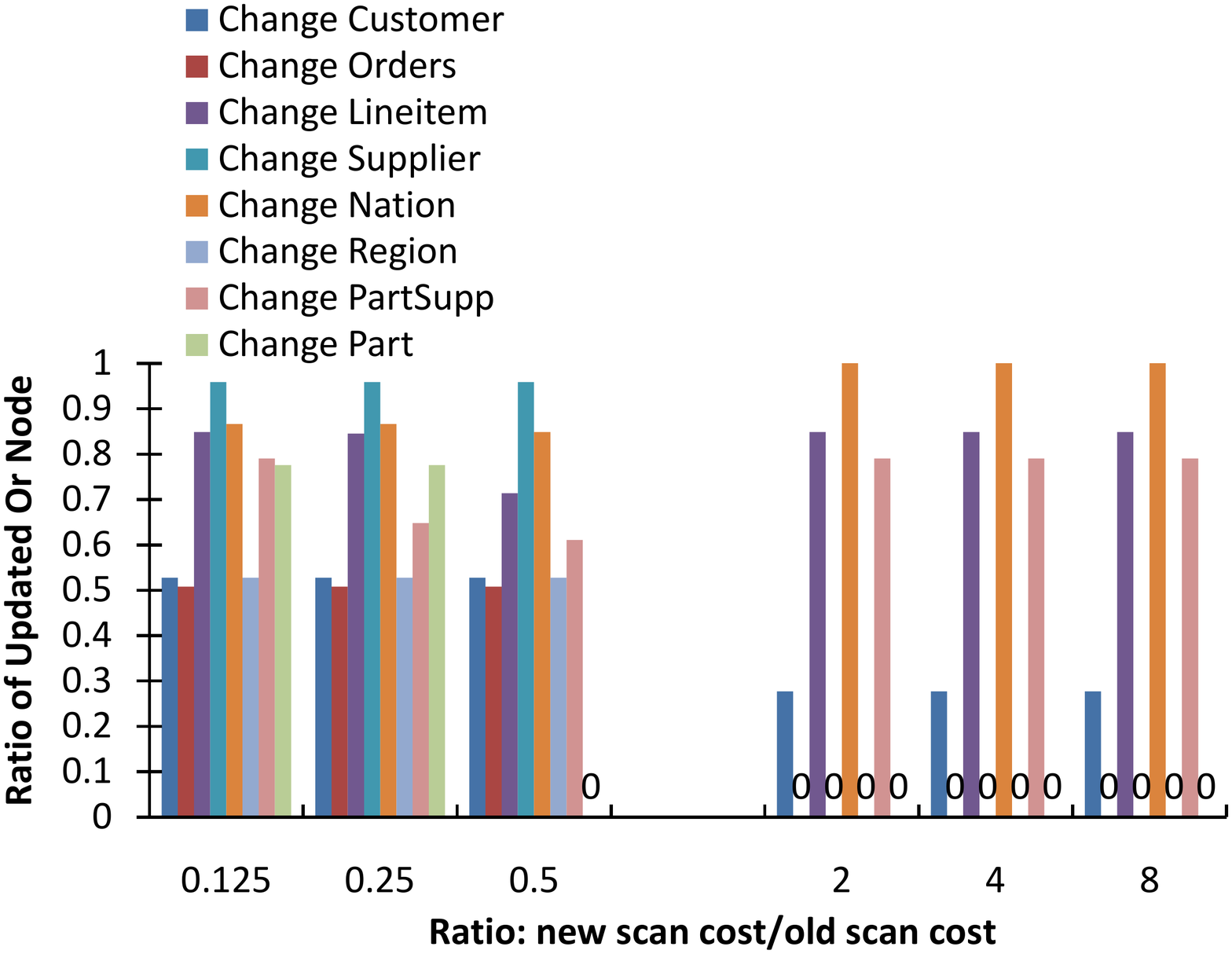}
\vspace{-3mm}
\small (b) Update ratio: plan table entries
\end{center}
\end{minipage}
\hfill
\begin{minipage}[t]{2.25in}
\begin{center}
\includegraphics[width=2.25in]{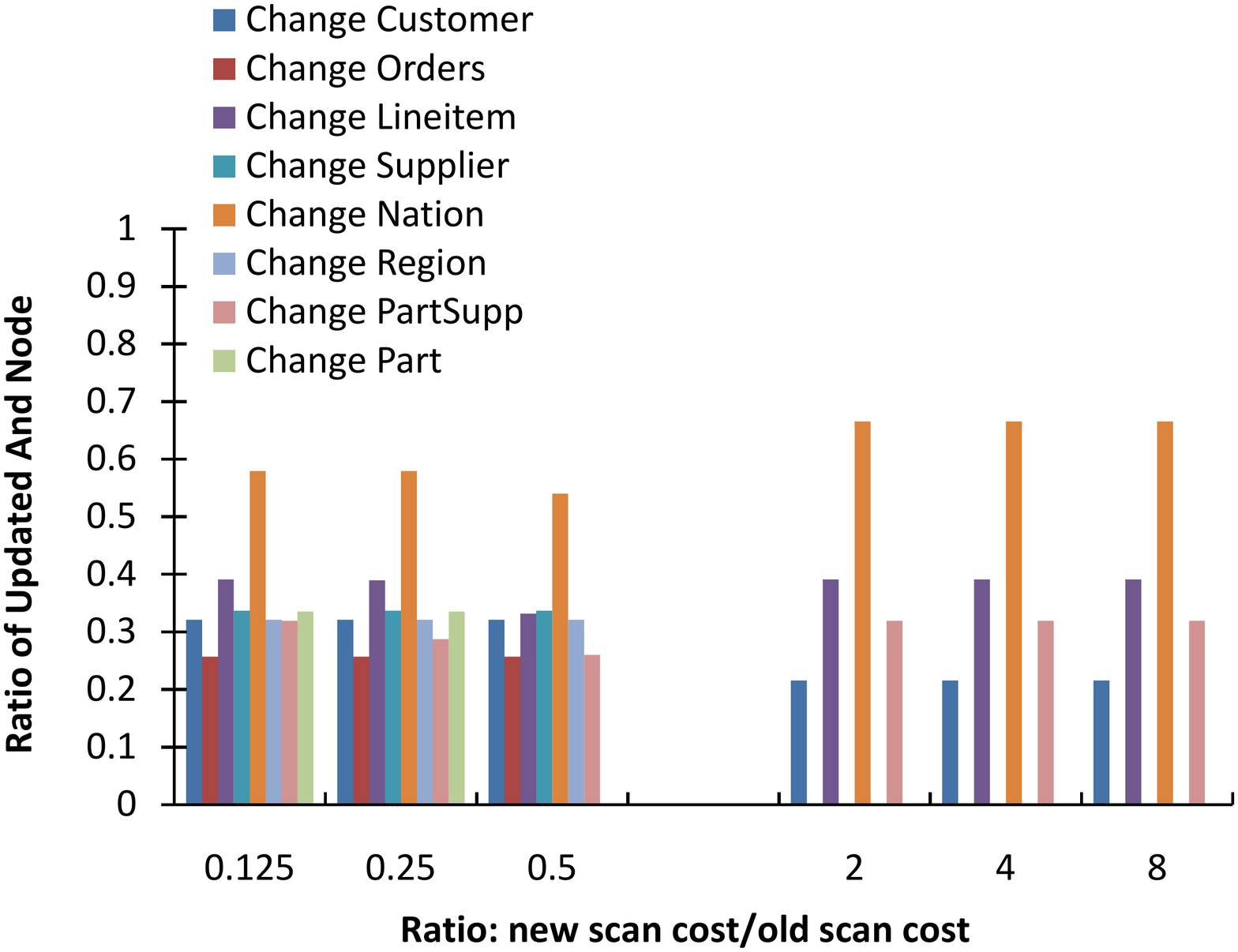}
\vspace{-3mm}
\small (c) Update ratio: plan alternatives
\end{center}
\end{minipage}
\caption{\small Declarative incremental re-optimization for query Q8Join when a base relation scan cost is changed \label{fig:incremental-Q8Join-scan-cost}}
\end{figure*}

\begin{figure*}[t]
\begin{minipage}[t]{2.25in}
\begin{center}
\includegraphics[width=2.25in]{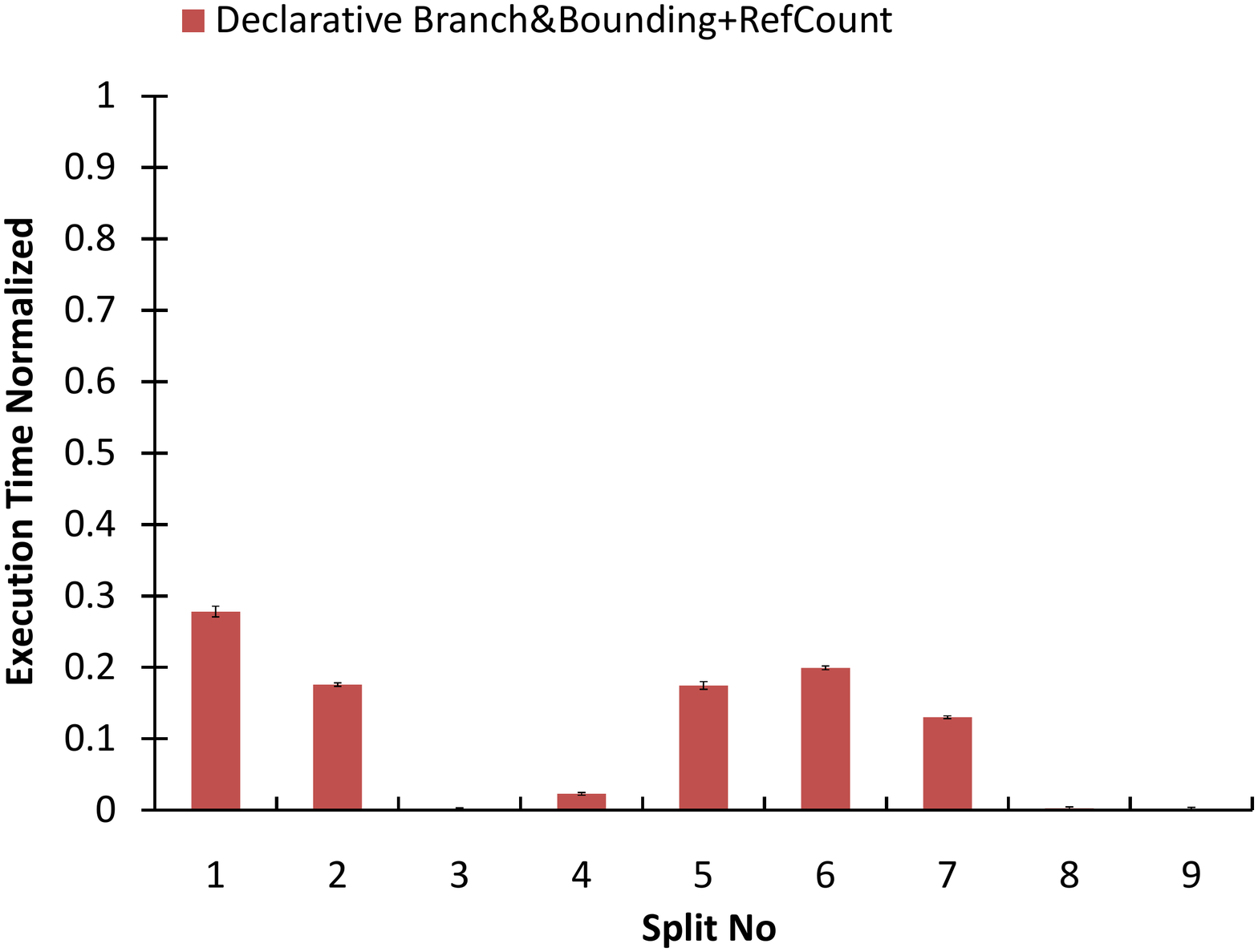}
\vspace{-3mm}
\small (a) Execution time (normalized to Volcano)
\end{center}
\end{minipage}
\hfill
\begin{minipage}[t]{2.25in}
\begin{center}
\includegraphics[width=2.25in]{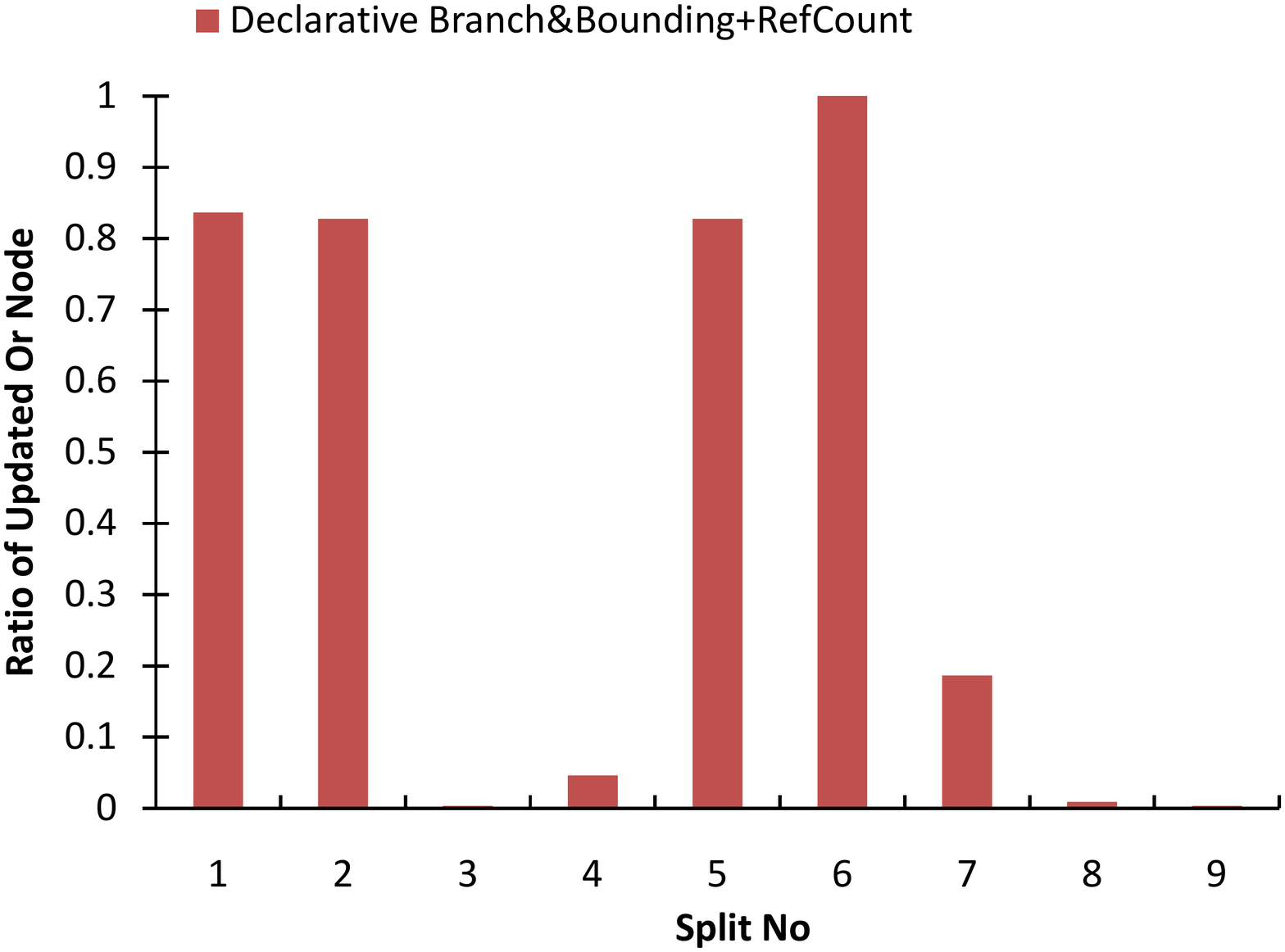}
\vspace{-3mm}
\small (b) Update ratio: plan table entries
\end{center}
\end{minipage}
\hfill
\begin{minipage}[t]{2.25in}
\begin{center}
\includegraphics[width=2.25in]{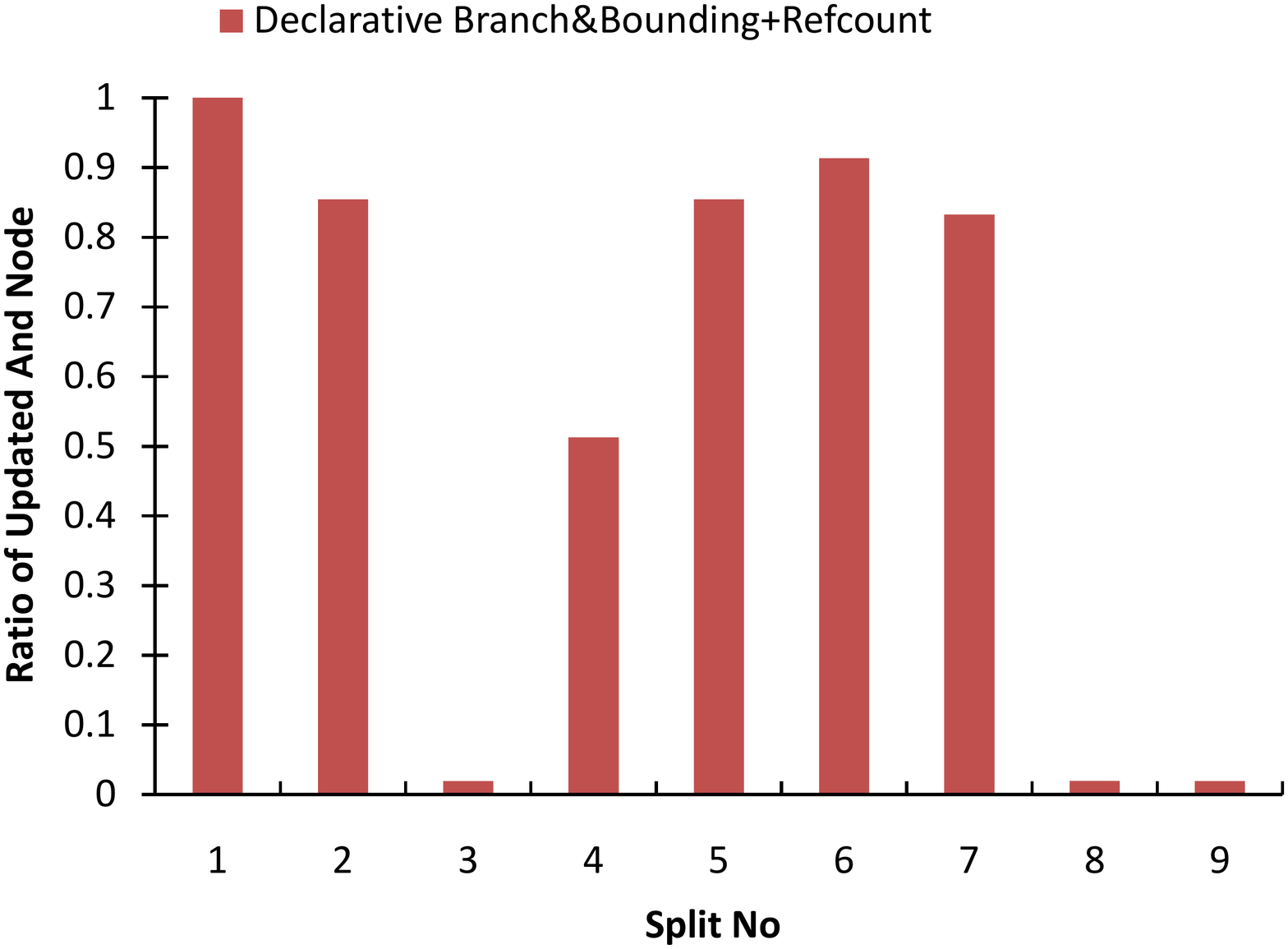}
\vspace{-3mm}
\small (c) Update ratio: plan alternatives
\end{center}
\end{minipage}
\caption{\small Declarative incremental re-optimization for TPCH Q8Join over real runtime conditions \label{fig:incremental-Q8Join-real-workload}}
\end{figure*}

\end{appendix}
}

\end{document}